\documentclass[%
 reprint,
superscriptaddress,
 amsmath,amssymb,
 aps,
]{revtex4-2}

\usepackage{graphicx}
\usepackage{hyperref}
\usepackage{color}
\usepackage{booktabs}
\usepackage{makecell}
\usepackage{soul}
\usepackage{amsmath,amsfonts,amsthm,bm} 
\usepackage{dcolumn}
\usepackage{bm}

\usepackage{verbatim}

\usepackage{lineno}

\begin{document}

\title{Topological phases, van Hove singularities, and spin texture in magic-angle twisted bilayer graphene in the presence of proximity-induced spin-orbit couplings}
\author{Yuting Tan}
\altaffiliation{ytan77@umd.edu}
\affiliation{Condensed Matter Theory Center and Joint Quantum Institute,
Department of Physics, University of Maryland, College Park, Maryland 20742, USA}
\author{Yang-Zhi Chou}
\affiliation{Condensed Matter Theory Center and Joint Quantum Institute,
Department of Physics, University of Maryland, College Park, Maryland 20742, USA}
\author{Fengcheng Wu}
\affiliation{School of Physics and Technology, Wuhan University, Wuhan 430072, China}
\affiliation{Wuhan Institute of Quantum Technology, Wuhan 430206, China}
\author{Sankar Das Sarma}

\affiliation{Condensed Matter Theory Center and Joint Quantum Institute,
Department of Physics, University of Maryland, College Park, Maryland 20742, USA}
\begin{abstract}

We investigate magic-angle twisted bilayer graphene (MATBG) with proximity-induced Ising and Rashba spin-orbit couplings (SOC) in the top layer, as recently achieved experimentally. Utilizing the Bistritzer-MacDonald model with SOCs, we reveal a rich single-particle topological phase diagram featuring topological flat bands across different twist angles and interlayer hopping energies. The evolution of Dirac cones and Chern numbers is examined to understand the topological phase transitions. We find that all phases can be achieved with an experimentally accessible SOC strength ($\sim$1 meV) in systems with angles very close to the magic angle. Furthermore, the van Hove singularity for each topological flat band splits in the presence of SOC, significantly altering the electronic properties. Additionally, we investigate the spin textures of each band in momentum space, discovering a skyrmion-like spin texture in the center of the moiré Brillouin zone, which is correlated with the topological phase transitions and can be tuned via the SOCs and an out-of-plane electric field. Our findings provide a comprehensive understanding of the topological flat bands, establishing a foundation for grasping the intrinsic and rich roles of SOCs in MATBG.

\end{abstract}

\maketitle


\begin{figure*}
    \centering
    \includegraphics[width=\textwidth]{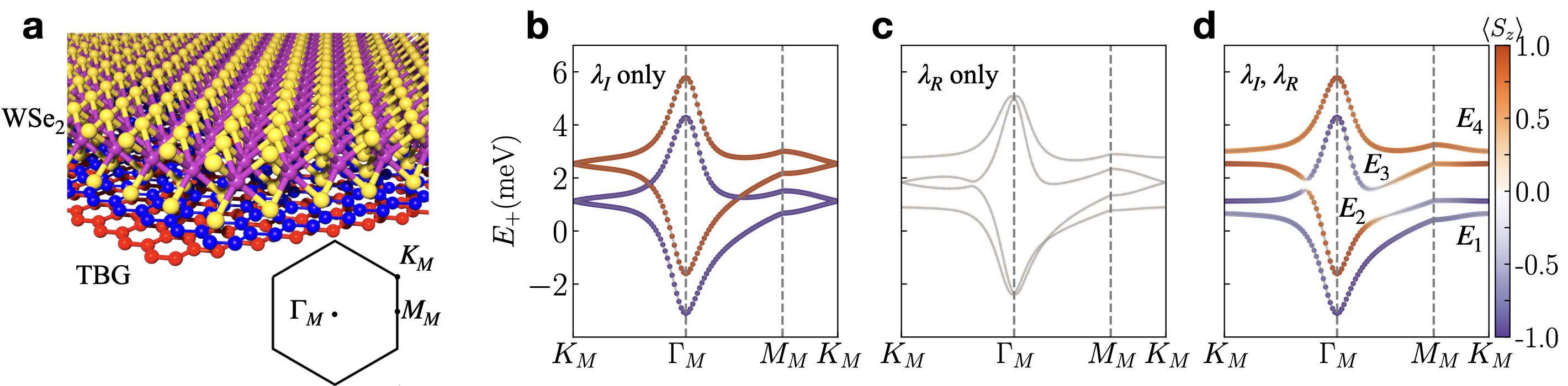}
\caption{\textbf{a}, Illustration depicting the crystal lattice of MATBG (blue and red) with a WSe$_2$ monolayer (yellow and purple) atop, capable of inducing SOC on the upper graphene layer (blue). \textbf{b-d} MATBG band structure of $\theta=1.05^\circ$. \textbf{b}, $\lambda_I = 3$ meV and $\lambda_R=0$. \textbf{c}, $\lambda_I = 0$ and $\lambda_R=3$ meV. \textbf{d}, $\lambda_I = \lambda_R=3$ meV. The color represents the expectation value of $S_z$, with orange indicating spin up and purple indicating spin down.}
\label{Fig:setup}
\end{figure*}


\section{\label{sec:introduction}Introduction}

Graphene materials, such as magic-angle twisted bilayer graphene (MATBG), bernal bilayer graphene, rhombohedral trilayer graphene, etc., possess strongly correlated and superconducting phases of matter \cite{CaoY2018a,CaoY2018,PolshynH2019,SharpeAL2019,JiangY2019,LuX2019,YankowitzM2019,KerelskyA2019,XieY2019,PolshynH2019,ChoiY2019,CaoY2020,SerlinM2020,AroraHS2020,WongD2020,ChoiY2021,OhM2021,LiuX2020,HaoZ2021,ZhouH2021,ZhouH2021a,ChoiY2021,ParkJM2021,ParkJM2021a,LiuX2021,ZhouH2022,KuiriM2022,ZhangY2023,HolleisL2023,SuR2023,PixleyJH2019,AndreiEY2020,BalentsL2020,AndreiEY2021,khosravian_moire-enabled_2024}. Their properties can be tuned through accessible external parameters, such as gating, straining, and twist angles, providing valuable opportunities to study topology and correlated physics \cite{LiY2019, parker_strain-induced_2021,NaimerT2021, ZollnerK2023a,zheng_gate-defined_2024}.   In addition, experiments have utilized the proximity effect between graphene and a transition metal dichalcogenide (TMD) layer (such as WSe$_2$), inducing proximity spin-orbit couplings (SOCs) in graphene and offering another method to control graphene-based materials \cite{avsar_spinorbit_2014,IslandJO2019,AroraHS2020,ZhangY2023,HolleisL2023,SuR2023,SunL2023,XieM2023,han_large_2024}. Interestingly, recent experiments have shown that observable superconductivity (SC) can be induced \cite{AroraHS2020,SuR2023,li2024tunable} or enhanced \cite{ZhangY2023,HolleisL2023} by the proximate $\text{WSe}_2$ layer. The interplay between SOC and SC in graphene is an active area of research \cite{chou2024topological,CurtisJB2023,Jimeno-PozoA2023,PantaleonPA2023,ParappurathA2023,WagnerG2023}.

In general, SOCs can significantly alter the electronic band structure in graphene and lead to many interesting features, such as quantum spin Hall \cite{kane_quantum_2005,kane_z_2_2005}, Rashba Edelstein effect \cite{lee_charge--spin_2022,ingla-aynes_omnidirectional_2022}, etc. The proximity-induced SOC effect is especially pronounced in MATBG, because its small bandwidth in the low-energy moir\'e bands.  By coupling the spin and orbital degrees of freedom of the electron, SOC can lift the spin degeneracies and create band splitting, making it an important ingredient in the construction of the topological phase diagram for MATBG. It can also significantly alter the electronic density of states (DOS) and potentially amplify the many-body correlation, which can induce interaction-driven phases as well as SC \cite{sherkunov_electronic_2018,isobe_unconventional_2018,liu_chiral_2018}. Furthermore, SOC fundamentally influences spin configurations in 2D materials, which have potential applications in spintronics and information storage due to their stability and manipulability by external fields \cite{han_graphene_2014,frank_emergence_2024,avsar_colloquium_2020,ahn_2d_2020}.

In this work, we study MATBG with proximity-induced SOCs in the top layer. In our previous work \cite{chou2024topological}, we focused on the formation of unconventional intervalley interband phonon-mediated superconductivity in topological flat bands induced by SOCs, in the presence of valley imbalance. In the current work, we further investigate the topological phase diagram at the single-particle level, revealing three distinct topological phases across different twist angles. We examine in detail the evolution of Dirac cones and Chern numbers across the topological phase transitions. Our object here is to provide a foundational understanding of MATBG with SOCs, aiming to lay the groundwork for further exploration into the intrinsic role of SOCs in this system.

Using the continuum Bistritzer-MacDonald (BM) model with Rashba and Ising SOCs \cite{WangT2020,haddad_twisted_2023}, our findings show that the SOCs significantly reconstruct the band structure, splitting the two flat minibands (without SOC) into four spin-split bands. Each band has its own pair of van Hove singularities (VHSs), leading to a total of eight VHSs in DOS per valley, while only four VHSs per valley are present without SOC. Additionally, we discover a skyrmion-like spin texture in momentum space, which can evolve when crossing three distinct topological phases. Furthermore, we demonstrate that the in-plane spin texture in momentum space can be tuned and, and the skyrmion-like feature can be further modified, by applying an out-of-plane electric field. Our results provide a systematic understanding of MATBG moir\'e bands in the presence of proximity-induced SOCs.

The rest of the paper is organized as follows: In Sec. \ref{Model}, we introduce the continuum moiré Hamiltonian, incorporating Ising and Rashba SOCs into the top layer. We then diagonalize the Hamiltonian to obtain the band structure across a range of SOCs and twist angles. The results are presented in Sec. \ref{SPD}, \ref{DOS} and \ref{ST}. Particularly, we construct the single-particle topological phase diagram across different angles and a range of values for SOCs in Sec. \ref{SPD}. We then show the effects of SOC on VHSs in Sec. \ref{DOS} and spin texture in Sec. \ref{ST_}, and how the spin texture can be tuned via applying an out-of-plane electric field to the graphene layers in Sec. \ref{STWE}. In Sec. \ref{Discussion}, we discuss the implications of our results. Appendices.~\ref{BC_CN},\ref{PD_different_hoping_energy},\ref{DOS_nosoc} complement the theory presented in the main text.


\begin{figure*}
    \centering
    \includegraphics[width=\textwidth]{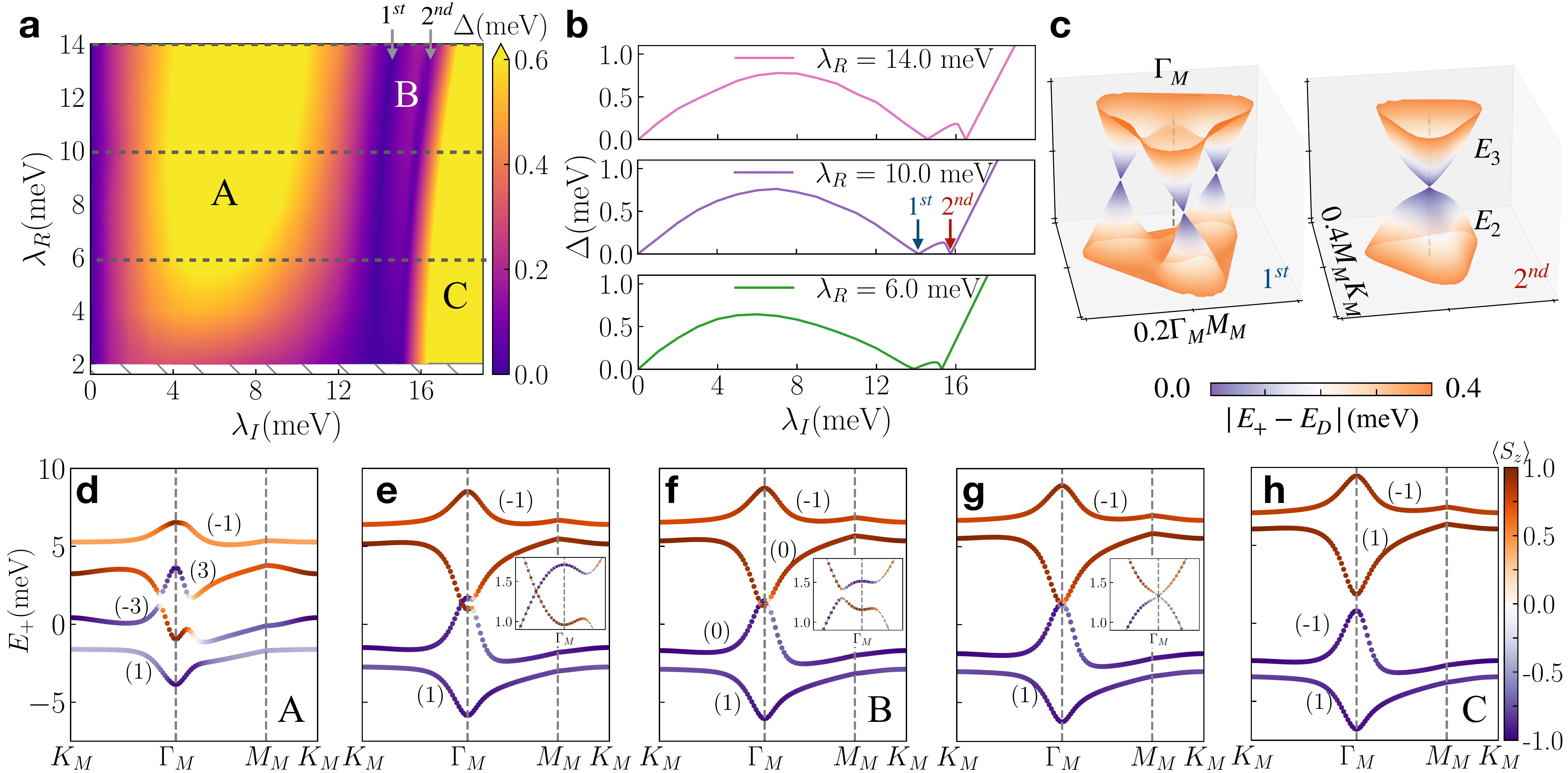}
\caption{\textbf{a}, Single-particle topological phase diagram at $\theta=1.05^\circ$, highlighting three distinct topological phases, A, B and C. Colors indicate the minimum direct gap between the two middle bands ($\Delta \equiv \min\left[E_3(\textbf{k}) - E_2(\textbf{k})\right]$) in the moiré Brillouin zone. \textbf{b}, Grey line-cuts from \textbf{a}, showing $\Delta$ reaching zero at the first and second phase transition points. \textbf{c}, Three Dirac cones between $E_2$ and $E_3$ emerge near $\Gamma_M$ at the first phase transition (left), while only one appears right at $\Gamma_M$ at the second transition (right), corresponding to the blue and red arrows in \textbf{b}, respectively. The grey dashed line indicates the $\Gamma_M$ point. \textbf{d-h}, Four flat minibands corresponding to $\lambda_I=6.00,14.16,15,15.76,18.00$ meV and $\lambda_R=10$ meV, respectively. The color represents the expectation value of $S_z$, with orange indicating spin up and purple indicating spin down. The inset plots in \textbf{e-g} enlarge the band structure around $\Gamma_M$.}
\label{Fig:phasediagram_1.05}
\end{figure*}


\section{Model} \label{Model}

As shown in Fig.~\ref{Fig:setup}, we consider MATBG-WSe$_2$ system, with proximity-induced Ising and Rashba SOCs on the top graphene layer, consistent with the experimental setup \cite{AroraHS2020}. In this system, the bottom and top graphene layers are rotated by angles $-\theta/2$ and $\theta/2$, respectively. A TMD layer, such as monolayer WSe$_2$, is placed on top of the upper graphene layer, inducing proximity effects that result in Ising and Rashba SOCs in the top layer of TBG \cite{wang_strong_2015,avsar_spinorbit_2014}. The single-particle physics of TBG with small $\theta$ can be described using a continuum moiré Hamiltonian—the Bistritzer-MacDonald model \cite{BistritzerR2011,WuF2018}—in which the low-energy Hamiltonian of the $+K$ valley is formulated as follows:

\begin{align}\label{Eq:H_0_BM}
    \mathcal{H}_{0,+}=\left[\begin{array}{cc}
        \hat{U}_{\theta/2}\left(\hat{h}_t^{(+)}(\textbf{k})+\hat{h}^{(+)}_{\text{SOC},t}\right)\hat{U}_{\theta/2}^{\dagger} & \hat{T}^{\dagger}(\textbf{x})\\[2mm]
        \hat{T}(\textbf{x}) & \hat{U}_{\theta/2}^{\dagger}\hat{h}_b^{(+)}(\textbf{k})\hat{U}_{\theta/2}
    \end{array}\right],
\end{align}
where $\theta$ represents the twist angle, while the subscripts $t$ and $b$ indicate the top and bottom layers, respectively. In the above expression, $\hat{h}_t^{(+)}$ and $\hat{h}_b^{(+)}$ stand for the isolated $+K$ valley Dirac Hamiltonian of the top and bottom layers, defined by $\hat{h}_l^{(+)}(\textbf{k})=v_F\left(\textbf{k}-\bm{\kappa}_l\right)\cdot\bm{\sigma}$ for $l=t,b$. $v_F\approx5.944 \text{eV\AA}$ denotes the Dirac velocity of monolayer graphene \cite{SongZ2019}. $ \bm{\sigma}=(\sigma_x,\sigma_y)$, where $\sigma_{\mu}$ represents the $\mu$-component Pauli matrix for the sublattice, and $\bm{\kappa}_t$ ($\bm{\kappa}_b$) stands for the rotated $+K$ valley point of the top (bottom) layer. To incorporate the rotation of spinors in the Dirac Hamiltonian, we apply the sublattice rotation matrix $\hat{U}_{\theta/2}=e^{i(\theta/4)\sigma_z}$. The interlayer tunneling between two twisted layers induces a spatially varying potential, described by $\hat{T}(\bm{x})=\hat{t}_0+\hat{t}_{1}e^{-i\bm{b}_+\cdot\bm{x}}+\hat{t}_{-1}e^{-i\bm{b}_-\cdot\bm{x}}$, where $\hat{t}_j=w_0\sigma_0+w_1[\cos(2\pi j/3)\sigma_x+\sin(2\pi j/3)\sigma_y]$, $\bm{b}_{\pm}=[4\pi/(\sqrt{3}a_M)]\left(\pm1/2,\sqrt{3}/2\right)$, and $a_M$ represents the moiré lattice constant. The interlayer hopping parameters, $w_1\approx 110$ meV, $w_0=0.8w_1$. The results with different $w_0/w_1$ values are also discussed. In our numerical calculations, we consider a $9 \times 9$ momentum grid in the plane-wave expansion for this continuum model, with $\Gamma_M$ at the center, and we explicitly checked that a larger grid does not change the band structure within our desired resolution. 

The proximity-induced SOC terms in the $\tau K$ valley are given by \cite{AroraHS2020,NaimerT2021,NaimerT2023,WangT2020,ScammellHD2023a,ScammellHD2024}

\begin{align}\label{Eq:h_SOC}
    \hat{h}_{\text{SOC},t}^{(\tau)}=&\frac{\lambda_I}{2}\tau \sigma_0s_z+\frac{\lambda_R}{2}\left(\tau\sigma_xs_y-\sigma_ys_x\right),
\end{align}
where $\lambda_I$ ($\lambda_R$) represents the strength of Ising (Rashba) SOC and $s_{\mu}$ is the $\mu$-component Pauli matrix for the spins.

The presence of SOCs alters the underlying symmetry of MATBG. The overall system (including both $+K$ and $-K$ valleys) obeys the spinful time-reversal symmetry, $\mathcal{T}_s=i\tau_xs_y\mathcal{K}$, where $\tau_x$ is the $x$-component Pauli matrix for the valley and $\mathcal{K}$ is the conjugation operator. Thus, we expect the moir\'e bands of the two valleys satisfy $\mathcal{E}_{+,b}(\bm{k})=\mathcal{E}_{-,b}(-\bm{k})$ and $\Psi_{-,b,-\bm{k}}=i\tau_xs_y\Psi_{+,b,\bm{k}}^*$ for the energies and wavefunctions of the $b$th band, respectively. The Hamiltonian also preserves $\mathcal{C}_{3z}$ rotation symmetry around the out-of-plane $z$ axis, but not the $\mathcal{C}_{2z}$ rotation symmetry, which is $i\tau_x\sigma_xs_z$ \cite{WangT2020}. Furthermore, to characterize the topology of the system, we calculate the Berry curvature $\Omega$ by numerically computing the Wilson loops in the momentum space with the rhombus grid, which is described in detail in the Appendix.~\ref{BC_CN}.

\begin{figure*}
    \centering
    \includegraphics[width=0.9\textwidth]{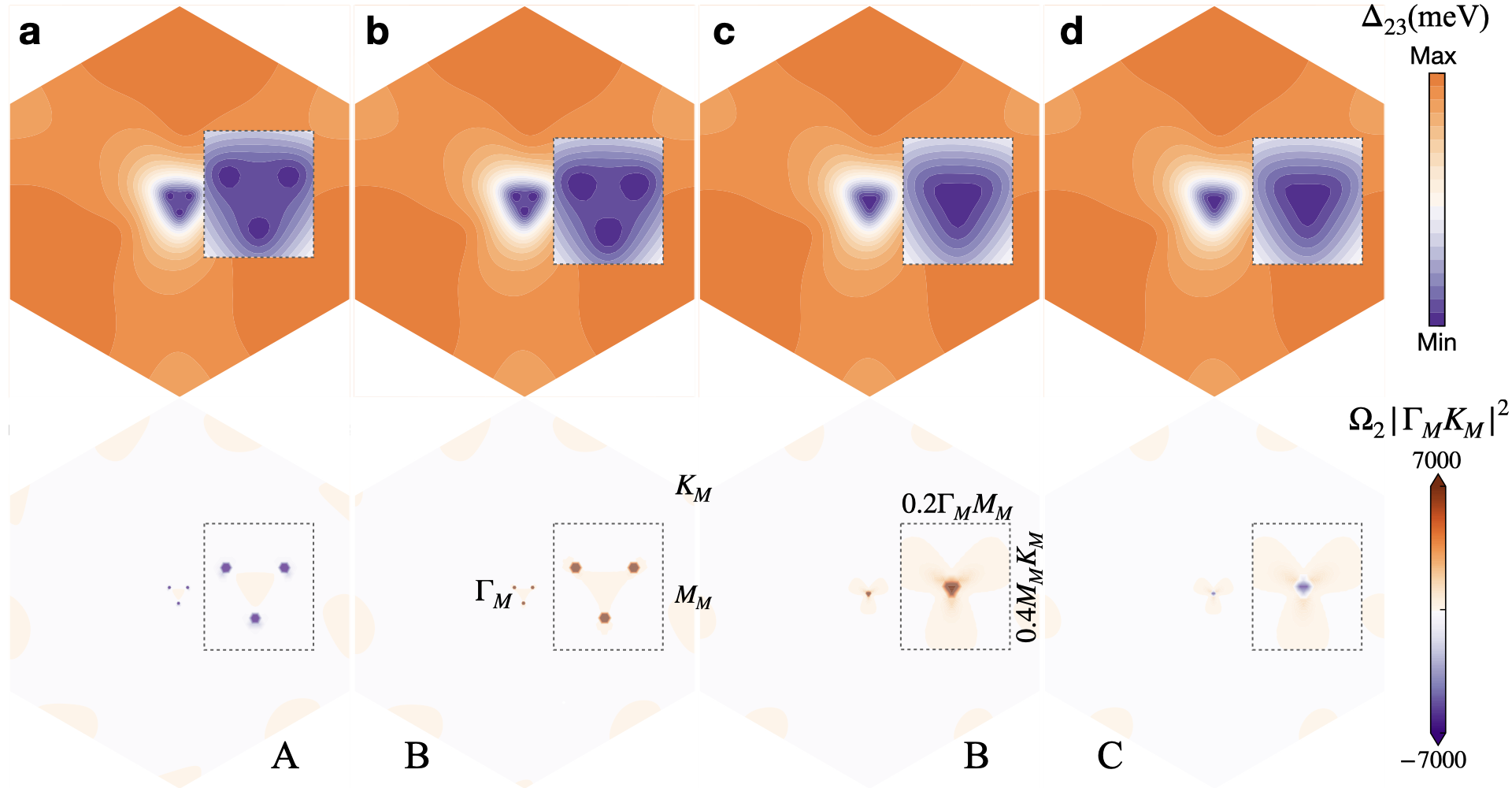}
\caption{Direct gap $\Delta_{23}(\textbf{k})=E_3(\textbf{k})-E_2(\textbf{k})$ (upper panel) and Berry curvature of the second band $\Omega_2$ (lower panel) near the first phase transition boundary $A\leftrightarrow B$ and the second boundary $B\leftrightarrow C$, at $\theta=1.05^\circ,\ \lambda_R=10$ meV. \textbf{a}, \textbf{b}, $\lambda_I=14.0, 14.2$ meV correspondingly.  \textbf{c}, \textbf{d}, $\lambda_I=15.72, 15.78$ meV correspondingly. $N_k=150^2$ in MBZ.}
\label{Fig:DeltaandBC}
\end{figure*}

We now address some subtleties concerning our model. First, the BM model is valid in the continuum limit and at low-energy and long wavelength. This is reasonable because, in this work, we only focus on the cases with isolated flat minibands with narrow bandwidth. Second, the large lattice constant mismatch between graphene (lattice constant 0.246 nm) and WSe$_2$ (lattice constant 0.353 nm) result in a very weak moire effect (moiré lattice constant $\sim 1$ nm \cite{AroraHS2020}), which we ignore. The proximity-induced SOC can have some spatial variations, but it averaging out quite rapidly within the TBG moire lattice constant $\sim 10$ nm. Third, in our model, we use a simplified expression for the $\lambda_R$ term \cite{NaimerT2021,NaimerT2023,ZollnerK2023a,frank_emergence_2024}. The Rashba SOC will naturally be present when the mirror symmetry is broken, such that it can not only be induced by the proximate layer, but also an electric field perpendicular to the graphene sample. The proximity-induced SOC strengths depend on the relative angle between the SOC layer and the top graphene layer \cite{LiY2019,NaimerT2021,DavidA2019,ChouYZ2022,ZollnerK2023a}, while in the latter case, it depends on the strength of the electric field \cite{avsar_colloquium_2020}. $\lambda_I$ and $\lambda_R$ can be as large as $\sim20$ meV in calculations and fitting to experiments \cite{NaimerT2023,LiY2019,wang_strong_2015,wang_origin_2016,IslandJO2019,wang_quantum_2019}, which is comparable to the bandwidth of MATBG. In general, it is well-known that an accurate knowledge of SOC from first principles calculations is a huge challenge, and it is more appropriate to obtain them by comparing with experiments. Here, we treat $\lambda_I$ and $\lambda_R$ as free parameters and study a range of the twist angles.

\section{Single particle phase diagram, Berry phase and Chern number\label{SPD}}

We diagonalize the single-particle Hamiltonian $\mathcal{H}_{0,+}$ [Eq.~(\ref{Eq:H_0_BM})] in momentum space. Without SOC, the isolated flat minibands can be found within twist angle $\theta=0.97^\circ\sim 2.5^\circ$, and the smallest bandwidth (approximately 1.1 meV) occurs around a twist angle of $\theta=1.08^{\circ}$, which we define as the magic angle. 
In the scenario where $\lambda_I\neq0$ and $\lambda_R=0$ (Fig. \ref{Fig:setup}\textbf{b}), the bands corresponding to spin-up (orange) and spin-down (purple) electrons are shifted according to the valley-spin Zeeman field described by Eq. (\ref{Eq:h_SOC}). The two Dirac cones at $K_M$ are approximately separated by $\lambda_I/2$. When $\lambda_I=0$ and $\lambda_R\neq0$ (Fig. \ref{Fig:setup}\textbf{c}), there is a single Dirac band touching at the $K_M$ ($K'_M$ ) point. In the Rashba-only case, the expectation value $\langle S_z\rangle$ of these four bands is exactly zero. This indicates that there is no net spin polarization in the $z$-direction due to the Rashba SOC. As illustrated in Fig. \ref{Fig:setup}\textbf{d}, the combination of both Ising and Rashba SOCs significantly reconstructs the moir\'e bands, generically resulting in four spin-split bands around charge neutrality, which are called $E_1, E_2, E_3, E_4$, from bottom to top. In this case, spin-up and spin-down states are mixed, leading to a more complex band structure compared to the individual SOC cases.

To characterize the topological properties of the moir\'e bands induced by SOCs, we further extract the Chern number $\mathcal{C}$ (Eq.~\ref{Eq:cn}) and Berry curvature $\Omega$ (Eq.~\ref{Eq:bc}) of each band by numerically computing Wilson loops in a momentum-space rhombus grid. The Chern number of the $b$th band in the $\pm K$ valley is denoted by $\mathcal{C}_{\pm, b}$. Due to the time-reversal symmetry, $\mathcal{C}_{-, b} = -\mathcal{C}_{+, b}$. For simplicity, we present our results for the $+K$ valley only.

\begin{figure*}
    \centering
    \includegraphics[width=0.9\textwidth]{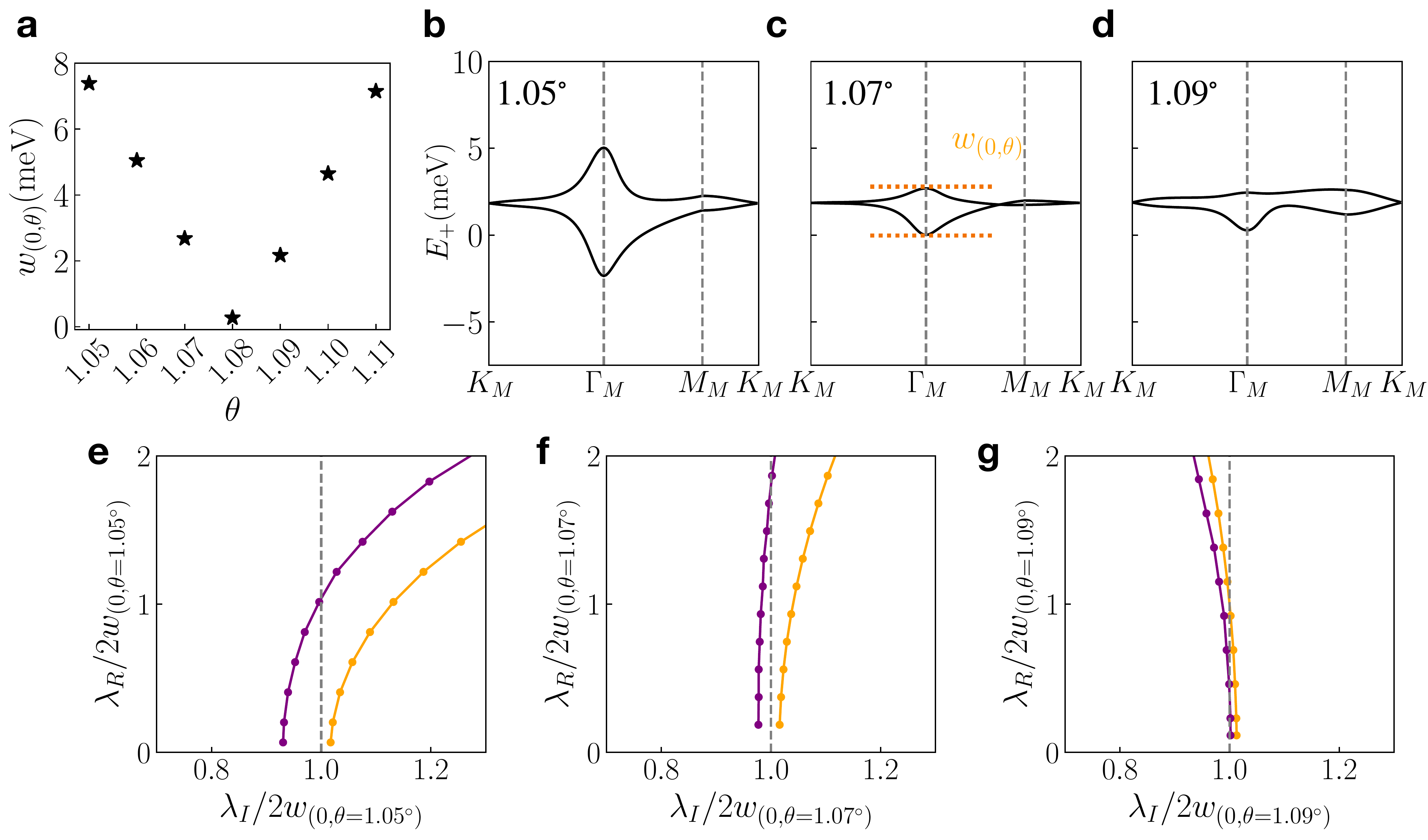}
\caption{\textbf{a}, Energy difference $w_{(0,\theta)}$ at the $\Gamma_M$ point between the conduction and valence bands for a given twist angle $\theta$ in the absence of SOC.  Band structures (\textbf{b-d}) without SOC,  and the phase diagrams (\textbf{e-g}) with rescaled SOCs, for $\theta=1.05^\circ$, $\theta=1.07^\circ$, and $\theta=1.09^\circ$, respectively.}
\label{Fig:phasediagrams}
\end{figure*}

For clarity and without loss of generality, we first present our results at $\theta=1.05^{\circ}$, a convenient choice for presenting the phase diagram. We will present the phase diagrams in a more compact way with proper rescaled SOC parameters for other angles $\theta$ (see Fig.~\ref{Fig:phasediagrams}) and for the other choice of $w_0/w_1=0.4$ (see Fig.~\ref{Fig:phasediagram_1.05_w0_w1_0_4}), where the exact locations of the phase boundaries are modified.  As shown in Fig.~\ref{Fig:phasediagram_1.05}\textbf{a}, we identify three distinct phases: A, B, and C, which are characterized by different sets of Chern numbers $(\mathcal{C}_{+,1}, \mathcal{C}_{+,2}, \mathcal{C}_{+,3}, \mathcal{C}_{+,4}) = (1, -3, 3, -1)$, $(1, 0, 0, -1)$, and $(1, -1, 1, -1)$, respectively. The Chern numbers change sign with a negative $\lambda_I$, while the sign of $\lambda_R$ does not affect the Chern numbers. It is important to emphasize that the critical Ising/Rashba SOCs for the topological phase transitions are parameter-dependent, e.g., twist angle $\theta$, $w_0/w_1$, etc. We provide representative results, but the numbers of parameters are simply too numerous to be completely comprehensive with respect to all the relevant parameters.

Phases A \cite{WangT2020,LinJX2022,BhowmikS2023}, and C \cite{WangT2020} have previously been reported, and phase B is also reported in \cite{WangT2020} in the presence of finite sublattice splitting due to the SOC layer, which we ignore in this study. Here and in our previous paper \cite{chou2024topological}, we point out that the phase B phase can be realized with only Ising and Rashba SOCs. By varying $\lambda_R$ and $\lambda_I$, $\mathcal{C}_{+,1}$ and $\mathcal{C}_{+,4}$ remain unchanged (as long as $\lambda_I>0$ and $\lambda_R\neq0$), and topological transitions occur only in the middle two bands, $E_2$ and $E_3$, associated with the emergence of Dirac nodes. Therefore, we plot the minimum of the direct gap between $E_2$ and $E_3$, $\Delta\equiv\text{min}\left[E_3(\textbf{k})-E_2(\textbf{k})\right]$ in moir\'e Brillouin zone (MBZ) in Fig.~\ref{Fig:phasediagram_1.05}\textbf{a}.
The A-B and B-C phase boundaries are where $\Delta$ reaches zero, shown as two deep-purple lines, and pinpointed by two gray arrows.  The critical Ising/Rashba SOCs here depend on the twist angle $\theta$, $w_0/w_1$, etc. As we show later, the critical Ising SOCs actually are roughly twice $w_{0,\theta}=\left[E_3(\textbf{k}=\Gamma_M)-E_2(\textbf{k}=\Gamma_M)\right]$, which is the energy difference between $E_2$ and $E_3$ at the $\Gamma_M$ point without SOC.
So the critical SOCs reduce significantly when the twist angle is close to the magic angle $\theta=1.08^\circ$, which implies that phases A, B and C are all experimentally observable within a realistic range of SOCs (0-3 meV) \cite{ZhangY2023}.

Figure.~\ref{Fig:phasediagram_1.05}\textbf{b} shows $\Delta$ as a function of $\lambda_I$ along the gray dashed lines, for three representative $\lambda_R$, in Fig.~\ref{Fig:phasediagram_1.05}\textbf{a}, which clearly displays two gap-closing points:$\lambda_I^l$ (blue arrow) and $\lambda_I^h$ (red arrow). The minimum of direct gap, $\Delta$, is generally pretty small in phase B. The maximum of $\Delta$ in phase B and the width of phase B ($\lambda_I^h-\lambda_I^l$) increase, with the increase of Rashba SOC, $\lambda_R$. But we do not see the trend of  $\lambda_I^h$ and $\lambda_I^l$ merging at $\lambda_R\rightarrow0$ limit. Thus, the two purple lines divide the phase diagram into three regions, with gaps vanishing at the individual topological quantum phase transition points separating the regimes A, B, C.

Specifically, for $\lambda_R=10$ meV (the middle plot of Fig.~\ref{Fig:phasediagram_1.05}\textbf{b}), $\Delta$ reaches zero at $\lambda_I=14.16$ meV (marked by the blue arrow) and $15.76$ meV (marked by the red arrow), which correspond to the A-B and B-C phase transitions respectively. At the first phase transition (blue arrow), three Dirac cones near the $\Gamma_M$ point are manifest, as shown in the left plot of Fig.~\ref{Fig:phasediagram_1.05}\textbf{c}. This corresponds to the fact that the changes in $\mathcal{C}_{+,2}$ and $\mathcal{C}_{+,3}$ are $\pm 3$ during the A-B phase transition. One of the three Dirac cones is located almost right on top of the $\Gamma_M K_M$ line. In contrast, only one Dirac cone right at $\Gamma_M$ appears in the B-C phase transition (red arrow), as shown in the right plot of Fig.~\ref{Fig:phasediagram_1.05}\textbf{c}, explaining why the changes in $\mathcal{C}_{+,2}$ and $\mathcal{C}_{+,3}$ are only $\pm 1$. Importantly, as long as the Dirac point is not located at $\Gamma_M$, the symmetry $C_3$ ensures that the number of Dirac cones is $3$ and the change in Chern number must be $\pm 3$ \cite{FangC2012}.

Figure.~\ref{Fig:phasediagram_1.05}\textbf{d}-\ref{Fig:phasediagram_1.05}\textbf{h} show the band structures at $\lambda_R=10$ meV, with different choices of $\lambda_I$, accrosing phase A, B and C. The effect of $\lambda_I$ on the bands is notable: smaller values of $\lambda_I$ lead to more significant spin mixing (Fig.~\ref{Fig:phasediagram_1.05}\textbf{d}). On the other hand, in the large $\lambda_I$ limit, the upper bands $E_3$ and $E_4$ are almost completely spin-up, whereas the lower bands $E_1$ and $E_2$ are almost completely spin-down (Fig.~\ref{Fig:phasediagram_1.05}\textbf{h}), which is similar to the $\lambda_I$-only case in Fig.~\ref{Fig:setup}\textbf{b}. There are always direct band gaps between the four mini bands, except at $\lambda_I=14.16$ meV (Fig.~\ref{Fig:phasediagram_1.05}\textbf{e}) and $15.76$ meV (Fig.~\ref{Fig:phasediagram_1.05}\textbf{g}), where the direct gap between $E_2$ and $E_3$, $\Delta_{23}$, closes. Fig.~\ref{Fig:phasediagram_1.05}\textbf{e} clearly shows that a Dirac cone (one of the three Dirac cones) is located almost right on top of the $\Gamma_M K_M$ line, while Fig.~\ref{Fig:phasediagram_1.05}\textbf{g} display one Dirac cone right on top of the $\Gamma_M$ point. In phase B (Fig.~\ref{Fig:phasediagram_1.05}\textbf{f}), the direct band gap $\Delta_{23}$ is quite small, which is enlarged in the inset to highlight the detailed band structure and the gap closure. 

\begin{figure}
    \centering
    \includegraphics[width=0.48\textwidth]{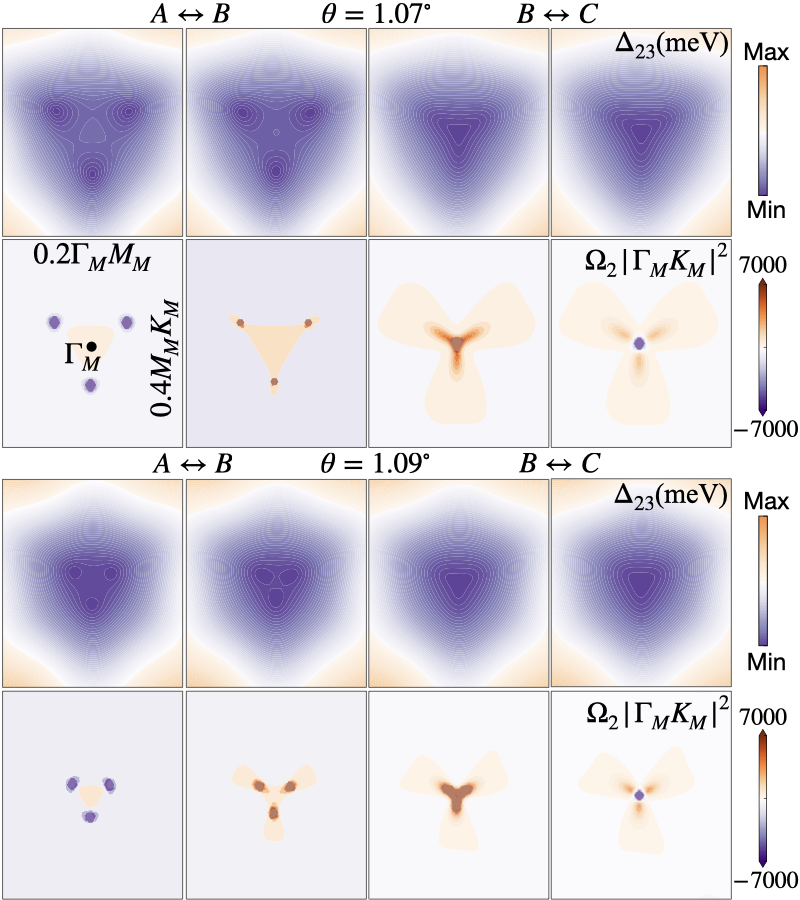}
\caption{Direct gap $\Delta_{23}(\textbf{k})=E_3(\textbf{k})-E_2(\textbf{k})$ (first panel) and Berry curvature of the second band $\Omega_2$ (second panel) around $\Gamma_M$ in momentum space,  near the first phase transition boundary $A\leftrightarrow B$ and the second boundary $B\leftrightarrow C$, at $\theta=1.07^\circ,\lambda_R=10$ meV. Here the critical $\lambda_I^l=5.37$ meV for A-B transition and $\lambda_I^h=5.91$ meV for B-C transition. $\lambda_I=5.30,5.39,5.90,5.94$ meV from left to right, correspondingly. The lower two panels are for $\theta=1.09^\circ,\lambda_R=10$ meV, with $\lambda_I^l=3.96$ meV and $\lambda_I^h=4.10$ meV. $\lambda_I=3.94,4.00,4.08,4.12$ meV from left to right.  }
\label{Fig:Diraccons107109}
\end{figure}

Figure.~\ref{Fig:DeltaandBC} shows the contour plots of $\Delta_{23}(\textbf{k})$ and the Berry curvature of the second band $\Omega_2(\textbf{k})$ in the MBZ, near the A-B (panels \textbf{a} and \textbf{b}) and B-C (panels \textbf{c} and \textbf{d}) phase transitions, respectively. Around the A-B phase transition, there are three minima in the direct gap $\Delta_{23}$ located near $\Gamma_M$. When $E_3\left(\textbf{k}\right)-E_2\left(\textbf{k}\right)\rightarrow0$ at a certain $\textbf{k}$ point, Berry curvature tends to diverge at that point. Therefore $\Omega_2$ in the lower panel tends to diverge negatively in phase A, but positively in phase B, confirming the existence of a topological phase transition. On the other hand, near the B-C phase transition, there is only one minimum of $\Delta_{23}$ located directly at $\Gamma_M$, and $\Omega_2$ at $\Gamma_M$ tends to diverge positively in phase B and negatively in phase C. Generally speaking, the Berry curvature, $\Omega_2$ and $\Omega_3$ are extremely localized around the Dirac points near the phase boundaries. On the other hand, the Berry curvature for the first and the fourth bands, $\Omega_1$ and $\Omega_4$, are much smaller in magnitude but still not flat enough to be considered similar to Landau levels, even though their variations are significantly less pronounced than in the second and third bands.

We also investigate the twist angle dependence of the single-particle phase diagrams. Similar to $\theta=1.05^\circ$, we identify three distinct topological phases—A, B, and C—for twist angles $\theta=1.07^\circ$ and $\theta=1.09^\circ$.  In fact, the critical $\lambda_I$ for B-C transition in $\lambda_R\rightarrow0$ limit should be the value that just separates two spin-up bands and two spin-down bands (as in Fig.~\ref{Fig:phasediagram_1.05}\textbf{g}). This means the transition between phases B and C roughly follows the relation $\lambda_I^h = 2w_{(0,\theta)}$, where $w_{(0,\theta)}$ is the energy difference at the $\Gamma_M$ point between the conduction and valence bands for a given twist angle $\theta$ in the absence of SOC. We first extract $w_{(0,\theta)}$ for different twist angels, as shown in Fig.~\ref{Fig:phasediagrams}\textbf{b}-\ref{Fig:phasediagrams}\textbf{d}, which are summarized in Fig.~\ref{Fig:phasediagrams}\textbf{a}. The minimum of $w_{(0,\theta)}$ occurs roughly near the magic angle $\theta=1.08^\circ$. We present the phase diagrams for twist angles $\theta=1.05^\circ$, $\theta=1.07^\circ$, and $\theta=1.09^\circ$ in Fig.~\ref{Fig:phasediagrams}\textbf{e}-\ref{Fig:phasediagrams}\textbf{g}, using rescaled Ising and Rashba SOCs, $\lambda_I / 2w_{0,\theta}$ and $\lambda_R / 2w_{0,\theta}$. In these phase diagrams, the B-C phase boundaries are determined by $\Delta < 10^{-4}$ meV, whereas the A-B phase boundaries are marked by changes in Chern numbers. This is because determining the exact location of Dirac points at the A-B boundaries is challenging, as opposed to the Dirac node being precisely at $\Gamma_M$ for the B-C boundaries. 
Nevertheless, we can see that the two boundaries in the $\lambda_R\rightarrow0$ limit trace the quantity $2w_{(0,\theta)}$.

When $\lambda_R$ increases, the width of phase B broadens, both for $\theta<1.08^\circ$ and when $\theta>1.08^\circ$. However, the width of phase B is narrower when $\theta$ is closer to the magic angle, and the "turning" of phase B at large Rashba SOCs behaves differently when passing through the magic angle. 
This feature may be related to the band inversion that occurs when crossing the magic angle, which needs further investigation. 
Exactly at the magic angle $\theta=1.08^\circ$, $2w_{(0,\theta)}=0.54$ meV, which is quite small, making phase C accessible in the experiment. On the other hand, around the magic angle, phase B is so narrow that it is difficult to observe both in experiment and theory. Similarly to $\theta=1.05^\circ$, for $\theta=1.07^\circ,1.09^\circ$, three Dirac cones near the A-B phase transition can be observed near $\Gamma_M$, while only one Dirac cone emerges directly at $\Gamma_M$ during the B-C phase transition, as shown in Fig.~\ref{Fig:Diraccons107109}. 
We also explore the phase diagram with $w_0/w_1=0.4$ and $\theta=1.05^\circ$, as shown in Fig.~\ref{Fig:phasediagram_1.05_w0_w1_0_4} in the Appendix.~\ref{PD_different_hoping_energy}. The width of phase B is much smaller than that in the $w_0/w_1=0.8$ case. In fact, phase B is extremely narrow in the chiral limit ($w_0/w_1=0$). The details for the topological phase diagram for other $w_0/w_1$ value are given in the Appendix.~\ref{PD_different_hoping_energy}.

In summary, we find three distinct topological phases—A, B, and C—across different twist angles and values of $w_0/w_1$. The transitions between these phases are marked by notable changes in the electronic properties, such as the number and position of Dirac cones near the $\Gamma_M$. Specifically, the transition between phases A and B is characterized by the presence of three Dirac cones near $\Gamma_M$, while the transition from B to C is marked by a single Dirac cone at $\Gamma_M$. The boundary between phases B and C is approximately determined by the energy difference of bands at $\Gamma_M$ without SOC. Phase A generally occurs with a small Ising SOC, while phase C can be observed with large Ising SOC or relatively small Ising SOC near the magic angle. Phase B is not easily found close to the magic angle or in the chiral limit due to its narrow width, but it can be more accessible in the presence of a large Rashba SOC. Nevertheless, the Ising SOC required for phases B and C is still experimentally accessible near the magic angle.

\section{Density of States, Van Hove singularity\label{DOS}}


\begin{figure*}
    \centering
    \includegraphics[width=0.9\textwidth]{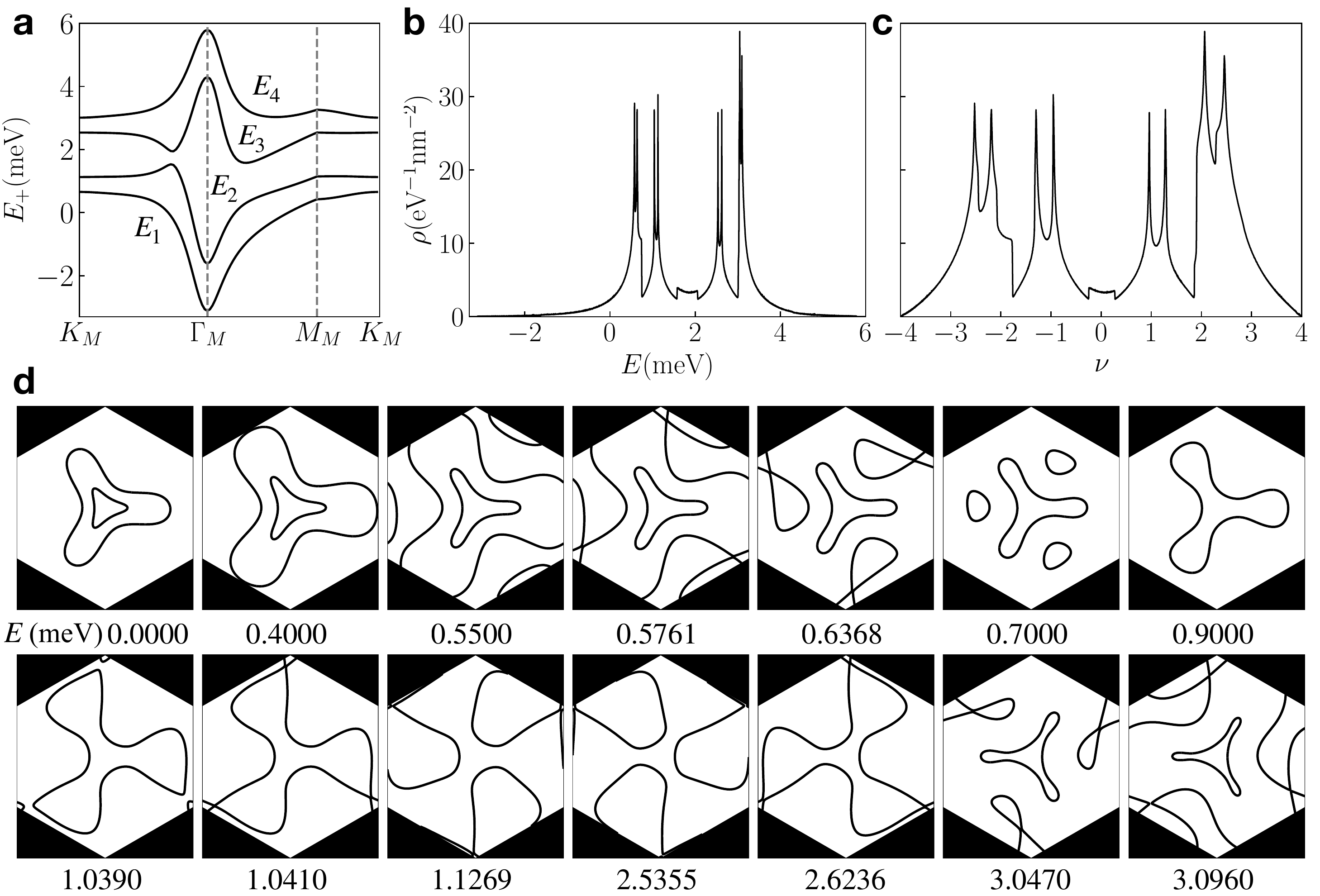}
\caption{\textbf{a}, Band structure for $\theta=1.05$ with $\lambda_I=\lambda_R=3$ meV. \textbf{b}, Density of state per valley as a function of energy $E$ and \textbf{c}, occupation number $v$. \textbf{d} Band structure contour plots at different energy $E$. The VHSs locate at $E=0.5761, 0.6368,1.0410,1.1269,2.5355,2.6236,3.0470,3.0960$ meV.}
\label{Fig:withsoc}
\end{figure*}


\begin{figure*}
    \centering
    \includegraphics[width=0.9\textwidth]{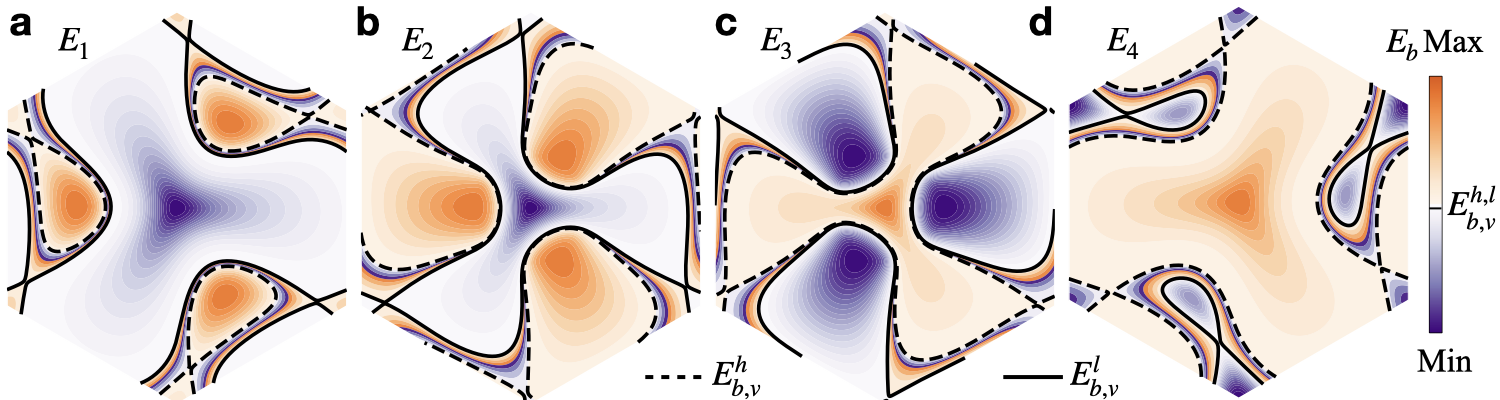}
\caption{Contour plots for $E_1,E_2,E_3,E_4$ in \textbf{a},\textbf{b},\textbf{c},\textbf{d}, respectively. The solid lines indicate the contour line plot at $E=E_{b,v}^l$, the low VHS for each band,  while the dashed lines indicate the contour line at $E=E_{b,v}^h$, the high VHS for each band. The orange and purple regions indicate the hole and electron pockets correspondingly. For each band, in between $E=E_{b,v}^l$ and $E=E_{b,v}^h$, there are multiple type of pockets.}
\label{Fig:withsoc_color}
\end{figure*}


In MATBG, turning on interlayer tunneling between the layers produces avoided crossings, leading to saddle points in the moir\'e minibands. Saddle points are locations in momentum space where an energy band reaches minima and maxima along orthogonal directions. These saddle points create significantly enhanced peaks in DOS, which are easily identified in scanning tunneling spectroscopy studies and are referred to as VHSs. 

In this section, we investigate the impact of SOC on the DOS and analyze its effect on the VHS in MATBG. When a VHS is close to the Fermi energy, the increased DOS amplifies the many-body correlation, resulting in various ordering instabilities, such as density waves and superconductivity at low temperatures. The DOS can be calculated using the following equation:

\begin{equation}
    \rho\left(E\right)=\sum_{i=1}^N\frac{1}{S}\sum_{j=1}^{N_k}\delta\left(E-E_i\left(k_j\right)\right)
\end{equation}
where $N$ is the number of bands; $N_k=3\times10^6$, the number of k points in the first MBZ; $S=N_k\times\Omega_0$, and $\Omega_0=\frac{\sqrt{3}}{2}a^2_M$, which is the real space moir\'e unit cell; and $\delta\left(E-E_i\left(k_j\right)\right)\approx\frac{1}{\pi}\frac{\gamma}{\left(E-E\left(k\right)\right)^2+\gamma^2}$, with $\gamma=0.0005$ meV, which is comparable to the mean energy level spacing $\sim0.00058$ meV. 


For comparison, we first discuss the DOS for the non-SOC case, fixing the twist angle at $\theta=1.05^\circ$. As shown in Fig.~\ref{Fig:nosoc} in the Appendix.~\ref{DOS_nosoc}, two minibands are present near the charge neutrality point.  But due to spin degeneracy, $E_1=E_2$ and $E_3=E_4$. There are actually four bands, and each band has one VHS. For the lower two bands, $E_1$ and $E_2$, the VHS is of the ordinary type with a logarithmically divergent DOS. At the VHS energy $E=1.7045$ 
 meV, the two Fermi pockets intersect at a finite angle, as shown in the second left plot in Fig.~\ref{Fig:nosoc}\textbf{d}. For $E_3$ and $E_4$, a higher-order VHS (with a power-law divergent DOS \cite{yuan_magic_2019}) appears at $E=2.0159$ meV, characterized by the tangential touching of the two Fermi pockets, as shown in the second right plot in Fig.~\ref{Fig:nosoc}\textbf{d}. As the energy surpasses the VHSs, the Fermi contour undergoes a transformation from electron-type (purple) pocket(s) , where the band reaches its minimum, to hole-type (orange) pocket(s), where the band reaches its maximum (Fig.~\ref{Fig:nosoc_color}). The details for the non-SOC case are given in the Appendix.~\ref{DOS_nosoc}.


\begin{figure*}
    \centering
    \includegraphics[width=1\textwidth]{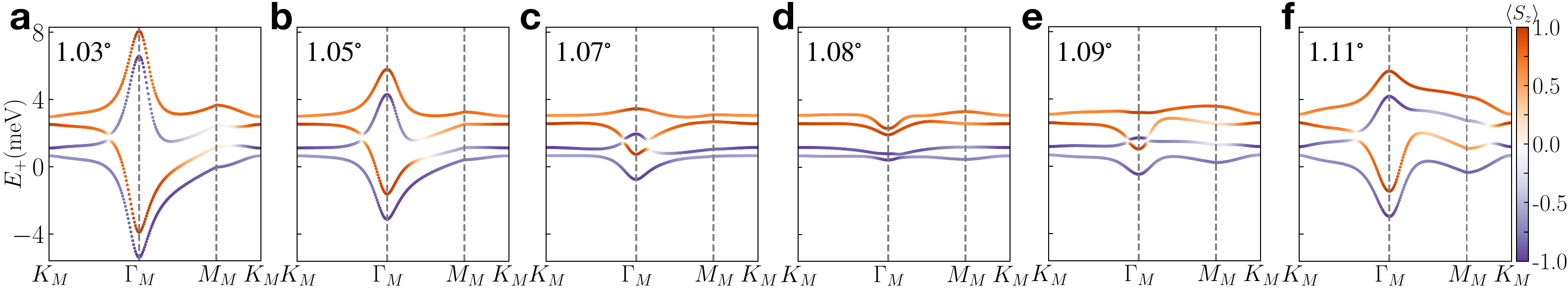}
\caption{The band structures for $\lambda_I=\lambda_R=3$ meV, across different twist angles, with color encodes $\langle S_z\rangle$. The skyrmion-like spin texture disappears in phase C in \textbf{d}.}
\label{Fig:spintextures}
\end{figure*}



\begin{figure*}
    \centering
    \includegraphics[width=1\textwidth]{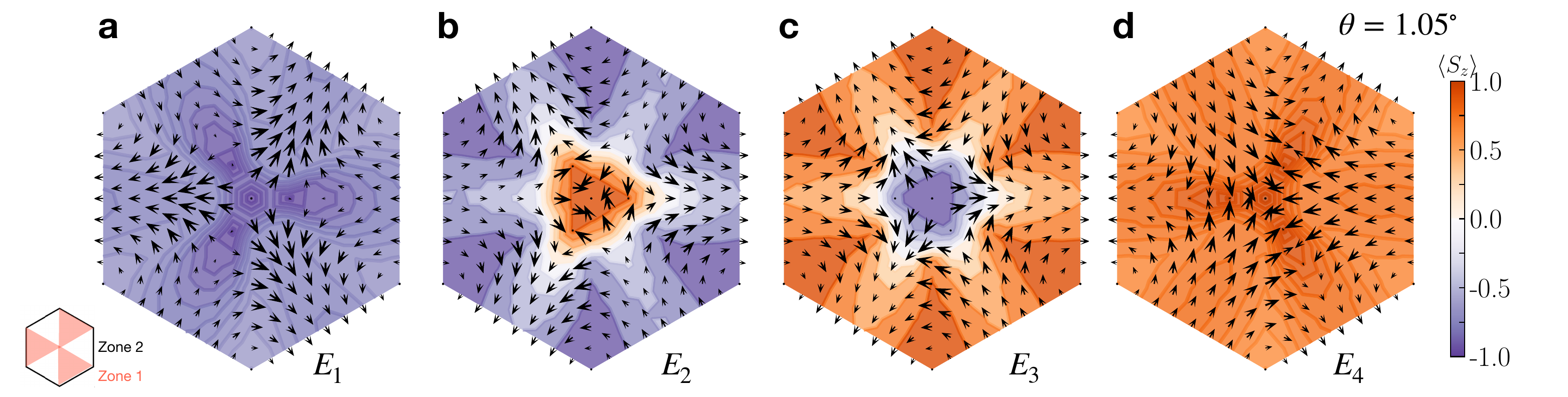}
\caption{\textbf{a-d}, Spin textures for $E_1,E_2,E_3,E_4$ correspondingly, with $\theta=1.05^\circ,\lambda_I=\lambda_R=3$ meV. The color code denotes $\langle S_z\rangle$. The vector field encodes $S_x$ and $S_y$ components of the spin expectation value.}
\label{Fig:spintexture_105}
\end{figure*}


Now we discuss the impact of SOC to these VHSs. Without loss of generality, we take the case with $\lambda_I=\lambda_R=3$ meV and $\theta=1.05^\circ$ for an example. The situation in all three phases is similar. The proximity-induced SOC lifts the spin degeneracy, resulting in four spin-split bands: $E_1,E_2,E_3,E_4$, as shown in Fig.~\ref{Fig:withsoc}\textbf{a}. The corresponding DOS are plotted in Fig.~\ref{Fig:withsoc}\textbf{b} and \ref{Fig:withsoc}\textbf{c}. Interestingly, the VHS for each band splits into a pair of VHS, with energies very close to each other. These splittings result in a total of eight ordinary VHSs per valley, each exhibiting a logarithmically divergent DOS. The higher-order VHS observed in the absence of SOC disappears in the presence of SOCs. The precise VHS energies are mentioned in the caption of Fig.~\ref{Fig:withsoc} and the corresponding energy contour plots are shown in Fig.~\ref{Fig:withsoc}\textbf{d}. Each VHS energy has three VHS points in k-space, where the Fermi pockets intersect at an angle. The VHS points are all close to the original Dirac points without SOC ($K_M$ and $K'_M$). The splitting of the VHS is attributed to the breaking of the mirror symmetry around $k_x$ by SOC (but the $\mathcal{C}_3$ symmetry remains). Consequently, the energy $E_k$ is not the same at $K_M$ and $K'_M$,  causing the Fermi pockets to intersect at slightly different energies while remaining close to $K_M$ and $K'_M$, unlike in Fig.~\ref{Fig:nosoc_color}. Therefore, for each pair of VHS, one is located near $K_M$ while the other is located near $K_M'$.

More interestingly, the splitting of VHS may affect how the Hall coefficient changes sign when passing VHS. Usually, the Hall coefficient will change from negative to positive when passing through a VHS from low energy, representing the effect of electron pocket or hole pocket, respectively. Does this mean that the Hall coefficient changes sign twice when passing through a pair of VHS for each band, meaning we have electron pocket both at the bottom and the top of the band? This is certainly not the case. In Fig.~\ref{Fig:withsoc_color}\textbf{a}-\ref{Fig:withsoc_color}\textbf{d}, we plot the energy contour for all four minibands, with color scheme represents the energy measured with respect to the low VHS energy $E_{b,v}^l$ (solid line) and the high VHS energy, $E_{b,v}^h$ (dashed line), where $b$ is the band index. For $E_1$ and $E_2$, when $E<E_{b,v}^l$, there is an electron pocket located at $\Gamma_M$, which is similar to the non-SOC case in Fig.~\ref{Fig:nosoc_color}\textbf{a}. For $E_3$ and $E_4$, when $E>E_{b,v}^h$, there is a hole pocket located at $\Gamma_M$, which is similar to the non-SOC case in Fig.~\ref{Fig:nosoc_color}\textbf{b}. Nevertheless, as illustrated in this figure, we have hole (orange) pockets when $E>E_{b,v}^{h}$ and electron (purple) pockets when $E<E_{b,v}^{l}$. But in between these two VHS energies, $E_{b,v}^l < E < E_{b,v}^h$, we point out that there may exist multiple pockets, both electron- and hole-type, colored by the purple and orange regime in between the dashed and solid lines. These multiple pockets may cause a cancellation effect, leading to a nearly zero or fluctuating Hall coefficient between these pairs of VHSs. It will be interesting to investigate this experimentally.

\section{Spin texture in momentum space\label{ST}}

In this section, we first examine the spin texture in momentum space of the four minibands for a range of twist angels, then show how a non-trivial spin texture evolves in the presence of an out-of-plane electric field. Understanding the spin texture is crucial for studying the pairing mechanism in superconductivity \cite{chou2024topological}, as well as the spintronics in MATBG.

\subsection{Emergent skyrmion-like spin texture in momentum space\label{ST_}}

We examine the spin texture of the four minibands for a range of twist angles, with $\lambda_I=\lambda_R=3$ meV. Figure.~\ref{Fig:spintextures} presents the band structures for $1.03^\circ\leq\theta\leq1.11^\circ$, where the color indicates the spin-$z$ expectation value: orange denotes spin-up, white denotes zero polarization, and purple denotes spin-down. For the angles presented, the system is in phase A, except for $1.08^\circ$, where the system is in phase C. Also, compared to $\theta=1.07^\circ$ and $1.09^\circ$, the cases with $\theta=1.03^\circ$ and $1.11^\circ$ are deeper in phase A because the energy difference between two middle bands at $\Gamma_M$, denoted by $w_{0,\theta}$, is larger. Consequently, $\lambda_I/2w_{0,\theta}$ is smaller, indicating that these angles are further away from the A-B phase boundary.

Interestingly, for the cases in phase A, the middle two bands, $E_2$ and $E_3$, exhibit skyrmion-like spin texture: the $\langle S_z\rangle$ near the $\Gamma_M$ and the $K_M$ points have different signs. The case with $\theta=1.08^\circ$ is in phase C, where two almost spin-up bands are completely separated from two almost spin-down bands (with the Rashba SOC still providing some spin mixing.) Therefore, the skyrmion feature disappears in Phase C. On the other hand, for other angles, the system is in phase A, where two spin-up bands still intersect with two spin-down bands with Ising-only SOC, but the Rashba SOC further avoids band crossings, generating this skyrmion-type spin texture around the $\Gamma_M$ point. Additionally, the spin-up (orange) region in momentum space is large for $\theta=1.11^\circ$ and $1.03^\circ$, as they are deep in phase A. When closer to the magic angle ($1.08^\circ$), the spin-up region becomes smaller, eventually disappearing at the magic angle.

The spin texture profiles for $E_1,E_2,E_3,E_4$, with $\theta=1.05^\circ$, are depicted in Figs.~\ref{Fig:spintexture_105}\textbf{a}-\ref{Fig:spintexture_105}\textbf{d}. The vector field represents the $S_x$ and $S_y$ components of the spin expectation value, and the spin texture adheres to $\mathcal{C}_3$ symmetry. For $E_1$, the spin vectors predominantly point downward (purple). At $\Gamma_M$,  the spin vector points directly downward. Moving away from $\Gamma_M$, the spins tilt slightly away from the perpendicular direction, pointing away from $\Gamma_M$ in Zone 1 while remaining mostly downward in Zone 2. Zone 1 (pink) and Zone 2 (white) in the momentum space are defined on the left-hand side of Figs.~\ref{Fig:spintexture_105}\textbf{a}. Similarly, for $E_4$,  the spins predominantly point upward (orange). At $\Gamma_M$, the spin vector points directly upward. In this case, the spins point toward $\Gamma_M$ in Zone 2 while remaining mostly upward in Zone 1. 
For $E_2$, the spins exhibit a skyrmion-like feature: the spin points purely upward at $\Gamma_M$ and downward at the corners of MBZ. Moving away from $\Gamma_M$, spins start to lie in-plane around the intersection of the orange and purple regions. Around $\Gamma_M$, the in-plane spin components point toward the center of the MBZ. 
On the other hand, for $E_3$, the trend is the opposite. At $\Gamma_M$, the spin points purely downward, while pointing upward at the corners of the MBZ, again indicative of a skyrmion-like feature. The spins also lie in-plane around the intersection of the orange and purple regions. However, in this scenario, the in-plane spin components around $\Gamma_M$ point away from the center of the MBZ. Moreover, on average, the in-plane spin components in Zone 1 are larger than that in Zone 2 for $E_1,E_3$, and opposive for $E_2,E_4$.


\begin{figure*}
    \centering
    \includegraphics[width=1\textwidth]{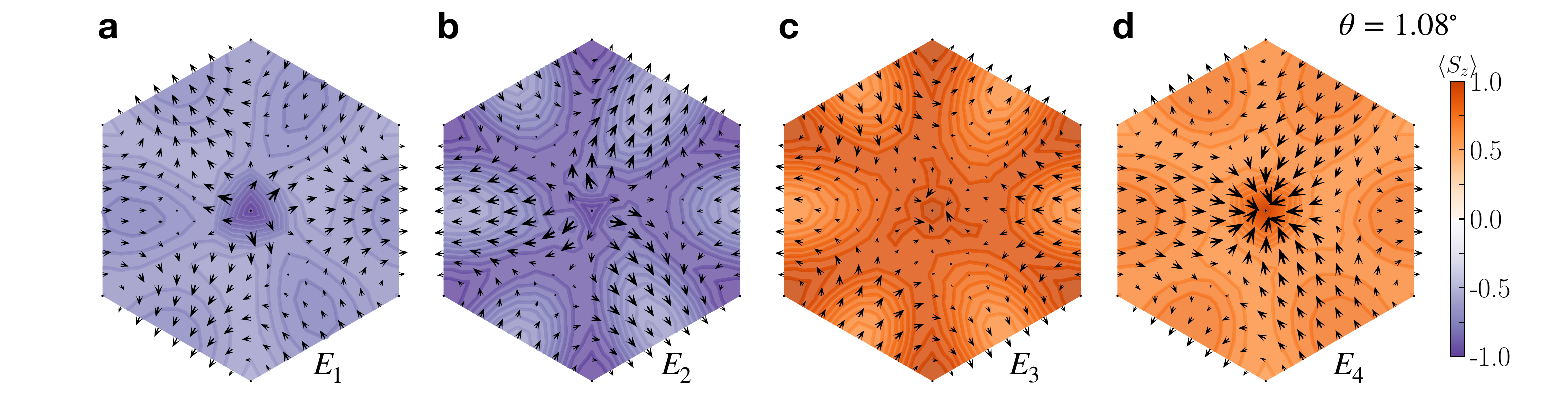}
\caption{\textbf{a-d}, Spin textures for $E_1,E_2,E_3,E_4$ correspondingly, with $\theta=1.08^\circ,\lambda_I=\lambda_R=3$ meV. The color code denotes $\langle S_z\rangle$. The vector field encodes $S_x$ and $S_y$ components of the spin expectation value.}
\label{Fig:spintexture_108}
\end{figure*}



\begin{figure*}
    \centering
    \includegraphics[width=1\textwidth]{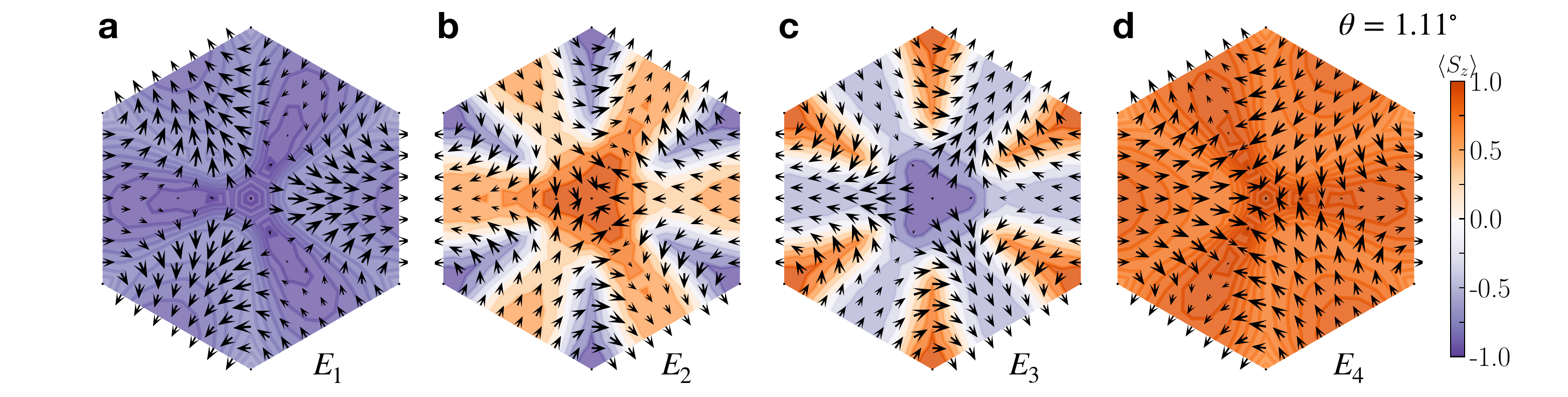}
\caption{\textbf{a-d}, Spin textures for $E_1,E_2,E_3,E_4$ correspondingly, with $\theta=1.11^\circ,\lambda_I=\lambda_R=3$ meV. The color code denotes $\langle S_z\rangle$. The vector field encodes $S_x$ and $S_y$ components of the spin expectation value.}
\label{Fig:spintexture_111}
\end{figure*}



\begin{figure*}
    \centering
    \includegraphics[width=1\textwidth]{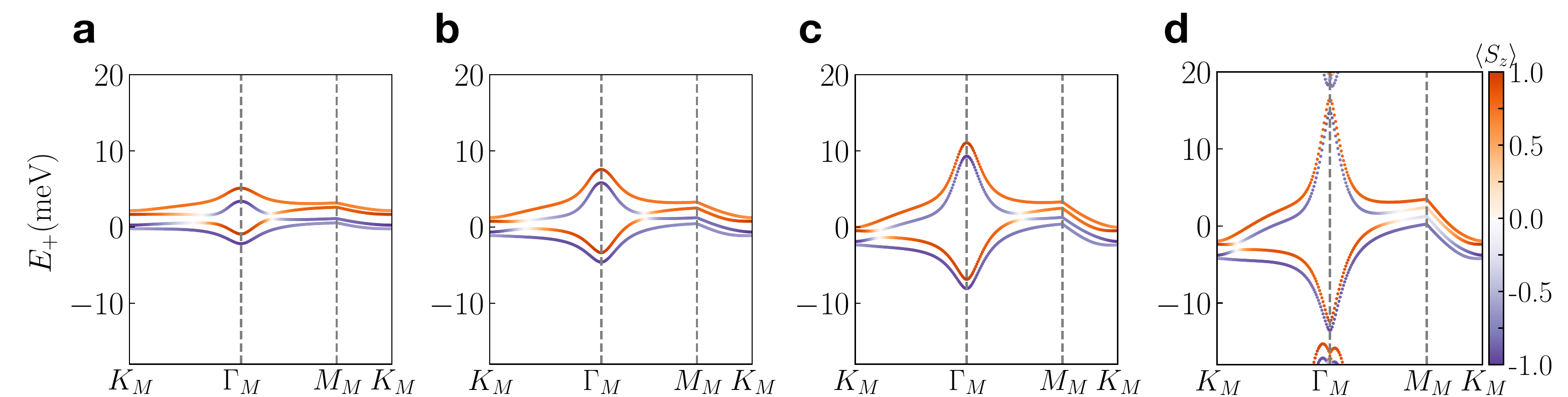}
\caption{Band structure for $\theta=1.07^\circ,\lambda_I=\lambda_R=3$ meV, and \textbf{a} $u=15$ meV, \textbf{b} $u=30$ meV, \textbf{c} $u=50$ meV, and \textbf{d} $u=80$ meV, respectively. The color code denotes $\langle S_z\rangle$. }
\label{FIG:BS_u}
\end{figure*}



\begin{figure*}
    \centering
    \includegraphics[width=0.85\textwidth]{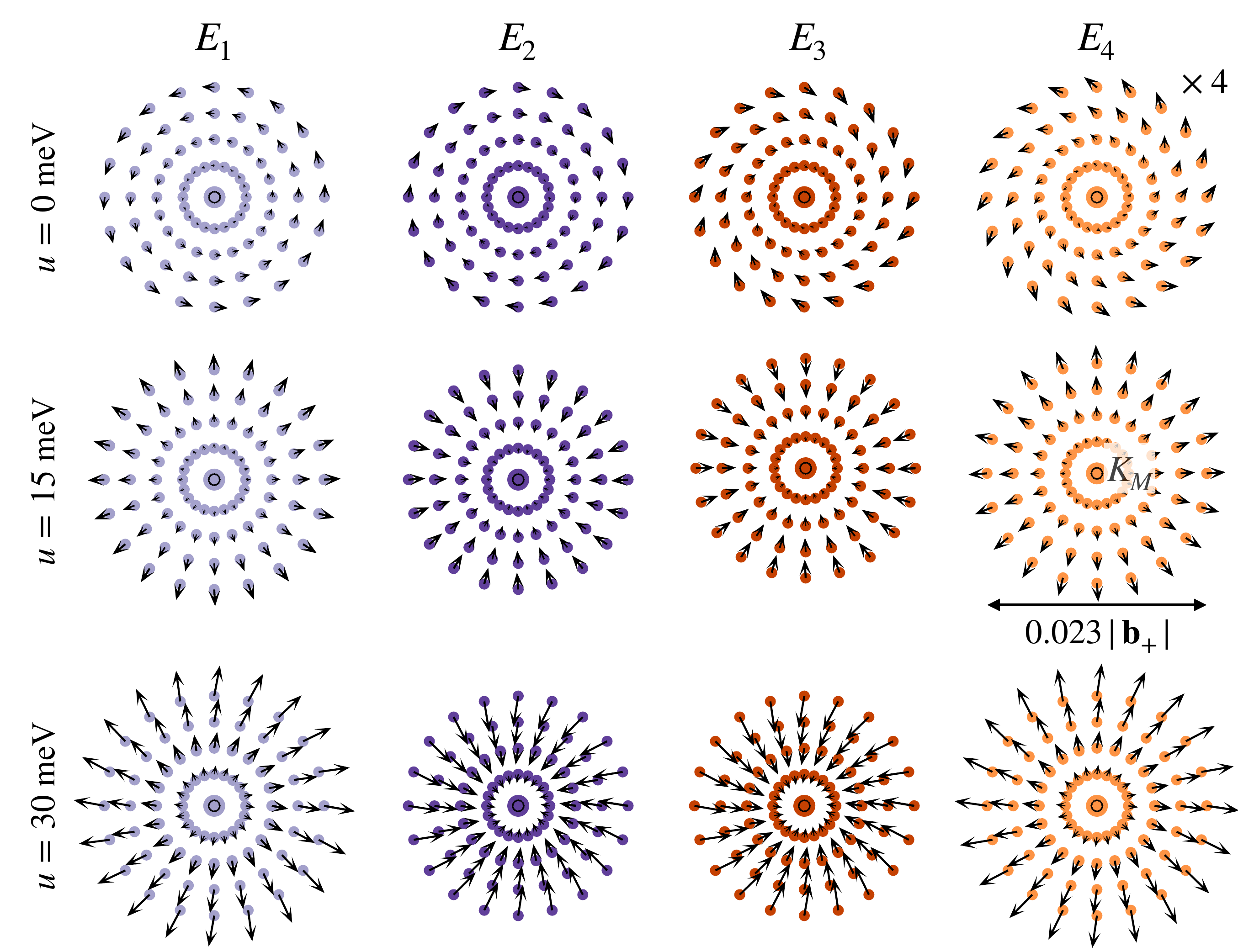}
\caption{The spin texture of four bands, $E_1,E_2,E_3,E_4$ in electric field near $K_M$ point in the electric field $u=0$ meV (top panel), $u=15$ meV (middle panel), $u=30$ meV (bottom panel). $\theta=1.07^\circ,\lambda_I=\lambda_R=3$ meV. The color code denotes $\langle S_z\rangle$, as in Fig.~\ref{FIG:BS_u}. The scale of $S_x,S_y$ in $u=0$ meV plot is quadrupled for visibility.}
\label{FIG:SOF_u_4bs}
\end{figure*}


For $\theta=1.08^\circ$, the spin textures for four minibands are shown in Fig.~\ref{Fig:spintexture_108}. Unlike $\theta=1.05^\circ$, $E_1$ and $E_2$ here are almost spin down, while $E_3$ and $E_4$ are almost spin up, consistent with Fig.~\ref{Fig:spintextures}\textbf{d}. The skyrmion-like feature completely disappears here. In addition, the in-plane spin components, on average, in Zone 1 and Zone 2 are less distinct compared to $\theta=1.05^\circ$ and $\theta=1.11^\circ$ cases (Fig.~\ref{Fig:spintexture_111}). For $\theta=1.11^\circ$, the skyrmion-like feature (spin-up region) for the middle two bands extends to the $M_M$ point in momentum space, which means that its size in momentum space is larger than in the $\theta=1.05^\circ$ case, because it is deeper inside phase A. Here, the in-plane spin components in Zone 2 are larger than those in Zone 1 for $E_1,E_3$, and the opposite is true for $E_2,E_4$. This occurs because the band inversion already takes place at this angle.

In summary, we observe a skyrmion-like spin texture for various twist angles, which disappears in phase C. Although most of the skyrmion-like features are shown here in phase A, they can also be observed in phase B, as shown in the inset of Fig.~\ref{Fig:phasediagram_1.05}\textbf{f}, although the spin-up region is extremely tiny, because it is very close to phase C. We conclude that this skyrmion-like spin texture is stable across a wide range of twist angles and SOCs, which is actually a crucial ingredient in the interband paring mechanism for superconductivity in MATBG \cite{chou2024topological}.


\begin{figure*}
    \centering
    \includegraphics[width=1\textwidth]{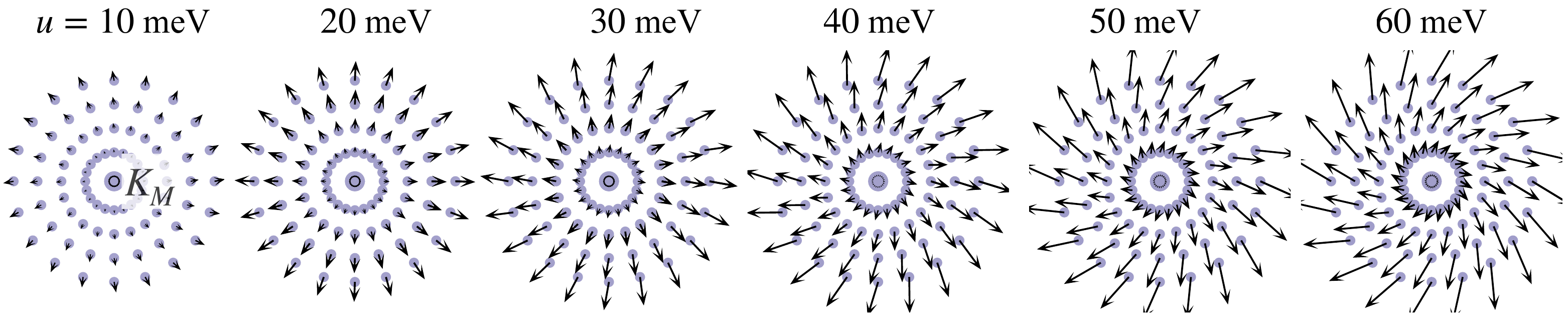}
\caption{The evolution of the spin texture of the first band $E_1$ near $K_M$ in various electric field $u=10,20,30,40,50,60$ meV. $\theta=1.07^\circ,\lambda_I=\lambda_R=3$ meV. The color code denotes $\langle S_z\rangle$, as in Fig.~\ref{FIG:BS_u}. }
\label{FIG:SOF_u_2b}
\end{figure*}



\begin{figure*}
    \centering
    \includegraphics[width=1\textwidth]{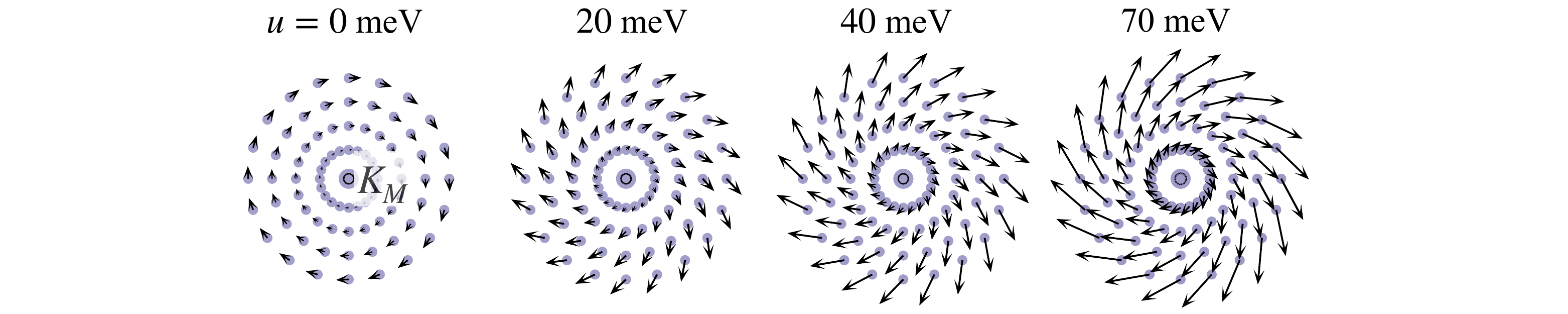}
\caption{The evolution of the spin texture of the first band $E_1$ near $K_M$ in various electric field $u=0,20,40,70$ meV. $\theta=1.05^\circ,\lambda_I=\lambda_R=3$ meV. The color code denotes $\langle S_z\rangle$, as in Fig.~\ref{FIG:BS_u}.}
\label{FIG:SOF_u_1b_theta105}
\end{figure*}


\subsection{Spin texture in presence of out-of-plane electric field\label{STWE}}
We now show how the spin texture evolves in the presence of an out-of-plane electric field. As mentioned before, the Rashba SOC is naturally present when the inversion symmetry is broken. Therefore, applying an out-of-plane electric field in MATBG also modifies the Rashba SOC. In this case, a dipolar coupling induces transitions between the $p_z$ and $s$ orbitals, flipping the spin \cite{huertas-hernando_spin-orbit_2006,min_intrinsic_2006,konschuh_tight-binding_2010,avsar_colloquium_2020}. In MATBG, the out-of-plane electric field also gives rise to layer polarization, which can make the in-plane spin texture tunable \cite{frank_emergence_2024}. Engineering SOC would influence the correlated phases \cite{XieM2023,zhumagulov_emergent_2024,zhumagulov_swapping_2023,koh_correlated_2024} and superconductivity \cite{ZhangY2023,HolleisL2023,Jimeno-PozoA2023,chou2024topological}. Specifically, engineering the Rashba SOC would control the polarization of spin accumulation and spin current \cite{ZollnerK2023a,frank_emergence_2024,ghiasi_charge--spin_2019}. In this section, we study how the in-plane spin components rotate around $\Gamma_M$ and $K_M$ points in the $+K$ valley. We find that, in the presence of electric field, a radial Rashba spin texture can be readily achieved around $K_M$, while around $\Gamma_M$, the spin texture can also be tuned to some extent in MATBG.

The general description of Rashba SOC requires the so-called Rashba angle $\psi$ (between the electron's momentum and spin). The general form of the Rashba SOC for $C_3$ symmetric systems at $K$ point, is \cite{NaimerT2021,NaimerT2023},
\begin{equation}
    h_R=\frac{\lambda_R}{2}e^{-i\psi s_z/2}\left(\tau\sigma_xs_y-\sigma_ys_x\right)e^{i\psi s_z/2},
\end{equation}
which is originally derived for the induced Rashba SOC in monolayer graphene. In our study, we assume that the SOCs are induced by the proximity effect from the WSe$_2$ layer, affecting only the top graphene layer. As a result, the Rashba SOC term follows the same form as in a monolayer due to the $\mathcal{C}_3$ symmetry. Although interlayer SOC terms involving layer pseudospin matrices could arise, they should be suppressed compared to the top-layer SOC terms in the proximity effect. This approach also aligns with experimental results, where SOC becomes significant only when carriers are polarized to the layer adjacent to the proximate TMD \cite{ZhangY2023,HolleisL2023,li2024tunable}. Futhermore, while the Rashba angle $\psi$ depends on the twist angle between the top graphene layer and the SOC layer \cite{NaimerT2021,lee_charge--spin_2022,Csaba2022,NaimerT2023}, we have neglected this effect in the previous section. Here, we set $\psi = 0$ as before but show that the effective Rashba angle can be tuned from radial ($\psi = 90^\circ$) to tangential ($\psi = 0^\circ$) around the $K_M$ point by a displacement field.

To incorporate an out-of-plane electric field, we put potentials of $\pm u$ onto the top and bottom layers (positive $u$ - positive field in $z$): $\mathcal{H}=\mathcal{H}_{0,+}+\mathcal{H}_u$, where
\begin{align}\label{Eq:Hu}
    \mathcal{H}_u=\left[\begin{array}{cc}
        u\textbf{I} & 0\\[2mm]
        0 & -u\textbf{I}
    \end{array}\right],
\end{align}
and $\textbf{I}$ is a four dimensional identity matrix. The outer blocks describe the layer degree of freedom (the first block is the top layer). The inner blocks describe the sublattice degree of freedom crossing the spin degree of freedom.

We present our results mainly at the twist angle $\theta=1.07^\circ$, because the effect of the out-of-plane electric field is more pronounced near the magic angle but right at magic angle, the skyrmion-like feature disappears (Fig.~\ref{Fig:spintextures}\textbf{d}). Later, we also comment on the results at other angles. As shown in Fig.~\ref{FIG:BS_u}, the bandwidth of the flat bands broadens as the displacement field increases. At large $u=80$ meV, the flat bands eventually merge into the higher energy bands, leading to the disappearance of the isolated flat bands (Fig.~\ref{FIG:BS_u}\textbf{d}). We limit our study to the vicinity of isolated flat bands, meaning $u<80$ meV. Overall, the top and bottom bands, $E_1$ and $E_4$, generally exhibit spin up/down characteristics even in the presence of an electric field. Interestingly, the spin-up/spin-down region for the $E_2$/$E_3$ in momentum space is significantly enhanced, as one increases the strength of the electric field: the position where $\langle S_z\rangle=0$ (white color) moves closer to $K_M$ instead of $\Gamma_M$ as $u$ increases.

We now focus on the evolution of the spin texture around $K_M$ in the presence of an electric field. The spin textures corresponding to the $u=0,15,30$ meV are shown in Fig.~\ref{FIG:SOF_u_4bs}. We find that the spin textures for $E_2$ and $E_3$ are similar, while those for $E_1$ and $E_4$ are also similar but opposite to $E_2$ and $E_3$. At $u=0$ meV, the spin textures are almost tangential to the circle centering $K_M$ for all four bands ( $E_1$ is the most tangential), while at $u=15$ meV, they are purely radial. At $u=30$ meV, the spin textures for all four bands deviate from being purely radial. The maximum in-plane spin expectation value for $u=0$ meV is 0.012, while the maximum for $u=30$ meV is 0.09. The scale of $S_x,S_y$ in $u=0$ meV plot is quadrupled for visibility. Despite the strong out-of-plane spins, our calculations clearly reveal the emergence of in-plane radial Rashba textures near $K_M$, in the presence of an experimental feasible out-of-plane electric field \cite{ChouYZ2022a}. The distance between two graphene layers is $\sim0.335$ nm \cite{cakmak_continuous_2019,guo_electrochemical_2013}, the experimental electric field is around 1V/nm \cite{ZhangY2023}, so energy difference between two layers $2u$ is around 33.5 meV if the dielectric constant is 10. The actual layer potential difference is hard to estimate and depends on the sample.

Figure.~\ref{FIG:SOF_u_2b} provides a more detailed description of how the electric field modulates the spin texture of $E_1$ around $K_M$. As the displacement field increases, the in-plane spin expectation values become larger, while the $z$-component remains roughly the same. The maximum in-plane spin expectation value for $u=10$ meV is 0.03, while the maximum for $u=60$ meV is 0.14. Without an electric field ($u=0$ meV), the spin texture of $E_1$ appears to be purely tangential. When $u\neq0$, the spins start to deviate from the tangential direction and develop a radial spin texture near $u=15$ meV. Remarkably, we find that in a range of $u$ from $10$ meV to $30$ meV, the spin texture near $K_M$ remains generally radial and is not highly susceptible to changes in the electric field. 

Here, we discuss the effect of the twist angle. At $\theta = 1.05^\circ$, the maximum $u$ at which isolated flat bands still exist is around 70 meV. As shown in Fig.~\ref{FIG:SOF_u_1b_theta105}, when $\theta=1.05^\circ$, the rotation of $E_1$ at $u=0$ meV is opposite to that in the $\theta=1.07^\circ$ case (top left plot in Fig.~\ref{FIG:SOF_u_4bs}). This difference arises because band inversion is already occurring at $K_M$. Generally speaking, the electric field here plays the same role, tuning the spin texure away from tangential. What differs is that within the range of electric field strength where isolated flat bands still exist, we do not observe a purely radial spin texture. This indicates that the out-of-plane electric field is more effective in tuning the spin texture when the twisted bilayer graphene is closer to the magic angle, which is $1.08^\circ$ in our case.

\begin{figure*}
    \centering
    \includegraphics[width=0.85\textwidth]{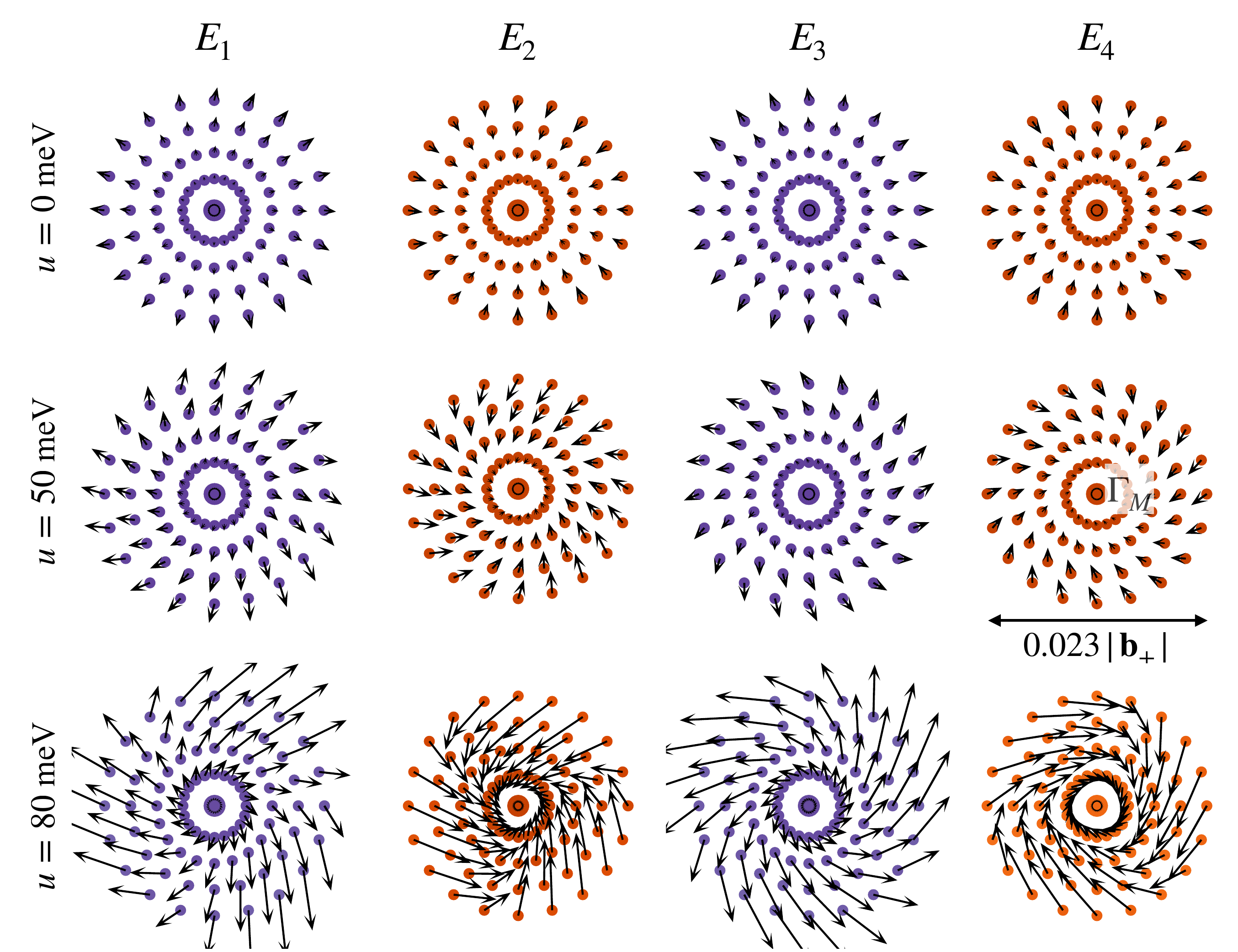}
\caption{The spin texture of four bands, $E_1,E_2,E_3,E_4$ in electric field near $\Gamma_M$ point in the electric field $u=0$ meV (top panel), $u=50$ meV (middle panel), $u=80$ meV (bottom panel). $\theta=1.07^\circ,\lambda_I=\lambda_R=3$ meV. The color code denotes $\langle S_z\rangle$, as in Fig.~\ref{FIG:BS_u}.}
\label{FIG:SOF_u_4bs_Gamma}
\end{figure*}

We also examine how spin textures around $\Gamma_M$ evolve in the presence of an electric field.  As shown in Fig.~\ref{FIG:SOF_u_4bs_Gamma}, without the electric field, the spin textures for all four bands around $\Gamma_M$ are radially oriented. The scale of $S_x,S_y$ in Fig.~\ref{FIG:SOF_u_4bs_Gamma} is decreased by a factor of 2.5, compared to the middle panel of Fig.~\ref{FIG:SOF_u_4bs}, for better illustration. When an electric field is applied, the spin texture deviates from being purely radial, and the tangential components start to appear. A larger electric field is needed to tune the spin texture around $\Gamma_M$, compared to the previous cases ($K_M$), meaning that we cannot tune the spin texture from purely tangential to purely radial at this twist angle, within the range of $u$ where the moir\'e bands are isolated from the remote bands.   

In summary, we examine the spin textures in the presence of an out-of-plane electric field for MABTG-WSe$_2$. We find that the electric field can tune the in-plane spin component from purely radial to tangential around $K_M$, while it has a less pronounced effect on the spin texture around the $\Gamma_M$ point. Tuning spins in plane or engineering the spin texture is crucial to achieve unconventional charge-to-spin conversion \cite{ghiasi_charge--spin_2019,ZollnerK2023a,lee_charge--spin_2022,zhao_unconventional_2020,veneri_twist_2022,ingla-aynes_omnidirectional_2022}, as well as influence correlated phases and superconductivity.

\section{Discussion\label{Discussion}}

Using the BM model, we construct the topological phase diagram of MATBG across different twist angles,  in the presence of Ising and Rashba SOCs. Our findings reveal that the introduction of SOCs into one layer of TBG significantly alters the band structure, leading to the emergence of three distinct topological phases in MATBG. Importantly, we find that the critical SOC strength depends on the twist angle, and all three phases can be realized with the experimentally accessible SOC strength ($\sim$ 1 
 meV) for systems with angles very close to the magic angle. We also find that the introduction of SOC splits the flat bands into four spin-split mini-bands, each featuring its own pair of VHSs, leading to a total of eight VHSs per valley in the DOS. The SOCs modify the DOS of MATBG, not only by introducing additional VHSs but also by altering the type of VHSs. The splitting of VHS for each band may significantly impact the Hall conductivity, which may fluctuate or remain nearly zero within each pair of VHS energies due to the possible canceling effect of multiple pockets. This should be experimentally investigated.
 
 Moreover, we discover a skyrmion-like spin texture in momentum space in phase A and B, and it eventually disappears as the system transitions to phase C. Additionally, we show that this skyrmion-like feature can be tuned by an out-of-plane electric field, along with the spin textures around the $K_M$ and $\Gamma_M$ points. This tunability opens up possibilities for controlling the spin texture in MATBG, which would potentially influence the correlated phases and superconductivity in the system. For example, the interband superconductivity in Ref.~\cite{chou2024topological} is more likely to happen below the magic angle because the spin textures do not favor interband Ising pairing above the magic angle. These skyrmionic features, in addition to being of intrinsic interest,  may also be useful to experimentally observe the topological phase transition to the phase C.

In this work, we focus only on the continuum model (the BM model), which is valid at low energies and long wavelengths. This is reasonable because in this work, we only focus on the twist angles that are close to the magic angle, leading to isolated flat bands with narrow bandwidth in all cases. In addition, the phase diagram of MATBG is constructed only at the single-particle level.  The continuum low-energy band description should be well-valid in these situations. It is expected that the interactions likely modify the low-temperature phase diagram presented in this paper, since flat bands typically enhance interaction effects. For example, when the Coulomb interaction is considered, valley polarization likely prevails over the entire doping region for a range of twist angles. Therefore, the topological insulators at integer fillings, which are predicted in Fig.~\ref{Fig:spintextures} and Ref.~\cite{WangT2020}, are likely absent due to the time-reversal breaking by the valley polarization. Moreover, the pairing between two time-reversal related bands is suppressed due to valley polarization, so the inter-band paring superconductivity phases emerge when the electron-phonon interactions are included \cite{chou2024topological}.

We now mention several future directions. One is investigating the effects of VHS splitting due to SOC, such as its influence on Hall conductance and quantum oscillations in MATBG. Another potential direction is studying how a vertical electric field affects VHS splitting, as varying the electric field and SOC may lead to multiple transitions in the number of VHSs.  In addition, one can also study the polarization of spin accumulation and spin current in the presence of SOC and out-of-plane electric field in MATBG, which should be able to serve as a platform with tailored electronic and spintronic properties. Inclusion of electron-electron interactions in the theory is also an important open and difficult challenge for the future.


\begin{figure}
    \centering
    \includegraphics[width=0.3\textwidth]{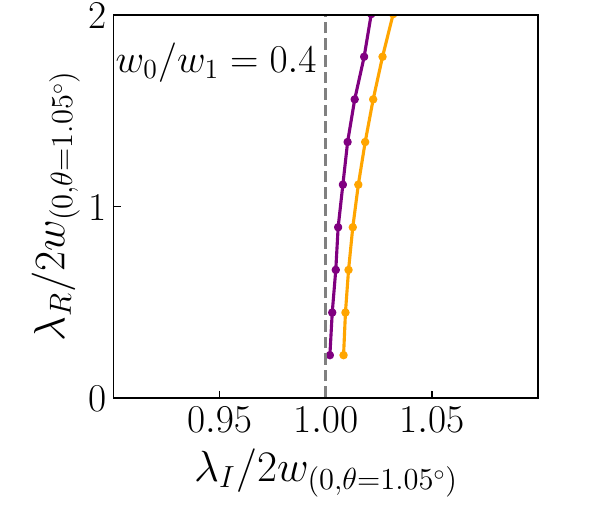}
\caption{Single particle phase diagram with rescaled Ising SOC $\lambda_I/2w_{(0,\theta=1.05^\circ)}$ and Rashba SOC $\lambda_R/2w_{(0,\theta=1.05^\circ)}$, with $w_0/w_1=0.4$, $w_1=110$ meV, $\theta=1.05^\circ$. $w_{(0,\theta=1.05^\circ)}=4.49$ meV in this case. }
\label{Fig:phasediagram_1.05_w0_w1_0_4}
\end{figure}



\begin{figure*}
    \centering
    \includegraphics[width=\textwidth]{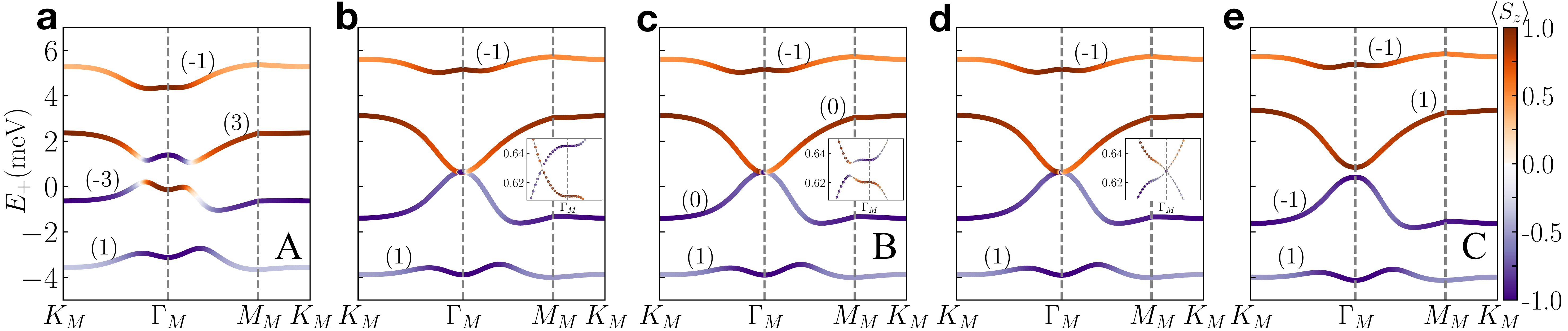}
\caption{\textbf{a-e}, Four mini flat bands corresponding to $\lambda_I=6.00,9.04,9.08,9.11,10.00$ meV (from left to right), with $\lambda_R=10$ meV, $w_0/w_1=0.4$. The color represents the expectation value of $S_z$, with orange indicating spin up and purple indicating spin down. The inset plots in \textbf{b-d} enlarge the band structure around $\Gamma_M$.}
\label{Fig:BS_105_w0_w1_0_4}
\end{figure*}


\section{Acknowledgements}
We are grateful to Jed Pixley, Zhentao Wang, Ming Xie, Jihang Zhu, Jay Sau, Silas Hoffman for useful discussion. This work is supported by the Laboratory for Physical Sciences (Y.T.,Y.-Z.C., and and S.D.S.) and F. W. is supported by National Key Research and Development Program of China (Grants No. 2022YFA1402401 and No. 2021YFA1401300) and National Natural Science Foundation of China (Grant No. 12274333).

\appendix

\section{Calulate Berry curvature and Chern number \label{BC_CN}}

To characterize the topology of the system, we calculate the Berry curvature $\Omega$ by numerically computing the Wilson loops in the momentum square with the rhombus grid, and each small grid spans a momentum space area $\mathcal{A}_0=\mathcal{A}_{\text{MBZ}}/\mathcal{N}^2$, where $\mathcal{A}_{\text{MBZ}}$ is the momentum-space area of MBZ and $\mathcal{N}=300$ in our calculations. The Berry curvature is approximated by \cite{ChouYZ2020}

\begin{align}\label{Eq:bc}
    &\Omega_b\left(\textbf{k}=\frac{\textbf{k}_1+\textbf{k}_2+\textbf{k}_3+\textbf{k}_4}{4}\right)\\ \nonumber
    &\approx\frac{\text{arg}\left[\langle u_{\textbf{k}_1,n}|u_{\textbf{k}_2,n}\rangle\langle u_{\textbf{k}_2,n}|u_{\textbf{k}_3,n}\rangle\langle u_{\textbf{k}_3,n}|u_{\textbf{k}_4,n}\rangle\langle u_{\textbf{k}_4,n}|u_{\textbf{k}_1,n}\rangle\right]}{\mathcal{A}_0},
\end{align}
where $\textbf{k}_1\rightarrow\textbf{k}_2\rightarrow\textbf{k}_3\rightarrow\textbf{k}_4\rightarrow\textbf{k}_1$ tracks in a counterclock-wise manner a small rhombus grid with the area $\mathcal{A}_0$. The Chern number $\mathcal{C}$ of the $b$th band can then be calculated via
\begin{align}\label{Eq:cn}
    \mathcal{C}_n=\frac{1}{2\pi}\int_{\text{MBZ}}d\textbf{k}\Omega_b\left(\textbf{k}\right).
\end{align}
The Chern numbers of two valleys are related by a minus sign. The overall Chern number is zero due to the time-reversal symmetry.

\begin{figure*}
    \centering
    \includegraphics[width=0.9\textwidth]{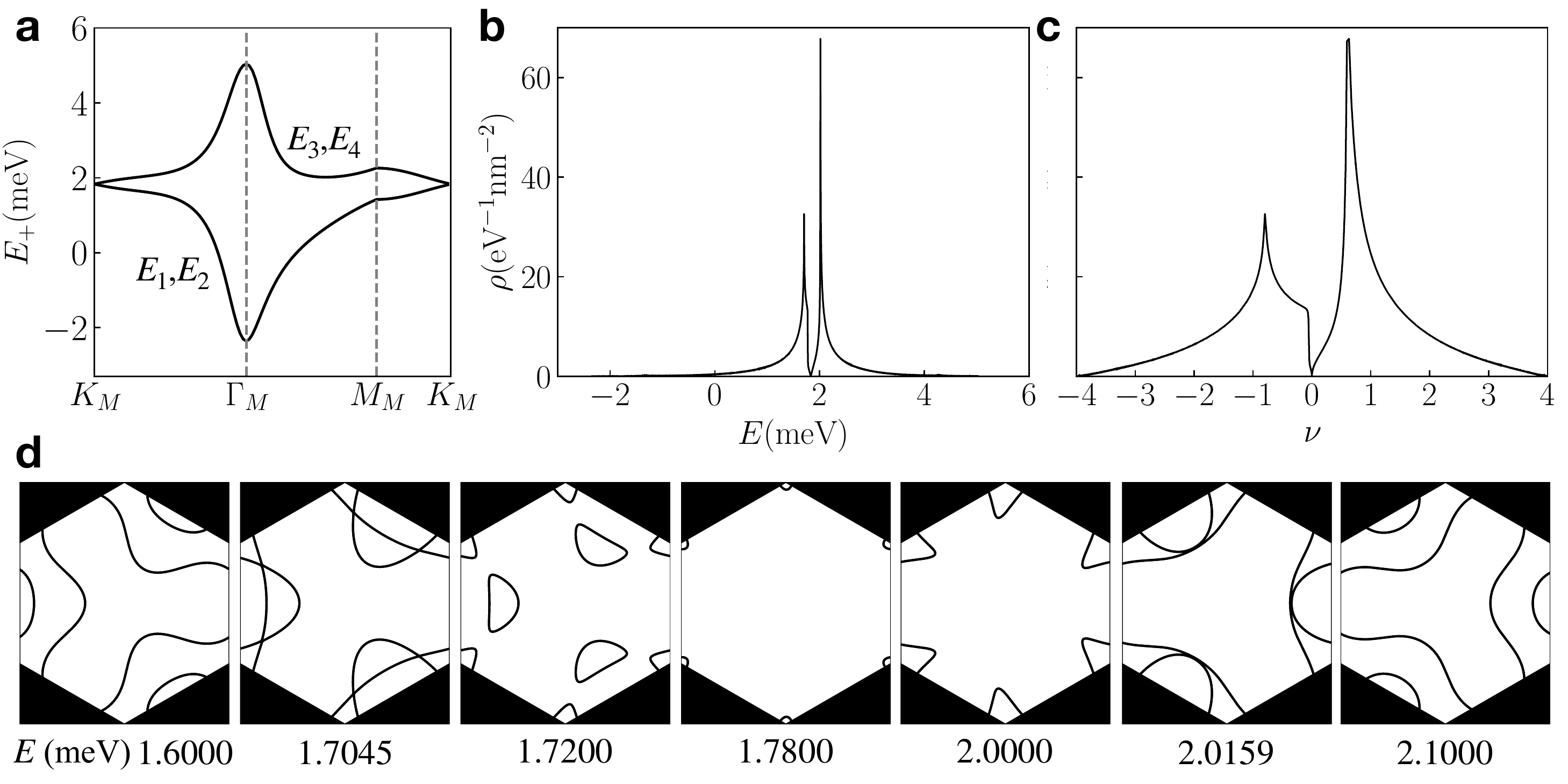}
\caption{\textbf{a}, Band structure for $\theta=1.05^\circ$ without SOC. Because of the spin degeneracy, $E_1=E_2$ and $E_3=E_4$. \textbf{b}, Density of State per spin per valley as a function of energy $E$ and \textbf{c}, occupation number $v$. \textbf{d} Band structure contour plots at different energy $E$. The VHSs locate at $E=1.7045$ meV and $E=2.0159$ meV.}
\label{Fig:nosoc}
\end{figure*}



\begin{figure}
    \centering
    \includegraphics[width=0.45\textwidth]{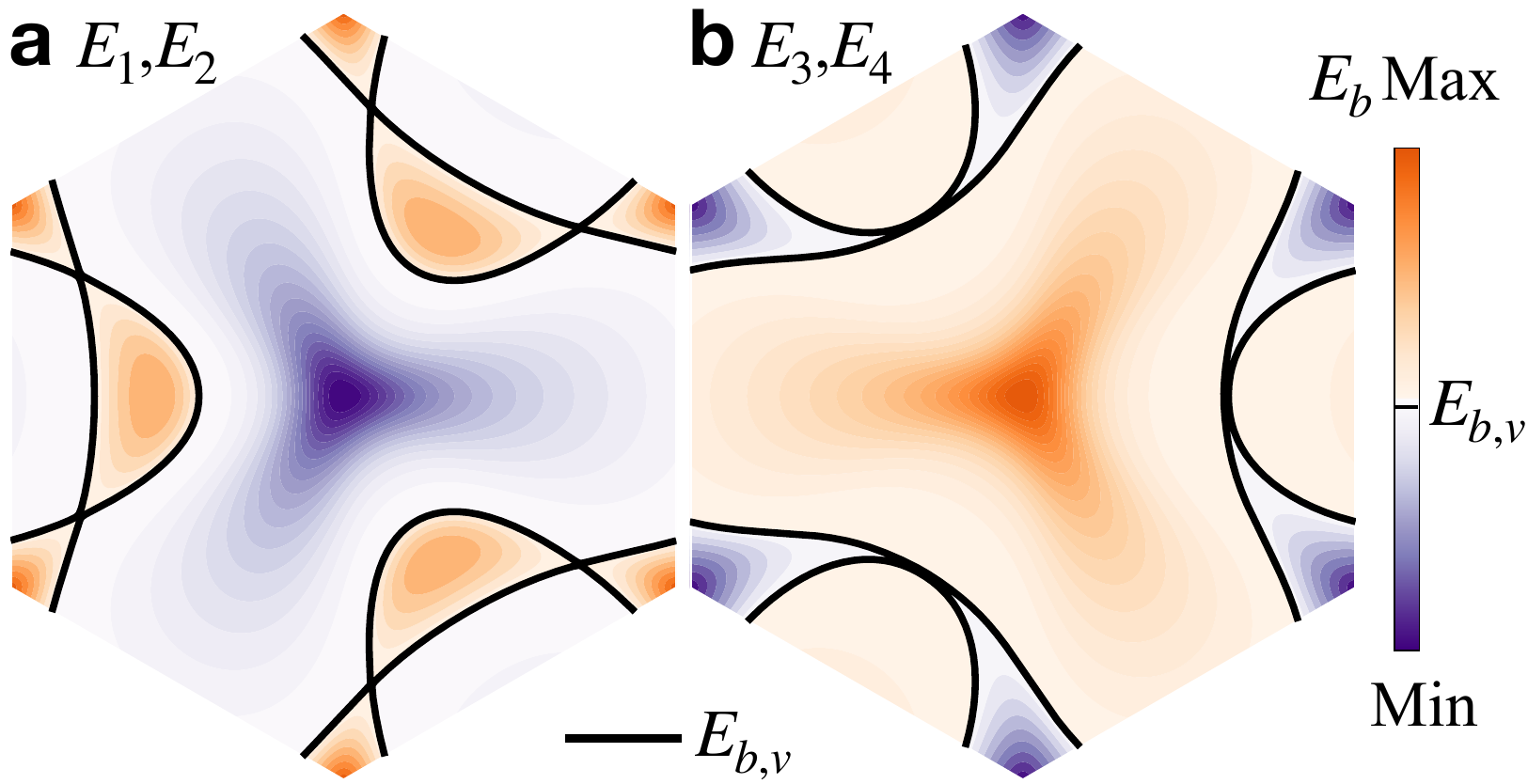}
\caption{Contour plots for $E_1,E_2$ in \textbf{a}, and $E_3,E_4$ in \textbf{b}, respectively. The black solid line indicate the contour line plot at \textbf{a}, an ordinary VHS, $E_{1,v}=1.7045$ meV,  and \textbf{b}, high-order VHS, $E_{1,v}=2.0159$ meV. The orange and purple regions indicate the hole and electron pockets correspondingly.} 
\label{Fig:nosoc_color}
\end{figure}


\section{Single-particle phase diagram with different interlayer hopping energy\label{PD_different_hoping_energy}}

In this Appendix, we use the interlayer hopping energy $w_1=110$ meV and $w_0/w_1=0.4$, to construct the single-particle phase diagram at twist angle $\theta=1.05^\circ$. As shown in Fig.~\ref{Fig:phasediagram_1.05_w0_w1_0_4}, we still find three distinct topological phases A, B, and C. The A-B boundary is marked by purple, while the B-C boundary is marked by orange. The energy difference between the lower and upper bands at $\Gamma_M$ without SOCs, $w_{(0,\theta=1.05^\circ)}=4.49$ meV here. We can see that with a smaller $w_0/w_1$ ratio (0.4), the width of phase B is smaller than that in the case with $w_0/w_1=0.8$ (Fig.~\ref{Fig:phasediagrams}\textbf{e}). In fact, in chiral limit, $w_0/w_1=0$, the phase B is extremely narrow. For example, when $\lambda_R=6$ meV, the width of phase B, $\lambda_I^h-\lambda_I^l\sim1.403$ meV for $w_0/w_1=0.8$; $\sim0.055$ meV for $w_0/w_1=0.4$; $\sim0.005$ meV for $w_0/w_1=0$.  

Figure.~\ref{Fig:BS_105_w0_w1_0_4} shows the band structure across three topological phases. The evolution of the band structure from phase A to C is similar to that in the case with $w_0/w_1=0.8$ in Fig.~\ref{Fig:phasediagram_1.05}\textbf{d-h}. At A-B boundary, one of the Dirac cones is located at $K_M\Gamma_M$ line (Fig.~\ref{Fig:BS_105_w0_w1_0_4}\textbf{b}), while the Dirac cone is located at $\Gamma_M$ at the B-C boundary (Fig.~\ref{Fig:BS_105_w0_w1_0_4}\textbf{d}).

\section{Density of states without SOC\label{DOS_nosoc}}

In 2D electron systems with an energy dispersion $E(\textbf{k})$, an VHS with diverging DOS occurs at a saddle point $K_v$, determined by 
\begin{equation}
    \nabla_\textbf{k} E=0
\end{equation}
In right choice of axes, the energy dispersion near $\textbf{k}_v$ can then be Taylor expanded as\cite{yuan_magic_2019}
\begin{equation}
    E-E_v=-\alpha p_x^2+\beta p_y^2+\gamma p_xp_y^2+\kappa p_yp_x^2+...,
\end{equation}
where $E_v$ is the VHS energy, the momentum $\textbf{p}=\textbf{k}-\textbf{k}_v$ is the momentum measured from the saddle point, and the coefficients $\alpha,\beta,\gamma,\kappa$ are the expansion coefficients. When $\alpha\beta<0$, the VHS is ordinary with a logarithmically diverging DOS. If $\alpha\beta=0$, a high-order VHS occurs. Specifically, If $\alpha=\beta=0$, the taylor expansion of $E_\textbf{k}$ then starts from at least the third order. A type-I higher-order VHS occurs, describing an intersection of three or more Fermi surfaces at a common \textbf{k} point\cite{yuan_magic_2019,biswas_su2-invariant_2011,shtyk_electrons_2017,efremov_multicritical_2019}, which is out of scope of this paper. When $\alpha=0,\beta\neq$0, or vice versa, a type-II higher-order VHS is present \cite{yuan_magic_2019}, characterized by a power-law divergence in the DOS, which enhances electron correlation significantly.

Here we present the DOS for the non-SOC case with twist angle at $\theta=1.05^\circ$. As shown in Fig.~\ref{Fig:nosoc}\textbf{a}, two minibands are present near the charge neutrality point, because due to spin degeneracy, $E_1=E_2$ and $E_3=E_4$. The Dirac cones are located at the corner of MBZ, labeled by $K_M$. Fig.~\ref{Fig:nosoc}\textbf{b} shows the corresponding DOS per spin per valley. Each band has one VHS.  But due to spin degeneracy, there are actually four VHSs per valley. The VHS in the lower energy bands $E_1,E_2$, located at $E=1.7045$ meV, is of the ordinary type with a logarithmically diverging DOS. In this scenario, the two Fermi pockets intersect at a finite angle, as shown in the second-left plot in Fig.~\ref{Fig:nosoc}\textbf{d}, having six VHS points in k-space.

Moreover, as shown in Fig.~\ref{Fig:nosoc_color}\textbf{a}, the Fermi contour undergoes a transformation from a single electron-type (purple) pocket at the center of the MBZ ($\Gamma_M$), where the band reaches its minimum, to five separate hole-type pockets where the band reaches its maximum, as the energy surpasses the VHS. Two Dirac pockets are located at the corners of the MBZ ($K_M$ and $K'_M$), while the other three are located inside the MBZ. 

In contrast, at $E=2.0159$ meV, higher-order VHSs appear for $E_3,E_4$, characterized by the tangential touching of the two Fermi pockets, as shown in the second right plot in Fig.~\ref{Fig:nosoc}\textbf{d}. In this scenario there are only three VHS points in the k-space. As shown in Fig.~\ref{Fig:nosoc_color}\textbf{b}, when the energy exceeds the VHS, the Fermi contour changes from two distinct electro-type Dirac pockets at the MBZ corners ($K_M$ and $K'_M$) to a single hole-type pocket that encompasses the center of the MBZ. These transformations result in a switch between electron and hole charge carriers, as indicated by a change in the sign of the Hall coefficient.




\bibliographystyle{myapsrev}
\bibliography{TBG_SOC}

\begin{thebibliography}{10}%
\makeatletter
\providecommand \@ifxundefined [1]{%
 \ifx #1\undefined \expandafter \@firstoftwo
 \else \expandafter \@secondoftwo
\fi
}%
\providecommand \@ifnum [1]{%
 \ifnum #1\expandafter \@firstoftwo
 \else \expandafter \@secondoftwo
\fi
}%
\providecommand \enquote [1]{``#1''}%
\providecommand \bibnamefont  [1]{#1}%
\providecommand \bibfnamefont [1]{#1}%
\providecommand \citenamefont [1]{#1}%
\providecommand\href[0]{\@sanitize\@href}%
\providecommand\@href[1]{\endgroup\@@startlink{#1}\endgroup\@@href}%
\providecommand\@@href[1]{#1\@@endlink}%
\providecommand \@sanitize [0]{\begingroup\catcode`\&12\catcode`\#12\relax}%
\@ifxundefined \pdfoutput {\@firstoftwo}{%
 \@ifnum{\z@=\pdfoutput}{\@firstoftwo}{\@secondoftwo}%
}{%
 \providecommand\@@startlink[1]{\leavevmode\special{html:<a href="#1">}}%
 \providecommand\@@endlink[0]{\special{html:</a>}}%
}{%
 \providecommand\@@startlink[1]{%
  \leavevmode
  \pdfstartlink
   attr{/Border[0 0 1 ]/H/I/C[0 1 1]}%
   user{/Subtype/Link/A<</Type/Action/S/URI/URI(#1)>>}%
  \relax
 }%
 \providecommand\@@endlink[0]{\pdfendlink}%
}%
\providecommand \url  [0]{\begingroup\@sanitize \@url }%
\providecommand \@url [1]{\endgroup\@href {#1}{\urlprefix}}%
\providecommand \urlprefix [0]{URL }%
\providecommand \Eprint[0]{\href }%
\@ifxundefined \urlstyle {%
  \providecommand \doi [1]{doi:\discretionary{}{}{}#1}%
}{%
  \providecommand \doi [0]{doi:\discretionary{}{}{}\begingroup \urlstyle{rm}\Url }%
}%
\providecommand \doibase [0]{http://dx.doi.org/}%
\providecommand \Doi[1]{\href{\doibase#1}}%
\providecommand \bibAnnote [3]{%
  \BibitemShut{#1}%
}%
\providecommand \bibAnnoteFile [2]{%
  \IfFileExists{#2}{\bibAnnote {#1} {#2} {\input{#2}}}{}%
}%
\providecommand \typeout [0]{\immediate \write \m@ne }%
\providecommand \bibinfo [0]{\@secondoftwo}%
\providecommand \bibfield [0]{\@secondoftwo}%
\providecommand \translation [1]{[#1]}%
\providecommand \BibitemOpen[0]{}%
\providecommand \bibitemStop [0]{}%
\providecommand \bibitemNoStop [0]{.\EOS\space}%
\providecommand \EOS [0]{\spacefactor3000\relax}%
\providecommand \BibitemShut [1]{\csname bibitem#1\endcsname}%
\bibitem{CaoY2018a}%
  \BibitemOpen
  \bibfield{author}{%
  \bibinfo {author} {\bibfnamefont{Y.}~\bibnamefont{Cao}}, \bibinfo {author} {\bibfnamefont{V.}~\bibnamefont{Fatemi}}, \bibinfo {author} {\bibfnamefont{A.}~\bibnamefont{Demir}}, \bibinfo {author} {\bibfnamefont{S.}~\bibnamefont{Fang}}, \bibinfo {author} {\bibfnamefont{S.~L.}\ \bibnamefont{Tomarken}}, \bibinfo {author} {\bibfnamefont{J.~Y.}\ \bibnamefont{Luo}}, \bibinfo {author} {\bibfnamefont{J.~D.}\ \bibnamefont{{Sanchez-Yamagishi}}}, \bibinfo {author} {\bibfnamefont{K.}~\bibnamefont{Watanabe}}, \bibinfo {author} {\bibfnamefont{T.}~\bibnamefont{Taniguchi}}, \bibinfo {author} {\bibfnamefont{E.}~\bibnamefont{Kaxiras}}, \bibinfo {author} {\bibfnamefont{R.~C.}\ \bibnamefont{Ashoori}},\ and\ \bibinfo {author} {\bibfnamefont{P.}~\bibnamefont{{Jarillo-Herrero}}},\ }%
  \Doi{10.1038/nature26154}{\emph{\bibinfo {title} {Correlated Insulator Behaviour at Half-Filling in Magic-Angle Graphene Superlattices}}},\ \bibinfo {journal} {Nature}\ \textbf{\bibinfo {volume} {556}},\ \bibinfo {pages} {80} (\bibinfo {year} {2018}).~%
  \bibAnnoteFile{Stop}{CaoY2018a}%
\bibitem{CaoY2018}%
  \BibitemOpen
  \bibfield{author}{%
  \bibinfo {author} {\bibfnamefont{Y.}~\bibnamefont{Cao}}, \bibinfo {author} {\bibfnamefont{V.}~\bibnamefont{Fatemi}}, \bibinfo {author} {\bibfnamefont{S.}~\bibnamefont{Fang}}, \bibinfo {author} {\bibfnamefont{K.}~\bibnamefont{Watanabe}}, \bibinfo {author} {\bibfnamefont{T.}~\bibnamefont{Taniguchi}}, \bibinfo {author} {\bibfnamefont{E.}~\bibnamefont{Kaxiras}},\ and\ \bibinfo {author} {\bibfnamefont{P.}~\bibnamefont{{Jarillo-Herrero}}},\ }%
  \Doi{10.1038/nature26160}{\emph{\bibinfo {title} {Unconventional Superconductivity in Magic-Angle Graphene Superlattices}}},\ \bibinfo {journal} {Nature}\ \textbf{\bibinfo {volume} {556}},\ \bibinfo {pages} {43} (\bibinfo {year} {2018}).~%
  \bibAnnoteFile{Stop}{CaoY2018}%
\bibitem{PolshynH2019}%
  \BibitemOpen
  \bibfield{author}{%
  \bibinfo {author} {\bibfnamefont{H.}~\bibnamefont{Polshyn}}, \bibinfo {author} {\bibfnamefont{M.}~\bibnamefont{Yankowitz}}, \bibinfo {author} {\bibfnamefont{S.}~\bibnamefont{Chen}}, \bibinfo {author} {\bibfnamefont{Y.}~\bibnamefont{Zhang}}, \bibinfo {author} {\bibfnamefont{K.}~\bibnamefont{Watanabe}}, \bibinfo {author} {\bibfnamefont{T.}~\bibnamefont{Taniguchi}}, \bibinfo {author} {\bibfnamefont{C.~R.}\ \bibnamefont{Dean}},\ and\ \bibinfo {author} {\bibfnamefont{A.~F.}\ \bibnamefont{Young}},\ }%
  \Doi{10.1038/s41567-019-0596-3}{\emph{\bibinfo {title} {Large Linear-in-Temperature Resistivity in Twisted Bilayer Graphene}}},\ \bibinfo {journal} {Nat. Phys.}\ \textbf{\bibinfo {volume} {15}},\ \bibinfo {pages} {1011} (\bibinfo {year} {2019}).~%
  \bibAnnoteFile{Stop}{PolshynH2019}%
\bibitem{SharpeAL2019}%
  \BibitemOpen
  \bibfield{author}{%
  \bibinfo {author} {\bibfnamefont{A.~L.}\ \bibnamefont{Sharpe}}, \bibinfo {author} {\bibfnamefont{E.~J.}\ \bibnamefont{Fox}}, \bibinfo {author} {\bibfnamefont{A.~W.}\ \bibnamefont{Barnard}}, \bibinfo {author} {\bibfnamefont{J.}~\bibnamefont{Finney}}, \bibinfo {author} {\bibfnamefont{K.}~\bibnamefont{Watanabe}}, \bibinfo {author} {\bibfnamefont{T.}~\bibnamefont{Taniguchi}}, \bibinfo {author} {\bibfnamefont{M.~A.}\ \bibnamefont{Kastner}},\ and\ \bibinfo {author} {\bibfnamefont{D.}~\bibnamefont{{Goldhaber-Gordon}}},\ }%
  \Doi{10.1126/science.aaw3780}{\emph{\bibinfo {title} {Emergent Ferromagnetism near Three-Quarters Filling in Twisted Bilayer Graphene}}},\ \bibinfo {journal} {Science}\ \textbf{\bibinfo {volume} {365}},\ \bibinfo {pages} {605} (\bibinfo {year} {2019}).~%
  \bibAnnoteFile{Stop}{SharpeAL2019}%
\bibitem{JiangY2019}%
  \BibitemOpen
  \bibfield{author}{%
  \bibinfo {author} {\bibfnamefont{Y.}~\bibnamefont{Jiang}}, \bibinfo {author} {\bibfnamefont{X.}~\bibnamefont{Lai}}, \bibinfo {author} {\bibfnamefont{K.}~\bibnamefont{Watanabe}}, \bibinfo {author} {\bibfnamefont{T.}~\bibnamefont{Taniguchi}}, \bibinfo {author} {\bibfnamefont{K.}~\bibnamefont{Haule}}, \bibinfo {author} {\bibfnamefont{J.}~\bibnamefont{Mao}},\ and\ \bibinfo {author} {\bibfnamefont{E.~Y.}\ \bibnamefont{Andrei}},\ }%
  \Doi{10.1038/s41586-019-1460-4}{\emph{\bibinfo {title} {Charge Order and Broken Rotational Symmetry in Magic-Angle Twisted Bilayer Graphene}}},\ \bibinfo {journal} {Nature}\ \textbf{\bibinfo {volume} {573}},\ \bibinfo {pages} {91} (\bibinfo {year} {2019}).~%
  \bibAnnoteFile{Stop}{JiangY2019}%
\bibitem{LuX2019}%
  \BibitemOpen
  \bibfield{author}{%
  \bibinfo {author} {\bibfnamefont{X.}~\bibnamefont{Lu}}, \bibinfo {author} {\bibfnamefont{P.}~\bibnamefont{Stepanov}}, \bibinfo {author} {\bibfnamefont{W.}~\bibnamefont{Yang}}, \bibinfo {author} {\bibfnamefont{M.}~\bibnamefont{Xie}}, \bibinfo {author} {\bibfnamefont{M.~A.}\ \bibnamefont{Aamir}}, \bibinfo {author} {\bibfnamefont{I.}~\bibnamefont{Das}}, \bibinfo {author} {\bibfnamefont{C.}~\bibnamefont{Urgell}}, \bibinfo {author} {\bibfnamefont{K.}~\bibnamefont{Watanabe}}, \bibinfo {author} {\bibfnamefont{T.}~\bibnamefont{Taniguchi}}, \bibinfo {author} {\bibfnamefont{G.}~\bibnamefont{Zhang}}, \bibinfo {author} {\bibfnamefont{A.}~\bibnamefont{Bachtold}}, \bibinfo {author} {\bibfnamefont{A.~H.}\ \bibnamefont{MacDonald}},\ and\ \bibinfo {author} {\bibfnamefont{D.~K.}\ \bibnamefont{Efetov}},\ }%
  \Doi{10.1038/s41586-019-1695-0}{\emph{\bibinfo {title} {Superconductors, Orbital Magnets and Correlated States in Magic-Angle Bilayer Graphene}}},\ \bibinfo {journal} {Nature}\ \textbf{\bibinfo {volume} {574}},\ \bibinfo {pages} {653} (\bibinfo {year} {2019}).~%
  \bibAnnoteFile{Stop}{LuX2019}%
\bibitem{YankowitzM2019}%
  \BibitemOpen
  \bibfield{author}{%
  \bibinfo {author} {\bibfnamefont{M.}~\bibnamefont{Yankowitz}}, \bibinfo {author} {\bibfnamefont{S.}~\bibnamefont{Chen}}, \bibinfo {author} {\bibfnamefont{H.}~\bibnamefont{Polshyn}}, \bibinfo {author} {\bibfnamefont{Y.}~\bibnamefont{Zhang}}, \bibinfo {author} {\bibfnamefont{K.}~\bibnamefont{Watanabe}}, \bibinfo {author} {\bibfnamefont{T.}~\bibnamefont{Taniguchi}}, \bibinfo {author} {\bibfnamefont{D.}~\bibnamefont{Graf}}, \bibinfo {author} {\bibfnamefont{A.~F.}\ \bibnamefont{Young}},\ and\ \bibinfo {author} {\bibfnamefont{C.~R.}\ \bibnamefont{Dean}},\ }%
  \Doi{10.1126/science.aav1910}{\emph{\bibinfo {title} {Tuning Superconductivity in Twisted Bilayer Graphene}}},\ \bibinfo {journal} {Science}\ \textbf{\bibinfo {volume} {363}},\ \bibinfo {pages} {1059} (\bibinfo {year} {2019}).~%
  \bibAnnoteFile{Stop}{YankowitzM2019}%
\bibitem{KerelskyA2019}%
  \BibitemOpen
  \bibfield{author}{%
  \bibinfo {author} {\bibfnamefont{A.}~\bibnamefont{Kerelsky}}, \bibinfo {author} {\bibfnamefont{L.~J.}\ \bibnamefont{McGilly}}, \bibinfo {author} {\bibfnamefont{D.~M.}\ \bibnamefont{Kennes}}, \bibinfo {author} {\bibfnamefont{L.}~\bibnamefont{Xian}}, \bibinfo {author} {\bibfnamefont{M.}~\bibnamefont{Yankowitz}}, \bibinfo {author} {\bibfnamefont{S.}~\bibnamefont{Chen}}, \bibinfo {author} {\bibfnamefont{K.}~\bibnamefont{Watanabe}}, \bibinfo {author} {\bibfnamefont{T.}~\bibnamefont{Taniguchi}}, \bibinfo {author} {\bibfnamefont{J.}~\bibnamefont{Hone}}, \bibinfo {author} {\bibfnamefont{C.}~\bibnamefont{Dean}}, \bibinfo {author} {\bibfnamefont{A.}~\bibnamefont{Rubio}},\ and\ \bibinfo {author} {\bibfnamefont{A.~N.}\ \bibnamefont{Pasupathy}},\ }%
  \Doi{10.1038/s41586-019-1431-9}{\emph{\bibinfo {title} {Maximized Electron Interactions at the Magic Angle in Twisted Bilayer Graphene}}},\ \bibinfo {journal} {Nature}\ \textbf{\bibinfo {volume} {572}},\ \bibinfo {pages} {95} (\bibinfo {year} {2019}).~%
  \bibAnnoteFile{Stop}{KerelskyA2019}%
\bibitem{XieY2019}%
  \BibitemOpen
  \bibfield{author}{%
  \bibinfo {author} {\bibfnamefont{Y.}~\bibnamefont{Xie}}, \bibinfo {author} {\bibfnamefont{B.}~\bibnamefont{Lian}}, \bibinfo {author} {\bibfnamefont{B.}~\bibnamefont{J{\"a}ck}}, \bibinfo {author} {\bibfnamefont{X.}~\bibnamefont{Liu}}, \bibinfo {author} {\bibfnamefont{C.-L.}\ \bibnamefont{Chiu}}, \bibinfo {author} {\bibfnamefont{K.}~\bibnamefont{Watanabe}}, \bibinfo {author} {\bibfnamefont{T.}~\bibnamefont{Taniguchi}}, \bibinfo {author} {\bibfnamefont{B.~A.}\ \bibnamefont{Bernevig}},\ and\ \bibinfo {author} {\bibfnamefont{A.}~\bibnamefont{Yazdani}},\ }%
  \Doi{10.1038/s41586-019-1422-x}{\emph{\bibinfo {title} {Spectroscopic Signatures of Many-Body Correlations in Magic-Angle Twisted Bilayer Graphene}}},\ \bibinfo {journal} {Nature}\ \textbf{\bibinfo {volume} {572}},\ \bibinfo {pages} {101} (\bibinfo {year} {2019}).~%
  \bibAnnoteFile{Stop}{XieY2019}%
\bibitem{ChoiY2019}%
  \BibitemOpen
  \bibfield{author}{%
  \bibinfo {author} {\bibfnamefont{Y.}~\bibnamefont{Choi}}, \bibinfo {author} {\bibfnamefont{J.}~\bibnamefont{Kemmer}}, \bibinfo {author} {\bibfnamefont{Y.}~\bibnamefont{Peng}}, \bibinfo {author} {\bibfnamefont{A.}~\bibnamefont{Thomson}}, \bibinfo {author} {\bibfnamefont{H.}~\bibnamefont{Arora}}, \bibinfo {author} {\bibfnamefont{R.}~\bibnamefont{Polski}}, \bibinfo {author} {\bibfnamefont{Y.}~\bibnamefont{Zhang}}, \bibinfo {author} {\bibfnamefont{H.}~\bibnamefont{Ren}}, \bibinfo {author} {\bibfnamefont{J.}~\bibnamefont{Alicea}}, \bibinfo {author} {\bibfnamefont{G.}~\bibnamefont{Refael}}, \bibinfo {author} {\bibfnamefont{F.}~\bibnamefont{{von Oppen}}}, \bibinfo {author} {\bibfnamefont{K.}~\bibnamefont{Watanabe}}, \bibinfo {author} {\bibfnamefont{T.}~\bibnamefont{Taniguchi}},\ and\ \bibinfo {author} {\bibfnamefont{S.}~\bibnamefont{{Nadj-Perge}}},\ }%
  \Doi{10.1038/s41567-019-0606-5}{\emph{\bibinfo {title} {Electronic Correlations in Twisted Bilayer Graphene near the Magic Angle}}},\ \bibinfo {journal} {Nat. Phys.}\ \textbf{\bibinfo {volume} {15}},\ \bibinfo {pages} {1174} (\bibinfo {year} {2019}).~%
  \bibAnnoteFile{Stop}{ChoiY2019}%
\bibitem{CaoY2020}%
  \BibitemOpen
  \bibfield{author}{%
  \bibinfo {author} {\bibfnamefont{Y.}~\bibnamefont{Cao}}, \bibinfo {author} {\bibfnamefont{D.}~\bibnamefont{Chowdhury}}, \bibinfo {author} {\bibfnamefont{D.}~\bibnamefont{{Rodan-Legrain}}}, \bibinfo {author} {\bibfnamefont{O.}~\bibnamefont{{Rubies-Bigorda}}}, \bibinfo {author} {\bibfnamefont{K.}~\bibnamefont{Watanabe}}, \bibinfo {author} {\bibfnamefont{T.}~\bibnamefont{Taniguchi}}, \bibinfo {author} {\bibfnamefont{T.}~\bibnamefont{Senthil}},\ and\ \bibinfo {author} {\bibfnamefont{P.}~\bibnamefont{{Jarillo-Herrero}}},\ }%
  \Doi{10.1103/PhysRevLett.124.076801}{\emph{\bibinfo {title} {Strange {{Metal}} in {{Magic-Angle Graphene}} with near {{Planckian Dissipation}}}}},\ \bibinfo {journal} {Phys. Rev. Lett.}\ \textbf{\bibinfo {volume} {124}},\ \bibinfo {pages} {076801} (\bibinfo {year} {2020}).~%
  \bibAnnoteFile{Stop}{CaoY2020}%
\bibitem{SerlinM2020}%
  \BibitemOpen
  \bibfield{author}{%
  \bibinfo {author} {\bibfnamefont{M.}~\bibnamefont{Serlin}}, \bibinfo {author} {\bibfnamefont{C.~L.}\ \bibnamefont{Tschirhart}}, \bibinfo {author} {\bibfnamefont{H.}~\bibnamefont{Polshyn}}, \bibinfo {author} {\bibfnamefont{Y.}~\bibnamefont{Zhang}}, \bibinfo {author} {\bibfnamefont{J.}~\bibnamefont{Zhu}}, \bibinfo {author} {\bibfnamefont{K.}~\bibnamefont{Watanabe}}, \bibinfo {author} {\bibfnamefont{T.}~\bibnamefont{Taniguchi}}, \bibinfo {author} {\bibfnamefont{L.}~\bibnamefont{Balents}},\ and\ \bibinfo {author} {\bibfnamefont{A.~F.}\ \bibnamefont{Young}},\ }%
  \Doi{10.1126/science.aay5533}{\emph{\bibinfo {title} {Intrinsic Quantized Anomalous {{Hall}} Effect in a Moir{\'e} Heterostructure}}},\ \bibinfo {journal} {Science}\ \textbf{\bibinfo {volume} {367}},\ \bibinfo {pages} {900} (\bibinfo {year} {2020}).~%
  \bibAnnoteFile{Stop}{SerlinM2020}%
\bibitem{AroraHS2020}%
  \BibitemOpen
  \bibfield{author}{%
  \bibinfo {author} {\bibfnamefont{H.~S.}\ \bibnamefont{Arora}}, \bibinfo {author} {\bibfnamefont{R.}~\bibnamefont{Polski}}, \bibinfo {author} {\bibfnamefont{Y.}~\bibnamefont{Zhang}}, \bibinfo {author} {\bibfnamefont{A.}~\bibnamefont{Thomson}}, \bibinfo {author} {\bibfnamefont{Y.}~\bibnamefont{Choi}}, \bibinfo {author} {\bibfnamefont{H.}~\bibnamefont{Kim}}, \bibinfo {author} {\bibfnamefont{Z.}~\bibnamefont{Lin}}, \bibinfo {author} {\bibfnamefont{I.~Z.}\ \bibnamefont{Wilson}}, \bibinfo {author} {\bibfnamefont{X.}~\bibnamefont{Xu}}, \bibinfo {author} {\bibfnamefont{J.-H.}\ \bibnamefont{Chu}}, \bibinfo {author} {\bibfnamefont{K.}~\bibnamefont{Watanabe}}, \bibinfo {author} {\bibfnamefont{T.}~\bibnamefont{Taniguchi}}, \bibinfo {author} {\bibfnamefont{J.}~\bibnamefont{Alicea}},\ and\ \bibinfo {author} {\bibfnamefont{S.}~\bibnamefont{{Nadj-Perge}}},\ }%
  \Doi{10.1038/s41586-020-2473-8}{\emph{\bibinfo {title} {Superconductivity in Metallic Twisted Bilayer Graphene Stabilized by {{WSe$_2$}}}}},\ \bibinfo {journal} {Nature}\ \textbf{\bibinfo {volume} {583}},\ \bibinfo {pages} {379} (\bibinfo {year} {2020}).~%
  \bibAnnoteFile{Stop}{AroraHS2020}%
\bibitem{WongD2020}%
  \BibitemOpen
  \bibfield{author}{%
  \bibinfo {author} {\bibfnamefont{D.}~\bibnamefont{Wong}}, \bibinfo {author} {\bibfnamefont{K.~P.}\ \bibnamefont{Nuckolls}}, \bibinfo {author} {\bibfnamefont{M.}~\bibnamefont{Oh}}, \bibinfo {author} {\bibfnamefont{B.}~\bibnamefont{Lian}}, \bibinfo {author} {\bibfnamefont{Y.}~\bibnamefont{Xie}}, \bibinfo {author} {\bibfnamefont{S.}~\bibnamefont{Jeon}}, \bibinfo {author} {\bibfnamefont{K.}~\bibnamefont{Watanabe}}, \bibinfo {author} {\bibfnamefont{T.}~\bibnamefont{Taniguchi}}, \bibinfo {author} {\bibfnamefont{B.~A.}\ \bibnamefont{Bernevig}},\ and\ \bibinfo {author} {\bibfnamefont{A.}~\bibnamefont{Yazdani}},\ }%
  \Doi{10.1038/s41586-020-2339-0}{\emph{\bibinfo {title} {Cascade of Electronic Transitions in Magic-Angle Twisted Bilayer Graphene}}},\ \bibinfo {journal} {Nature}\ \textbf{\bibinfo {volume} {582}},\ \bibinfo {pages} {198} (\bibinfo {year} {2020}).~%
  \bibAnnoteFile{Stop}{WongD2020}%
\bibitem{ChoiY2021}%
  \BibitemOpen
  \bibfield{author}{%
  \bibinfo {author} {\bibfnamefont{Y.}~\bibnamefont{Choi}}, \bibinfo {author} {\bibfnamefont{H.}~\bibnamefont{Kim}}, \bibinfo {author} {\bibfnamefont{Y.}~\bibnamefont{Peng}}, \bibinfo {author} {\bibfnamefont{A.}~\bibnamefont{Thomson}}, \bibinfo {author} {\bibfnamefont{C.}~\bibnamefont{Lewandowski}}, \bibinfo {author} {\bibfnamefont{R.}~\bibnamefont{Polski}}, \bibinfo {author} {\bibfnamefont{Y.}~\bibnamefont{Zhang}}, \bibinfo {author} {\bibfnamefont{H.~S.}\ \bibnamefont{Arora}}, \bibinfo {author} {\bibfnamefont{K.}~\bibnamefont{Watanabe}}, \bibinfo {author} {\bibfnamefont{T.}~\bibnamefont{Taniguchi}}, \bibinfo {author} {\bibfnamefont{J.}~\bibnamefont{Alicea}},\ and\ \bibinfo {author} {\bibfnamefont{S.}~\bibnamefont{{Nadj-Perge}}},\ }%
  \Doi{10.1038/s41586-020-03159-7}{\emph{\bibinfo {title} {Correlation-Driven Topological Phases in Magic-Angle Twisted Bilayer Graphene}}},\ \bibinfo {journal} {Nature}\ \textbf{\bibinfo {volume} {589}},\ \bibinfo {pages} {536} (\bibinfo {year} {2021}).~%
  \bibAnnoteFile{Stop}{ChoiY2021}%
\bibitem{OhM2021}%
  \BibitemOpen
  \bibfield{author}{%
  \bibinfo {author} {\bibfnamefont{M.}~\bibnamefont{Oh}}, \bibinfo {author} {\bibfnamefont{K.~P.}\ \bibnamefont{Nuckolls}}, \bibinfo {author} {\bibfnamefont{D.}~\bibnamefont{Wong}}, \bibinfo {author} {\bibfnamefont{R.~L.}\ \bibnamefont{Lee}}, \bibinfo {author} {\bibfnamefont{X.}~\bibnamefont{Liu}}, \bibinfo {author} {\bibfnamefont{K.}~\bibnamefont{Watanabe}}, \bibinfo {author} {\bibfnamefont{T.}~\bibnamefont{Taniguchi}},\ and\ \bibinfo {author} {\bibfnamefont{A.}~\bibnamefont{Yazdani}},\ }%
  \Doi{10.1038/s41586-021-04121-x}{\emph{\bibinfo {title} {Evidence for Unconventional Superconductivity in Twisted Bilayer Graphene}}},\ \bibinfo {journal} {Nature}\ \textbf{\bibinfo {volume} {600}},\ \bibinfo {pages} {240} (\bibinfo {year} {2021}).~%
  \bibAnnoteFile{Stop}{OhM2021}%
\bibitem{LiuX2020}%
  \BibitemOpen
  \bibfield{author}{%
  \bibinfo {author} {\bibfnamefont{X.}~\bibnamefont{Liu}}, \bibinfo {author} {\bibfnamefont{Z.}~\bibnamefont{Hao}}, \bibinfo {author} {\bibfnamefont{E.}~\bibnamefont{Khalaf}}, \bibinfo {author} {\bibfnamefont{J.~Y.}\ \bibnamefont{Lee}}, \bibinfo {author} {\bibfnamefont{Y.}~\bibnamefont{Ronen}}, \bibinfo {author} {\bibfnamefont{H.}~\bibnamefont{Yoo}}, \bibinfo {author} {\bibfnamefont{D.}~\bibnamefont{Haei~Najafabadi}}, \bibinfo {author} {\bibfnamefont{K.}~\bibnamefont{Watanabe}}, \bibinfo {author} {\bibfnamefont{T.}~\bibnamefont{Taniguchi}}, \bibinfo {author} {\bibfnamefont{A.}~\bibnamefont{Vishwanath}},\ and\ \bibinfo {author} {\bibfnamefont{P.}~\bibnamefont{Kim}},\ }%
  \Doi{10.1038/s41586-020-2458-7}{\emph{\bibinfo {title} {Tunable Spin-Polarized Correlated States in Twisted Double Bilayer Graphene}}},\ \bibinfo {journal} {Nature}\ \textbf{\bibinfo {volume} {583}},\ \bibinfo {pages} {221} (\bibinfo {year} {2020}).~%
  \bibAnnoteFile{Stop}{LiuX2020}%
\bibitem{HaoZ2021}%
  \BibitemOpen
  \bibfield{author}{%
  \bibinfo {author} {\bibfnamefont{Z.}~\bibnamefont{Hao}}, \bibinfo {author} {\bibfnamefont{A.~M.}\ \bibnamefont{Zimmerman}}, \bibinfo {author} {\bibfnamefont{P.}~\bibnamefont{Ledwith}}, \bibinfo {author} {\bibfnamefont{E.}~\bibnamefont{Khalaf}}, \bibinfo {author} {\bibfnamefont{D.~H.}\ \bibnamefont{Najafabadi}}, \bibinfo {author} {\bibfnamefont{K.}~\bibnamefont{Watanabe}}, \bibinfo {author} {\bibfnamefont{T.}~\bibnamefont{Taniguchi}}, \bibinfo {author} {\bibfnamefont{A.}~\bibnamefont{Vishwanath}},\ and\ \bibinfo {author} {\bibfnamefont{P.}~\bibnamefont{Kim}},\ }%
  \Doi{10.1126/science.abg0399}{\emph{\bibinfo {title} {Electric Field{\textendash}Tunable Superconductivity in Alternating-Twist Magic-Angle Trilayer Graphene}}},\ \bibinfo {journal} {Science}\ \textbf{\bibinfo {volume} {371}},\ \bibinfo {pages} {1133} (\bibinfo {year} {2021}).~%
  \bibAnnoteFile{Stop}{HaoZ2021}%
\bibitem{ZhouH2021}%
  \BibitemOpen
  \bibfield{author}{%
  \bibinfo {author} {\bibfnamefont{H.}~\bibnamefont{Zhou}}, \bibinfo {author} {\bibfnamefont{T.}~\bibnamefont{Xie}}, \bibinfo {author} {\bibfnamefont{T.}~\bibnamefont{Taniguchi}}, \bibinfo {author} {\bibfnamefont{K.}~\bibnamefont{Watanabe}},\ and\ \bibinfo {author} {\bibfnamefont{A.~F.}\ \bibnamefont{Young}},\ }%
  \Doi{10.1038/s41586-021-03926-0}{\emph{\bibinfo {title} {Superconductivity in Rhombohedral Trilayer Graphene}}},\ \bibinfo {journal} {Nature}\ \textbf{\bibinfo {volume} {598}},\ \bibinfo {pages} {434} (\bibinfo {year} {2021}).~%
  \bibAnnoteFile{Stop}{ZhouH2021}%
\bibitem{ZhouH2021a}%
  \BibitemOpen
  \bibfield{author}{%
  \bibinfo {author} {\bibfnamefont{H.}~\bibnamefont{Zhou}}, \bibinfo {author} {\bibfnamefont{T.}~\bibnamefont{Xie}}, \bibinfo {author} {\bibfnamefont{A.}~\bibnamefont{Ghazaryan}}, \bibinfo {author} {\bibfnamefont{T.}~\bibnamefont{Holder}}, \bibinfo {author} {\bibfnamefont{J.~R.}\ \bibnamefont{Ehrets}}, \bibinfo {author} {\bibfnamefont{E.~M.}\ \bibnamefont{Spanton}}, \bibinfo {author} {\bibfnamefont{T.}~\bibnamefont{Taniguchi}}, \bibinfo {author} {\bibfnamefont{K.}~\bibnamefont{Watanabe}}, \bibinfo {author} {\bibfnamefont{E.}~\bibnamefont{Berg}}, \bibinfo {author} {\bibfnamefont{M.}~\bibnamefont{Serbyn}},\ and\ \bibinfo {author} {\bibfnamefont{A.~F.}\ \bibnamefont{Young}},\ }%
  \Doi{10.1038/s41586-021-03938-w}{\emph{\bibinfo {title} {Half- and Quarter-Metals in Rhombohedral Trilayer Graphene}}},\ \bibinfo {journal} {Nature}\ \textbf{\bibinfo {volume} {598}},\ \bibinfo {pages} {429} (\bibinfo {year} {2021}).~%
  \bibAnnoteFile{Stop}{ZhouH2021a}%
\bibitem{ParkJM2021}%
  \BibitemOpen
  \bibfield{author}{%
  \bibinfo {author} {\bibfnamefont{J.~M.}\ \bibnamefont{Park}}, \bibinfo {author} {\bibfnamefont{Y.}~\bibnamefont{Cao}}, \bibinfo {author} {\bibfnamefont{K.}~\bibnamefont{Watanabe}}, \bibinfo {author} {\bibfnamefont{T.}~\bibnamefont{Taniguchi}},\ and\ \bibinfo {author} {\bibfnamefont{P.}~\bibnamefont{{Jarillo-Herrero}}},\ }%
  \Doi{10.1038/s41586-021-03192-0}{\emph{\bibinfo {title} {Tunable Strongly Coupled Superconductivity in Magic-Angle Twisted Trilayer Graphene}}},\ \bibinfo {journal} {Nature}\ \textbf{\bibinfo {volume} {590}},\ \bibinfo {pages} {249} (\bibinfo {year} {2021}).~%
  \bibAnnoteFile{Stop}{ParkJM2021}%
\bibitem{ParkJM2021a}%
  \BibitemOpen
  \bibfield{author}{%
  \bibinfo {author} {\bibfnamefont{J.~M.}\ \bibnamefont{Park}}, \bibinfo {author} {\bibfnamefont{Y.}~\bibnamefont{Cao}}, \bibinfo {author} {\bibfnamefont{K.}~\bibnamefont{Watanabe}}, \bibinfo {author} {\bibfnamefont{T.}~\bibnamefont{Taniguchi}},\ and\ \bibinfo {author} {\bibfnamefont{P.}~\bibnamefont{{Jarillo-Herrero}}},\ }%
  \Doi{10.1038/s41586-021-03366-w}{\emph{\bibinfo {title} {Flavour {{Hund}}'s Coupling, {{Chern}} Gaps and Charge Diffusivity in Moir{\'e} Graphene}}},\ \bibinfo {journal} {Nature}\ \textbf{\bibinfo {volume} {592}},\ \bibinfo {pages} {43} (\bibinfo {year} {2021}).~%
  \bibAnnoteFile{Stop}{ParkJM2021a}%
\bibitem{LiuX2021}%
  \BibitemOpen
  \bibfield{author}{%
  \bibinfo {author} {\bibfnamefont{X.}~\bibnamefont{Liu}}, \bibinfo {author} {\bibfnamefont{Z.}~\bibnamefont{Wang}}, \bibinfo {author} {\bibfnamefont{K.}~\bibnamefont{Watanabe}}, \bibinfo {author} {\bibfnamefont{T.}~\bibnamefont{Taniguchi}}, \bibinfo {author} {\bibfnamefont{O.}~\bibnamefont{Vafek}},\ and\ \bibinfo {author} {\bibfnamefont{J.~I.~A.}\ \bibnamefont{Li}},\ }%
  \Doi{10.1126/science.abb8754}{\emph{\bibinfo {title} {Tuning Electron Correlation in Magic-Angle Twisted Bilayer Graphene Using {{Coulomb}} Screening}}},\ \bibinfo {journal} {Science}\ \textbf{\bibinfo {volume} {371}},\ \bibinfo {pages} {1261} (\bibinfo {year} {2021}).~%
  \bibAnnoteFile{Stop}{LiuX2021}%
\bibitem{ZhouH2022}%
  \BibitemOpen
  \bibfield{author}{%
  \bibinfo {author} {\bibfnamefont{H.}~\bibnamefont{Zhou}}, \bibinfo {author} {\bibfnamefont{L.}~\bibnamefont{Holleis}}, \bibinfo {author} {\bibfnamefont{Y.}~\bibnamefont{Saito}}, \bibinfo {author} {\bibfnamefont{L.}~\bibnamefont{Cohen}}, \bibinfo {author} {\bibfnamefont{W.}~\bibnamefont{Huynh}}, \bibinfo {author} {\bibfnamefont{C.~L.}\ \bibnamefont{Patterson}}, \bibinfo {author} {\bibfnamefont{F.}~\bibnamefont{Yang}}, \bibinfo {author} {\bibfnamefont{T.}~\bibnamefont{Taniguchi}}, \bibinfo {author} {\bibfnamefont{K.}~\bibnamefont{Watanabe}},\ and\ \bibinfo {author} {\bibfnamefont{A.~F.}\ \bibnamefont{Young}},\ }%
  \Doi{10.1126/science.abm8386}{\emph{\bibinfo {title} {Isospin Magnetism and Spin-Polarized Superconductivity in {{Bernal}} Bilayer Graphene}}},\ \bibinfo {journal} {Science}\ \textbf{\bibinfo {volume} {375}},\ \bibinfo {pages} {774} (\bibinfo {year} {2022}).~%
  \bibAnnoteFile{Stop}{ZhouH2022}%
\bibitem{KuiriM2022}%
  \BibitemOpen
  \bibfield{author}{%
  \bibinfo {author} {\bibfnamefont{M.}~\bibnamefont{Kuiri}}, \bibinfo {author} {\bibfnamefont{C.}~\bibnamefont{Coleman}}, \bibinfo {author} {\bibfnamefont{Z.}~\bibnamefont{Gao}}, \bibinfo {author} {\bibfnamefont{A.}~\bibnamefont{Vishnuradhan}}, \bibinfo {author} {\bibfnamefont{K.}~\bibnamefont{Watanabe}}, \bibinfo {author} {\bibfnamefont{T.}~\bibnamefont{Taniguchi}}, \bibinfo {author} {\bibfnamefont{J.}~\bibnamefont{Zhu}}, \bibinfo {author} {\bibfnamefont{A.~H.}\ \bibnamefont{MacDonald}},\ and\ \bibinfo {author} {\bibfnamefont{J.}~\bibnamefont{Folk}},\ }%
  \Doi{10.1038/s41467-022-34192-x}{\emph{\bibinfo {title} {Spontaneous Time-Reversal Symmetry Breaking in Twisted Double Bilayer Graphene}}},\ \bibinfo {journal} {Nat Commun}\ \textbf{\bibinfo {volume} {13}},\ \bibinfo {pages} {6468} (\bibinfo {year} {2022}).~%
  \bibAnnoteFile{Stop}{KuiriM2022}%
\bibitem{ZhangY2023}%
  \BibitemOpen
  \bibfield{author}{%
  \bibinfo {author} {\bibfnamefont{Y.}~\bibnamefont{Zhang}}, \bibinfo {author} {\bibfnamefont{R.}~\bibnamefont{Polski}}, \bibinfo {author} {\bibfnamefont{A.}~\bibnamefont{Thomson}}, \bibinfo {author} {\bibfnamefont{{\'E}.}~\bibnamefont{{Lantagne-Hurtubise}}}, \bibinfo {author} {\bibfnamefont{C.}~\bibnamefont{Lewandowski}}, \bibinfo {author} {\bibfnamefont{H.}~\bibnamefont{Zhou}}, \bibinfo {author} {\bibfnamefont{K.}~\bibnamefont{Watanabe}}, \bibinfo {author} {\bibfnamefont{T.}~\bibnamefont{Taniguchi}}, \bibinfo {author} {\bibfnamefont{J.}~\bibnamefont{Alicea}},\ and\ \bibinfo {author} {\bibfnamefont{S.}~\bibnamefont{{Nadj-Perge}}},\ }%
  \Doi{10.1038/s41586-022-05446-x}{\emph{\bibinfo {title} {Enhanced Superconductivity in Spin{\textendash}Orbit Proximitized Bilayer Graphene}}},\ \bibinfo {journal} {Nature}\ \textbf{\bibinfo {volume} {613}},\ \bibinfo {pages} {268} (\bibinfo {year} {2023}).~%
  \bibAnnoteFile{Stop}{ZhangY2023}%
\bibitem{HolleisL2023}%
  \BibitemOpen
  \bibfield{author}{%
  \bibinfo {author} {\bibfnamefont{L.}~\bibnamefont{Holleis}}, \bibinfo {author} {\bibfnamefont{C.~L.}\ \bibnamefont{Patterson}}, \bibinfo {author} {\bibfnamefont{Y.}~\bibnamefont{Zhang}}, \bibinfo {author} {\bibfnamefont{Y.}~\bibnamefont{Vituri}}, \bibinfo {author} {\bibfnamefont{H.~M.}\ \bibnamefont{Yoo}}, \bibinfo {author} {\bibfnamefont{H.}~\bibnamefont{Zhou}}, \bibinfo {author} {\bibfnamefont{T.}~\bibnamefont{Taniguchi}}, \bibinfo {author} {\bibfnamefont{K.}~\bibnamefont{Watanabe}}, \bibinfo {author} {\bibfnamefont{E.}~\bibnamefont{Berg}}, \bibinfo {author} {\bibfnamefont{S.}~\bibnamefont{Nadj-Perge}},\ and\ \bibinfo {author} {\bibfnamefont{A.~F.}\ \bibnamefont{Young}},\ }%
  {\bibinfo {title} {Nematicity and orbital depairing in superconducting bernal bilayer graphene with strong spin orbit coupling},}\  (\bibinfo {year} {2024}),\ \Eprint{http://arxiv.org/abs/2303.00742}{arXiv:2303.00742 [cond-mat.supr-con]}~%
  \bibAnnoteFile{NoStop}{HolleisL2023}%
\bibitem{SuR2023}%
  \BibitemOpen
  \bibfield{author}{%
  \bibinfo {author} {\bibfnamefont{R.}~\bibnamefont{Su}}, \bibinfo {author} {\bibfnamefont{M.}~\bibnamefont{Kuiri}}, \bibinfo {author} {\bibfnamefont{K.}~\bibnamefont{Watanabe}}, \bibinfo {author} {\bibfnamefont{T.}~\bibnamefont{Taniguchi}},\ and\ \bibinfo {author} {\bibfnamefont{J.}~\bibnamefont{Folk}},\ }%
  \Doi{10.1038/s41563-023-01653-7}{\emph{\bibinfo {title} {Superconductivity in Twisted Double Bilayer Graphene Stabilized by {{WSe$_2$}}}}},\ \bibinfo {journal} {Nat. Mater.}\ \textbf{\bibinfo {volume} {22}},\ \bibinfo {pages} {1332} (\bibinfo {year} {2023}).~%
  \bibAnnoteFile{Stop}{SuR2023}%
\bibitem{PixleyJH2019}%
  \BibitemOpen
  \bibfield{author}{%
  \bibinfo {author} {\bibfnamefont{J.~H.}\ \bibnamefont{Pixley}}\ and\ \bibinfo {author} {\bibfnamefont{E.~Y.}\ \bibnamefont{Andrei}},\ }%
  \Doi{10.1126/science.aay3409}{\emph{\bibinfo {title} {Ferromagnetism in Magic-Angle Graphene}}},\ \bibinfo {journal} {Science}\ \textbf{\bibinfo {volume} {365}},\ \bibinfo {pages} {543} (\bibinfo {year} {2019}).~%
  \bibAnnoteFile{Stop}{PixleyJH2019}%
\bibitem{AndreiEY2020}%
  \BibitemOpen
  \bibfield{author}{%
  \bibinfo {author} {\bibfnamefont{E.~Y.}\ \bibnamefont{Andrei}}\ and\ \bibinfo {author} {\bibfnamefont{A.~H.}\ \bibnamefont{MacDonald}},\ }%
  \Doi{10.1038/s41563-020-00840-0}{\emph{\bibinfo {title} {Graphene Bilayers with a Twist}}},\ \bibinfo {journal} {Nat. Mater.}\ \textbf{\bibinfo {volume} {19}},\ \bibinfo {pages} {1265} (\bibinfo {year} {2020}).~%
  \bibAnnoteFile{Stop}{AndreiEY2020}%
\bibitem{BalentsL2020}%
  \BibitemOpen
  \bibfield{author}{%
  \bibinfo {author} {\bibfnamefont{L.}~\bibnamefont{Balents}}, \bibinfo {author} {\bibfnamefont{C.~R.}\ \bibnamefont{Dean}}, \bibinfo {author} {\bibfnamefont{D.~K.}\ \bibnamefont{Efetov}},\ and\ \bibinfo {author} {\bibfnamefont{A.~F.}\ \bibnamefont{Young}},\ }%
  \Doi{10.1038/s41567-020-0906-9}{\emph{\bibinfo {title} {Superconductivity and Strong Correlations in Moir{\'e} Flat Bands}}},\ \bibinfo {journal} {Nat. Phys.}\ \textbf{\bibinfo {volume} {16}},\ \bibinfo {pages} {725} (\bibinfo {year} {2020}).~%
  \bibAnnoteFile{Stop}{BalentsL2020}%
\bibitem{AndreiEY2021}%
  \BibitemOpen
  \bibfield{author}{%
  \bibinfo {author} {\bibfnamefont{E.~Y.}\ \bibnamefont{Andrei}}, \bibinfo {author} {\bibfnamefont{D.~K.}\ \bibnamefont{Efetov}}, \bibinfo {author} {\bibfnamefont{P.}~\bibnamefont{{Jarillo-Herrero}}}, \bibinfo {author} {\bibfnamefont{A.~H.}\ \bibnamefont{MacDonald}}, \bibinfo {author} {\bibfnamefont{K.~F.}\ \bibnamefont{Mak}}, \bibinfo {author} {\bibfnamefont{T.}~\bibnamefont{Senthil}}, \bibinfo {author} {\bibfnamefont{E.}~\bibnamefont{Tutuc}}, \bibinfo {author} {\bibfnamefont{A.}~\bibnamefont{Yazdani}},\ and\ \bibinfo {author} {\bibfnamefont{A.~F.}\ \bibnamefont{Young}},\ }%
  \Doi{10.1038/s41578-021-00284-1}{\emph{\bibinfo {title} {The Marvels of Moir{\'e} Materials}}},\ \bibinfo {journal} {Nat Rev Mater}\ \textbf{\bibinfo {volume} {6}},\ \bibinfo {pages} {201} (\bibinfo {year} {2021}).~%
  \bibAnnoteFile{Stop}{AndreiEY2021}%
\bibitem{khosravian_moire-enabled_2024}%
  \BibitemOpen
  \bibfield{author}{%
  \bibinfo {author} {\bibfnamefont{M.}~\bibnamefont{Khosravian}}, \bibinfo {author} {\bibfnamefont{E.}~\bibnamefont{Bascones}},\ and\ \bibinfo {author} {\bibfnamefont{J.~L.}\ \bibnamefont{Lado}},\ }%
  \Doi{10.1088/2053-1583/ad3b0c}{\emph{\bibinfo {title} {Moiré-enabled topological superconductivity in twisted bilayer graphene}}},\ \bibinfo {journal} {2D Materials}\ \textbf{\bibinfo {volume} {11}},\ \bibinfo {pages} {035012} (\bibinfo {year} {2024}).~%
  \bibAnnoteFile{Stop}{khosravian_moire-enabled_2024}%
\bibitem{LiY2019}%
  \BibitemOpen
  \bibfield{author}{%
  \bibinfo {author} {\bibfnamefont{Y.}~\bibnamefont{Li}}\ and\ \bibinfo {author} {\bibfnamefont{M.}~\bibnamefont{Koshino}},\ }%
  \Doi{10.1103/PhysRevB.99.075438}{\emph{\bibinfo {title} {Twist-Angle Dependence of the Proximity Spin-Orbit Coupling in Graphene on Transition-Metal Dichalcogenides}}},\ \bibinfo {journal} {Phys. Rev. B}\ \textbf{\bibinfo {volume} {99}},\ \bibinfo {pages} {075438} (\bibinfo {year} {2019}).~%
  \bibAnnoteFile{Stop}{LiY2019}%
\bibitem{parker_strain-induced_2021}%
  \BibitemOpen
  \bibfield{author}{%
  \bibinfo {author} {\bibfnamefont{D.~E.}\ \bibnamefont{Parker}}, \bibinfo {author} {\bibfnamefont{T.}~\bibnamefont{Soejima}}, \bibinfo {author} {\bibfnamefont{J.}~\bibnamefont{Hauschild}}, \bibinfo {author} {\bibfnamefont{M.~P.}\ \bibnamefont{Zaletel}},\ and\ \bibinfo {author} {\bibfnamefont{N.}~\bibnamefont{Bultinck}},\ }%
  \Doi{10.1103/PhysRevLett.127.027601}{\emph{\bibinfo {title} {Strain-{Induced} {Quantum} {Phase} {Transitions} in {Magic}-{Angle} {Graphene}}}},\ \bibinfo {journal} {Physical Review Letters}\ \textbf{\bibinfo {volume} {127}},\ \bibinfo {pages} {027601} (\bibinfo {year} {2021}).~%
  \bibAnnoteFile{Stop}{parker_strain-induced_2021}%
\bibitem{NaimerT2021}%
  \BibitemOpen
  \bibfield{author}{%
  \bibinfo {author} {\bibfnamefont{T.}~\bibnamefont{Naimer}}, \bibinfo {author} {\bibfnamefont{K.}~\bibnamefont{Zollner}}, \bibinfo {author} {\bibfnamefont{M.}~\bibnamefont{Gmitra}},\ and\ \bibinfo {author} {\bibfnamefont{J.}~\bibnamefont{Fabian}},\ }%
  \Doi{10.1103/PhysRevB.104.195156}{\emph{\bibinfo {title} {Twist-Angle Dependent Proximity Induced Spin-Orbit Coupling in Graphene/Transition Metal Dichalcogenide Heterostructures}}},\ \bibinfo {journal} {Phys. Rev. B}\ \textbf{\bibinfo {volume} {104}},\ \bibinfo {pages} {195156} (\bibinfo {year} {2021}).~%
  \bibAnnoteFile{Stop}{NaimerT2021}%
\bibitem{ZollnerK2023a}%
  \BibitemOpen
  \bibfield{author}{%
  \bibinfo {author} {\bibfnamefont{K.}~\bibnamefont{Zollner}}, \bibinfo {author} {\bibfnamefont{S.~M.}\ \bibnamefont{Jo{\~a}o}}, \bibinfo {author} {\bibfnamefont{B.~K.}\ \bibnamefont{Nikoli{\'c}}},\ and\ \bibinfo {author} {\bibfnamefont{J.}~\bibnamefont{Fabian}},\ }%
  \Doi{10.1103/PhysRevB.108.235166}{\emph{\bibinfo {title} {Twist- and Gate-Tunable Proximity Spin-Orbit Coupling, Spin Relaxation Anisotropy, and Charge-to-Spin Conversion in Heterostructures of Graphene and Transition Metal Dichalcogenides}}},\ \bibinfo {journal} {Phys. Rev. B}\ \textbf{\bibinfo {volume} {108}},\ \bibinfo {pages} {235166} (\bibinfo {year} {2023}).~%
  \bibAnnoteFile{Stop}{ZollnerK2023a}%
\bibitem{zheng_gate-defined_2024}%
  \BibitemOpen
  \bibfield{author}{%
  \bibinfo {author} {\bibfnamefont{G.}~\bibnamefont{Zheng}}, \bibinfo {author} {\bibfnamefont{E.}~\bibnamefont{Portolés}}, \bibinfo {author} {\bibfnamefont{A.}~\bibnamefont{Mestre-Torà}}, \bibinfo {author} {\bibfnamefont{M.}~\bibnamefont{Perego}}, \bibinfo {author} {\bibfnamefont{T.}~\bibnamefont{Taniguchi}}, \bibinfo {author} {\bibfnamefont{K.}~\bibnamefont{Watanabe}}, \bibinfo {author} {\bibfnamefont{P.}~\bibnamefont{Rickhaus}}, \bibinfo {author} {\bibfnamefont{F.~K.}\ \bibnamefont{de~Vries}}, \bibinfo {author} {\bibfnamefont{T.}~\bibnamefont{Ihn}}, \bibinfo {author} {\bibfnamefont{K.}~\bibnamefont{Ensslin}},\ and\ \bibinfo {author} {\bibfnamefont{S.}~\bibnamefont{Iwakiri}},\ }%
  \Doi{10.1103/PhysRevResearch.6.L012051}{\emph{\bibinfo {title} {Gate-defined superconducting channel in magic-angle twisted bilayer graphene}}},\ \bibinfo {journal} {Physical Review Research}\ \textbf{\bibinfo {volume} {6}},\ \bibinfo {pages} {L012051} (\bibinfo {year} {2024}).~%
  \bibAnnoteFile{Stop}{zheng_gate-defined_2024}%
\bibitem{avsar_spinorbit_2014}%
  \BibitemOpen
  \bibfield{author}{%
  \bibinfo {author} {\bibfnamefont{A.}~\bibnamefont{Avsar}}, \bibinfo {author} {\bibfnamefont{J.~Y.}\ \bibnamefont{Tan}}, \bibinfo {author} {\bibfnamefont{T.}~\bibnamefont{Taychatanapat}}, \bibinfo {author} {\bibfnamefont{J.}~\bibnamefont{Balakrishnan}}, \bibinfo {author} {\bibfnamefont{G.~K.~W.}\ \bibnamefont{Koon}}, \bibinfo {author} {\bibfnamefont{Y.}~\bibnamefont{Yeo}}, \bibinfo {author} {\bibfnamefont{J.}~\bibnamefont{Lahiri}}, \bibinfo {author} {\bibfnamefont{A.}~\bibnamefont{Carvalho}}, \bibinfo {author} {\bibfnamefont{A.~S.}\ \bibnamefont{Rodin}}, \bibinfo {author} {\bibfnamefont{E.~C.~T.}\ \bibnamefont{O’Farrell}}, \bibinfo {author} {\bibfnamefont{G.}~\bibnamefont{Eda}}, \bibinfo {author} {\bibfnamefont{A.~H.}\ \bibnamefont{Castro~Neto}},\ and\ \bibinfo {author} {\bibfnamefont{B.}~\bibnamefont{Ozyilmaz}},\ }%
  \Doi{10.1038/ncomms5875}{\emph{\bibinfo {title} {Spin–orbit proximity effect in graphene}}},\ \bibinfo {journal} {Nature Communications}\ \textbf{\bibinfo {volume} {5}},\ \bibinfo {pages} {4875} (\bibinfo {year} {2014}).~%
  \bibAnnoteFile{Stop}{avsar_spinorbit_2014}%
\bibitem{IslandJO2019}%
  \BibitemOpen
  \bibfield{author}{%
  \bibinfo {author} {\bibfnamefont{J.~O.}\ \bibnamefont{Island}}, \bibinfo {author} {\bibfnamefont{X.}~\bibnamefont{Cui}}, \bibinfo {author} {\bibfnamefont{C.}~\bibnamefont{Lewandowski}}, \bibinfo {author} {\bibfnamefont{J.~Y.}\ \bibnamefont{Khoo}}, \bibinfo {author} {\bibfnamefont{E.~M.}\ \bibnamefont{Spanton}}, \bibinfo {author} {\bibfnamefont{H.}~\bibnamefont{Zhou}}, \bibinfo {author} {\bibfnamefont{D.}~\bibnamefont{Rhodes}}, \bibinfo {author} {\bibfnamefont{J.~C.}\ \bibnamefont{Hone}}, \bibinfo {author} {\bibfnamefont{T.}~\bibnamefont{Taniguchi}}, \bibinfo {author} {\bibfnamefont{K.}~\bibnamefont{Watanabe}}, \bibinfo {author} {\bibfnamefont{L.~S.}\ \bibnamefont{Levitov}}, \bibinfo {author} {\bibfnamefont{M.~P.}\ \bibnamefont{Zaletel}},\ and\ \bibinfo {author} {\bibfnamefont{A.~F.}\ \bibnamefont{Young}},\ }%
  \Doi{10.1038/s41586-019-1304-2}{\emph{\bibinfo {title} {Spin{\textendash}Orbit-Driven Band Inversion in Bilayer Graphene by the van Der {{Waals}} Proximity Effect}}},\ \bibinfo {journal} {Nature}\ \textbf{\bibinfo {volume} {571}},\ \bibinfo {pages} {85} (\bibinfo {year} {2019}).~%
  \bibAnnoteFile{Stop}{IslandJO2019}%
\bibitem{SunL2023}%
  \BibitemOpen
  \bibfield{author}{%
  \bibinfo {author} {\bibfnamefont{L.}~\bibnamefont{Sun}}, \bibinfo {author} {\bibfnamefont{L.}~\bibnamefont{Rademaker}}, \bibinfo {author} {\bibfnamefont{D.}~\bibnamefont{Mauro}}, \bibinfo {author} {\bibfnamefont{A.}~\bibnamefont{Scarfato}}, \bibinfo {author} {\bibfnamefont{{\'A}.}~\bibnamefont{P{\'a}sztor}}, \bibinfo {author} {\bibfnamefont{I.}~\bibnamefont{{Guti{\'e}rrez-Lezama}}}, \bibinfo {author} {\bibfnamefont{Z.}~\bibnamefont{Wang}}, \bibinfo {author} {\bibfnamefont{J.}~\bibnamefont{{Martinez-Castro}}}, \bibinfo {author} {\bibfnamefont{A.~F.}\ \bibnamefont{Morpurgo}},\ and\ \bibinfo {author} {\bibfnamefont{C.}~\bibnamefont{Renner}},\ }%
  \Doi{10.1038/s41467-023-39453-x}{\emph{\bibinfo {title} {Determining Spin-Orbit Coupling in Graphene by Quasiparticle Interference Imaging}}},\ \bibinfo {journal} {Nat Commun}\ \textbf{\bibinfo {volume} {14}},\ \bibinfo {pages} {3771} (\bibinfo {year} {2023}).~%
  \bibAnnoteFile{Stop}{SunL2023}%
\bibitem{XieM2023}%
  \BibitemOpen
  \bibfield{author}{%
  \bibinfo {author} {\bibfnamefont{M.}~\bibnamefont{Xie}}\ and\ \bibinfo {author} {\bibfnamefont{S.}~\bibnamefont{Das~Sarma}},\ }%
  \Doi{10.1103/PhysRevB.107.L201119}{\emph{\bibinfo {title} {Flavor Symmetry Breaking in Spin-Orbit Coupled Bilayer Graphene}}},\ \bibinfo {journal} {Phys. Rev. B}\ \textbf{\bibinfo {volume} {107}},\ \bibinfo {pages} {L201119} (\bibinfo {year} {2023}).~%
  \bibAnnoteFile{Stop}{XieM2023}%
\bibitem{han_large_2024}%
  \BibitemOpen
  \bibfield{author}{%
  \bibinfo {author} {\bibfnamefont{T.}~\bibnamefont{Han}}, \bibinfo {author} {\bibfnamefont{Z.}~\bibnamefont{Lu}}, \bibinfo {author} {\bibfnamefont{Y.}~\bibnamefont{Yao}}, \bibinfo {author} {\bibfnamefont{J.}~\bibnamefont{Yang}}, \bibinfo {author} {\bibfnamefont{J.}~\bibnamefont{Seo}}, \bibinfo {author} {\bibfnamefont{C.}~\bibnamefont{Yoon}}, \bibinfo {author} {\bibfnamefont{K.}~\bibnamefont{Watanabe}}, \bibinfo {author} {\bibfnamefont{T.}~\bibnamefont{Taniguchi}}, \bibinfo {author} {\bibfnamefont{L.}~\bibnamefont{Fu}}, \bibinfo {author} {\bibfnamefont{F.}~\bibnamefont{Zhang}},\ and\ \bibinfo {author} {\bibfnamefont{L.}~\bibnamefont{Ju}},\ }%
  \Doi{10.1126/science.adk9749}{\emph{\bibinfo {title} {Large quantum anomalous {Hall} effect in spin-orbit proximitized rhombohedral graphene}}},\ \bibinfo {journal} {Science}\ \textbf{\bibinfo {volume} {384}},\ \bibinfo {pages} {647} (\bibinfo {year} {2024}).~%
  \bibAnnoteFile{Stop}{han_large_2024}%
\bibitem{li2024tunable}%
  \BibitemOpen
  \bibfield{author}{%
  \bibinfo {author} {\bibfnamefont{C.}~\bibnamefont{Li}}, \bibinfo {author} {\bibfnamefont{F.}~\bibnamefont{Xu}}, \bibinfo {author} {\bibfnamefont{B.}~\bibnamefont{Li}}, \bibinfo {author} {\bibfnamefont{J.}~\bibnamefont{Li}}, \bibinfo {author} {\bibfnamefont{G.}~\bibnamefont{Li}}, \bibinfo {author} {\bibfnamefont{K.}~\bibnamefont{Watanabe}}, \bibinfo {author} {\bibfnamefont{T.}~\bibnamefont{Taniguchi}}, \bibinfo {author} {\bibfnamefont{B.}~\bibnamefont{Tong}}, \bibinfo {author} {\bibfnamefont{J.}~\bibnamefont{Shen}}, \bibinfo {author} {\bibfnamefont{L.}~\bibnamefont{Lu}}, \bibinfo {author} {\bibfnamefont{J.}~\bibnamefont{Jia}}, \bibinfo {author} {\bibfnamefont{F.}~\bibnamefont{Wu}}, \bibinfo {author} {\bibfnamefont{X.}~\bibnamefont{Liu}},\ and\ \bibinfo {author} {\bibfnamefont{T.}~\bibnamefont{Li}},\ }%
  \Doi{10.1038/s41586-024-07584-w}{\emph{\bibinfo {title} {Tunable superconductivity in electron- and hole-doped {Bernal} bilayer graphene}}},\ \bibinfo {journal} {Nature}\ \textbf{\bibinfo {volume} {631}},\ \bibinfo {pages} {300} (\bibinfo {year} {2024}).~%
  \bibAnnoteFile{Stop}{li2024tunable}%
\bibitem{chou2024topological}%
  \BibitemOpen
  \bibfield{author}{%
  \bibinfo {author} {\bibfnamefont{Y.-Z.}\ \bibnamefont{Chou}}, \bibinfo {author} {\bibfnamefont{Y.}~\bibnamefont{Tan}}, \bibinfo {author} {\bibfnamefont{F.}~\bibnamefont{Wu}},\ and\ \bibinfo {author} {\bibfnamefont{S.}~\bibnamefont{Das~Sarma}},\ }%
  \Doi{10.1103/PhysRevB.110.L041108}{\emph{\bibinfo {title} {Topological flat bands, valley polarization, and interband superconductivity in magic-angle twisted bilayer graphene with proximitized spin-orbit couplings}}},\ \bibinfo {journal} {Physical Review B}\ \textbf{\bibinfo {volume} {110}},\ \bibinfo {pages} {L041108} (\bibinfo {year} {2024}).~%
  \bibAnnoteFile{Stop}{chou2024topological}%
\bibitem{CurtisJB2023}%
  \BibitemOpen
  \bibfield{author}{%
  \bibinfo {author} {\bibfnamefont{J.~B.}\ \bibnamefont{Curtis}}, \bibinfo {author} {\bibfnamefont{N.~R.}\ \bibnamefont{Poniatowski}}, \bibinfo {author} {\bibfnamefont{Y.}~\bibnamefont{Xie}}, \bibinfo {author} {\bibfnamefont{A.}~\bibnamefont{Yacoby}}, \bibinfo {author} {\bibfnamefont{E.}~\bibnamefont{Demler}},\ and\ \bibinfo {author} {\bibfnamefont{P.}~\bibnamefont{Narang}},\ }%
  \Doi{10.1103/PhysRevLett.130.196001}{\emph{\bibinfo {title} {Stabilizing {{Fluctuating Spin-Triplet Superconductivity}} in {{Graphene}} via {{Induced Spin-Orbit Coupling}}}}},\ \bibinfo {journal} {Phys. Rev. Lett.}\ \textbf{\bibinfo {volume} {130}},\ \bibinfo {pages} {196001} (\bibinfo {year} {2023}).~%
  \bibAnnoteFile{Stop}{CurtisJB2023}%
\bibitem{Jimeno-PozoA2023}%
  \BibitemOpen
  \bibfield{author}{%
  \bibinfo {author} {\bibfnamefont{A.}~\bibnamefont{{Jimeno-Pozo}}}, \bibinfo {author} {\bibfnamefont{H.}~\bibnamefont{{Sainz-Cruz}}}, \bibinfo {author} {\bibfnamefont{T.}~\bibnamefont{Cea}}, \bibinfo {author} {\bibfnamefont{P.~A.}\ \bibnamefont{Pantale{\'o}n}},\ and\ \bibinfo {author} {\bibfnamefont{F.}~\bibnamefont{Guinea}},\ }%
  \Doi{10.1103/PhysRevB.107.L161106}{\emph{\bibinfo {title} {Superconductivity from Electronic Interactions and Spin-Orbit Enhancement in Bilayer and Trilayer Graphene}}},\ \bibinfo {journal} {Phys. Rev. B}\ \textbf{\bibinfo {volume} {107}},\ \bibinfo {pages} {L161106} (\bibinfo {year} {2023}).~%
  \bibAnnoteFile{Stop}{Jimeno-PozoA2023}%
\bibitem{PantaleonPA2023}%
  \BibitemOpen
  \bibfield{author}{%
  \bibinfo {author} {\bibfnamefont{P.~A.}\ \bibnamefont{Pantale{\'o}n}}, \bibinfo {author} {\bibfnamefont{A.}~\bibnamefont{{Jimeno-Pozo}}}, \bibinfo {author} {\bibfnamefont{H.}~\bibnamefont{{Sainz-Cruz}}}, \bibinfo {author} {\bibfnamefont{V.~T.}\ \bibnamefont{Phong}}, \bibinfo {author} {\bibfnamefont{T.}~\bibnamefont{Cea}},\ and\ \bibinfo {author} {\bibfnamefont{F.}~\bibnamefont{Guinea}},\ }%
  \Doi{10.1038/s42254-023-00575-2}{\emph{\bibinfo {title} {Superconductivity and Correlated Phases in Non-Twisted Bilayer and Trilayer Graphene}}},\ \bibinfo {journal} {Nat Rev Phys}\ \textbf{\bibinfo {volume} {5}},\ \bibinfo {pages} {304} (\bibinfo {year} {2023}).~%
  \bibAnnoteFile{Stop}{PantaleonPA2023}%
\bibitem{ParappurathA2023}%
  \BibitemOpen
  \bibfield{author}{%
  \bibinfo {author} {\bibfnamefont{A.}~\bibnamefont{Parappurath}}, \bibinfo {author} {\bibfnamefont{B.}~\bibnamefont{Ghawri}}, \bibinfo {author} {\bibfnamefont{S.}~\bibnamefont{Bhowmik}}, \bibinfo {author} {\bibfnamefont{A.}~\bibnamefont{Singha}}, \bibinfo {author} {\bibfnamefont{K.}~\bibnamefont{Watanabe}}, \bibinfo {author} {\bibfnamefont{T.}~\bibnamefont{Taniguchi}},\ and\ \bibinfo {author} {\bibfnamefont{A.}~\bibnamefont{Ghosh}},\ }%
  \Doi{10.1039/D3NR04864K}{\emph{\bibinfo {title} {Band Structure Sensitive Photoresponse in Twisted Bilayer Graphene Proximitized with {{WSe$_2$}}}}},\ \bibinfo {journal} {Nanoscale}\ \textbf{\bibinfo {volume} {15}},\ \bibinfo {pages} {18818} (\bibinfo {year} {2023}).~%
  \bibAnnoteFile{Stop}{ParappurathA2023}%
\bibitem{WagnerG2023}%
  \BibitemOpen
  \bibfield{author}{%
  \bibinfo {author} {\bibfnamefont{G.}~\bibnamefont{Wagner}}, \bibinfo {author} {\bibfnamefont{Y.~H.}\ \bibnamefont{Kwan}}, \bibinfo {author} {\bibfnamefont{N.}~\bibnamefont{Bultinck}}, \bibinfo {author} {\bibfnamefont{S.~H.}\ \bibnamefont{Simon}},\ and\ \bibinfo {author} {\bibfnamefont{S.~A.}\ \bibnamefont{Parameswaran}},\ }%
  {\bibinfo {title} {Superconductivity from repulsive interactions in {{Bernal-stacked}} bilayer graphene},}\  (\bibinfo {year} {2023}),\ \Eprint{http://arxiv.org/abs/2302.00682}{arxiv:2302.00682 [cond-mat]}~%
  \bibAnnoteFile{NoStop}{WagnerG2023}%
\bibitem{kane_quantum_2005}%
  \BibitemOpen
  \bibfield{author}{%
  \bibinfo {author} {\bibfnamefont{C.~L.}\ \bibnamefont{Kane}}\ and\ \bibinfo {author} {\bibfnamefont{E.~J.}\ \bibnamefont{Mele}},\ }%
  \Doi{10.1103/PhysRevLett.95.226801}{\emph{\bibinfo {title} {Quantum {Spin} {Hall} {Effect} in {Graphene}}}},\ \bibinfo {journal} {Physical Review Letters}\ \textbf{\bibinfo {volume} {95}},\ \bibinfo {pages} {226801} (\bibinfo {year} {2005}).~%
  \bibAnnoteFile{Stop}{kane_quantum_2005}%
\bibitem{kane_z_2_2005}%
  \BibitemOpen
  \bibfield{author}{%
  \bibinfo {author} {\bibfnamefont{C.~L.}\ \bibnamefont{Kane}}\ and\ \bibinfo {author} {\bibfnamefont{E.~J.}\ \bibnamefont{Mele}},\ }%
  \Doi{10.1103/PhysRevLett.95.146802}{\emph{\bibinfo {title} {\$\{{Z}\}\_\{2\}\$ {Topological} {Order} and the {Quantum} {Spin} {Hall} {Effect}}}},\ \bibinfo {journal} {Physical Review Letters}\ \textbf{\bibinfo {volume} {95}},\ \bibinfo {pages} {146802} (\bibinfo {year} {2005}).~%
  \bibAnnoteFile{Stop}{kane_z_2_2005}%
\bibitem{lee_charge--spin_2022}%
  \BibitemOpen
  \bibfield{author}{%
  \bibinfo {author} {\bibfnamefont{S.}~\bibnamefont{Lee}}, \bibinfo {author} {\bibfnamefont{D.~J.~P.}\ \bibnamefont{de~Sousa}}, \bibinfo {author} {\bibfnamefont{Y.-K.}\ \bibnamefont{Kwon}}, \bibinfo {author} {\bibfnamefont{F.}~\bibnamefont{de~Juan}}, \bibinfo {author} {\bibfnamefont{Z.}~\bibnamefont{Chi}}, \bibinfo {author} {\bibfnamefont{F.}~\bibnamefont{Casanova}},\ and\ \bibinfo {author} {\bibfnamefont{T.}~\bibnamefont{Low}},\ }%
  \Doi{10.1103/PhysRevB.106.165420}{\emph{\bibinfo {title} {Charge-to-spin conversion in twisted $\mathrm{graphene}/{\mathrm{WSe}}_{2}$ heterostructures}}},\ \bibinfo {journal} {Phys. Rev. B}\ \textbf{\bibinfo {volume} {106}},\ \bibinfo {pages} {165420} (\bibinfo {year} {2022}).~%
  \bibAnnoteFile{Stop}{lee_charge--spin_2022}%
\bibitem{ingla-aynes_omnidirectional_2022}%
  \BibitemOpen
  \bibfield{author}{%
  \bibinfo {author} {\bibfnamefont{J.}~\bibnamefont{Ingla-Aynés}}, \bibinfo {author} {\bibfnamefont{I.}~\bibnamefont{Groen}}, \bibinfo {author} {\bibfnamefont{F.}~\bibnamefont{Herling}}, \bibinfo {author} {\bibfnamefont{N.}~\bibnamefont{Ontoso}}, \bibinfo {author} {\bibfnamefont{C.~K.}\ \bibnamefont{Safeer}}, \bibinfo {author} {\bibfnamefont{F.}~\bibnamefont{De~Juan}}, \bibinfo {author} {\bibfnamefont{L.~E.}\ \bibnamefont{Hueso}}, \bibinfo {author} {\bibfnamefont{M.}~\bibnamefont{Gobbi}},\ and\ \bibinfo {author} {\bibfnamefont{F.}~\bibnamefont{Casanova}},\ }%
  \Doi{10.1088/2053-1583/ac76d1}{\emph{\bibinfo {title} {Omnidirectional spin-to-charge conversion in graphene/{NbSe} $_{\textrm{2}}$ van der {Waals} heterostructures}}},\ \bibinfo {journal} {2D Materials}\ \textbf{\bibinfo {volume} {9}},\ \bibinfo {pages} {045001} (\bibinfo {year} {2022}).~%
  \bibAnnoteFile{Stop}{ingla-aynes_omnidirectional_2022}%
\bibitem{sherkunov_electronic_2018}%
  \BibitemOpen
  \bibfield{author}{%
  \bibinfo {author} {\bibfnamefont{Y.}~\bibnamefont{Sherkunov}}\ and\ \bibinfo {author} {\bibfnamefont{J.~J.}\ \bibnamefont{Betouras}},\ }%
  \Doi{10.1103/PhysRevB.98.205151}{\emph{\bibinfo {title} {Electronic phases in twisted bilayer graphene at magic angles as a result of {Van} {Hove} singularities and interactions}}},\ \bibinfo {journal} {Physical Review B}\ \textbf{\bibinfo {volume} {98}},\ \bibinfo {pages} {205151} (\bibinfo {year} {2018}).~%
  \bibAnnoteFile{Stop}{sherkunov_electronic_2018}%
\bibitem{isobe_unconventional_2018}%
  \BibitemOpen
  \bibfield{author}{%
  \bibinfo {author} {\bibfnamefont{H.}~\bibnamefont{Isobe}}, \bibinfo {author} {\bibfnamefont{N.~F.}\ \bibnamefont{Yuan}},\ and\ \bibinfo {author} {\bibfnamefont{L.}~\bibnamefont{Fu}},\ }%
  \Doi{10.1103/PhysRevX.8.041041}{\emph{\bibinfo {title} {Unconventional {Superconductivity} and {Density} {Waves} in {Twisted} {Bilayer} {Graphene}}}},\ \bibinfo {journal} {Physical Review X}\ \textbf{\bibinfo {volume} {8}},\ \bibinfo {pages} {041041} (\bibinfo {year} {2018}).~%
  \bibAnnoteFile{Stop}{isobe_unconventional_2018}%
\bibitem{liu_chiral_2018}%
  \BibitemOpen
  \bibfield{author}{%
  \bibinfo {author} {\bibfnamefont{C.-C.}\ \bibnamefont{Liu}}, \bibinfo {author} {\bibfnamefont{L.-D.}\ \bibnamefont{Zhang}}, \bibinfo {author} {\bibfnamefont{W.-Q.}\ \bibnamefont{Chen}},\ and\ \bibinfo {author} {\bibfnamefont{F.}~\bibnamefont{Yang}},\ }%
  \Doi{10.1103/PhysRevLett.121.217001}{\emph{\bibinfo {title} {Chiral {Spin} {Density} {Wave} and \$d+id\$ {Superconductivity} in the {Magic}-{Angle}-{Twisted} {Bilayer} {Graphene}}}},\ \bibinfo {journal} {Physical Review Letters}\ \textbf{\bibinfo {volume} {121}},\ \bibinfo {pages} {217001} (\bibinfo {year} {2018}).~%
  \bibAnnoteFile{Stop}{liu_chiral_2018}%
\bibitem{han_graphene_2014}%
  \BibitemOpen
  \bibfield{author}{%
  \bibinfo {author} {\bibfnamefont{W.}~\bibnamefont{Han}}, \bibinfo {author} {\bibfnamefont{R.~K.}\ \bibnamefont{Kawakami}}, \bibinfo {author} {\bibfnamefont{M.}~\bibnamefont{Gmitra}},\ and\ \bibinfo {author} {\bibfnamefont{J.}~\bibnamefont{Fabian}},\ }%
  \Doi{10.1038/nnano.2014.214}{\emph{\bibinfo {title} {Graphene spintronics}}},\ \bibinfo {journal} {Nature Nanotechnology}\ \textbf{\bibinfo {volume} {9}},\ \bibinfo {pages} {794} (\bibinfo {year} {2014}).~%
  \bibAnnoteFile{Stop}{han_graphene_2014}%
\bibitem{frank_emergence_2024}%
  \BibitemOpen
  \bibfield{author}{%
  \bibinfo {author} {\bibfnamefont{T.}~\bibnamefont{Frank}}, \bibinfo {author} {\bibfnamefont{P.~E.~F.}\ \bibnamefont{Junior}}, \bibinfo {author} {\bibfnamefont{K.}~\bibnamefont{Zollner}},\ and\ \bibinfo {author} {\bibfnamefont{J.}~\bibnamefont{Fabian}},\ }%
  {\bibinfo {title} {Emergence of radial {Rashba} spin-orbit fields in twisted van der {Waals} heterostructures},}\  (\bibinfo {year} {2024}),\ \bibinfo {note} {arXiv:2402.12353 [cond-mat]}~%
  \bibAnnoteFile{NoStop}{frank_emergence_2024}%
\bibitem{avsar_colloquium_2020}%
  \BibitemOpen
  \bibfield{author}{%
  \bibinfo {author} {\bibfnamefont{A.}~\bibnamefont{Avsar}}, \bibinfo {author} {\bibfnamefont{H.}~\bibnamefont{Ochoa}}, \bibinfo {author} {\bibfnamefont{F.}~\bibnamefont{Guinea}}, \bibinfo {author} {\bibfnamefont{B.}~\bibnamefont{Özyilmaz}}, \bibinfo {author} {\bibfnamefont{B.}~\bibnamefont{van Wees}},\ and\ \bibinfo {author} {\bibfnamefont{I.}~\bibnamefont{Vera-Marun}},\ }%
  \Doi{10.1103/RevModPhys.92.021003}{\emph{\bibinfo {title} {Colloquium: {Spintronics} in graphene and other two-dimensional materials}}},\ \bibinfo {journal} {Reviews of Modern Physics}\ \textbf{\bibinfo {volume} {92}},\ \bibinfo {pages} {021003} (\bibinfo {year} {2020}).~%
  \bibAnnoteFile{Stop}{avsar_colloquium_2020}%
\bibitem{ahn_2d_2020}%
  \BibitemOpen
  \bibfield{author}{%
  \bibinfo {author} {\bibfnamefont{E.~C.}\ \bibnamefont{Ahn}},\ }%
  \Doi{10.1038/s41699-020-0152-0}{\emph{\bibinfo {title} {2D materials for spintronic devices}}},\ \bibinfo {journal} {npj 2D Materials and Applications}\ \textbf{\bibinfo {volume} {4}},\ \bibinfo {pages} {1} (\bibinfo {year} {2020}).~%
  \bibAnnoteFile{Stop}{ahn_2d_2020}%
\bibitem{WangT2020}%
  \BibitemOpen
  \bibfield{author}{%
  \bibinfo {author} {\bibfnamefont{T.}~\bibnamefont{Wang}}, \bibinfo {author} {\bibfnamefont{N.}~\bibnamefont{Bultinck}},\ and\ \bibinfo {author} {\bibfnamefont{M.~P.}\ \bibnamefont{Zaletel}},\ }%
  \Doi{10.1103/PhysRevB.102.235146}{\emph{\bibinfo {title} {Flat-Band Topology of Magic Angle Graphene on a Transition Metal Dichalcogenide}}},\ \bibinfo {journal} {Phys. Rev. B}\ \textbf{\bibinfo {volume} {102}},\ \bibinfo {pages} {235146} (\bibinfo {year} {2020}).~%
  \bibAnnoteFile{Stop}{WangT2020}%
\bibitem{haddad_twisted_2023}%
  \BibitemOpen
  \bibfield{author}{%
  \bibinfo {author} {\bibfnamefont{S.}~\bibnamefont{Haddad}}, \bibinfo {author} {\bibfnamefont{T.}~\bibnamefont{Kato}}, \bibinfo {author} {\bibfnamefont{J.}~\bibnamefont{Zhu}},\ and\ \bibinfo {author} {\bibfnamefont{L.}~\bibnamefont{Mandhour}},\ }%
  \Doi{10.1103/PhysRevB.108.L121101}{\emph{\bibinfo {title} {Twisted bilayer graphene reveals its flat bands under spin pumping}}},\ \bibinfo {journal} {Physical Review B}\ \textbf{\bibinfo {volume} {108}},\ \bibinfo {pages} {L121101} (\bibinfo {year} {2023}).~%
  \bibAnnoteFile{Stop}{haddad_twisted_2023}%
\bibitem{wang_strong_2015}%
  \BibitemOpen
  \bibfield{author}{%
  \bibinfo {author} {\bibfnamefont{Z.}~\bibnamefont{Wang}}, \bibinfo {author} {\bibfnamefont{D.-K.}\ \bibnamefont{Ki}}, \bibinfo {author} {\bibfnamefont{H.}~\bibnamefont{Chen}}, \bibinfo {author} {\bibfnamefont{H.}~\bibnamefont{Berger}}, \bibinfo {author} {\bibfnamefont{A.~H.}\ \bibnamefont{MacDonald}},\ and\ \bibinfo {author} {\bibfnamefont{A.~F.}\ \bibnamefont{Morpurgo}},\ }%
  \Doi{10.1038/ncomms9339}{\emph{\bibinfo {title} {Strong interface-induced spin–orbit interaction in graphene on {WS2}}}},\ \bibinfo {journal} {Nature Communications}\ \textbf{\bibinfo {volume} {6}},\ \bibinfo {pages} {8339} (\bibinfo {year} {2015}).~%
  \bibAnnoteFile{Stop}{wang_strong_2015}%
\bibitem{BistritzerR2011}%
  \BibitemOpen
  \bibfield{author}{%
  \bibinfo {author} {\bibfnamefont{R.}~\bibnamefont{Bistritzer}}\ and\ \bibinfo {author} {\bibfnamefont{A.~H.}\ \bibnamefont{MacDonald}},\ }%
  \Doi{10.1073/pnas.1108174108}{\emph{\bibinfo {title} {Moir{\'e} Bands in Twisted Double-Layer Graphene}}},\ \bibinfo {journal} {Proceedings of the National Academy of Sciences}\ \textbf{\bibinfo {volume} {108}},\ \bibinfo {pages} {12233} (\bibinfo {year} {2011}).~%
  \bibAnnoteFile{Stop}{BistritzerR2011}%
\bibitem{WuF2018}%
  \BibitemOpen
  \bibfield{author}{%
  \bibinfo {author} {\bibfnamefont{F.}~\bibnamefont{Wu}}, \bibinfo {author} {\bibfnamefont{A.~H.}\ \bibnamefont{MacDonald}},\ and\ \bibinfo {author} {\bibfnamefont{I.}~\bibnamefont{Martin}},\ }%
  \Doi{10.1103/PhysRevLett.121.257001}{\emph{\bibinfo {title} {Theory of {{Phonon-Mediated Superconductivity}} in {{Twisted Bilayer Graphene}}}}},\ \bibinfo {journal} {Phys. Rev. Lett.}\ \textbf{\bibinfo {volume} {121}},\ \bibinfo {pages} {257001} (\bibinfo {year} {2018}).~%
  \bibAnnoteFile{Stop}{WuF2018}%
\bibitem{SongZ2019}%
  \BibitemOpen
  \bibfield{author}{%
  \bibinfo {author} {\bibfnamefont{Z.}~\bibnamefont{Song}}, \bibinfo {author} {\bibfnamefont{Z.}~\bibnamefont{Wang}}, \bibinfo {author} {\bibfnamefont{W.}~\bibnamefont{Shi}}, \bibinfo {author} {\bibfnamefont{G.}~\bibnamefont{Li}}, \bibinfo {author} {\bibfnamefont{C.}~\bibnamefont{Fang}},\ and\ \bibinfo {author} {\bibfnamefont{B.~A.}\ \bibnamefont{Bernevig}},\ }%
  \Doi{10.1103/PhysRevLett.123.036401}{\emph{\bibinfo {title} {All {{Magic Angles}} in {{Twisted Bilayer Graphene}} Are {{Topological}}}}},\ \bibinfo {journal} {Phys. Rev. Lett.}\ \textbf{\bibinfo {volume} {123}},\ \bibinfo {pages} {036401} (\bibinfo {year} {2019}).~%
  \bibAnnoteFile{Stop}{SongZ2019}%
\bibitem{NaimerT2023}%
  \BibitemOpen
  \bibfield{author}{%
  \bibinfo {author} {\bibfnamefont{T.}~\bibnamefont{Naimer}}\ and\ \bibinfo {author} {\bibfnamefont{J.}~\bibnamefont{Fabian}},\ }%
  \Doi{10.1103/PhysRevB.107.195144}{\emph{\bibinfo {title} {Twist-Angle Dependent Proximity Induced Spin-Orbit Coupling in Graphene/Topological Insulator Heterostructures}}},\ \bibinfo {journal} {Phys. Rev. B}\ \textbf{\bibinfo {volume} {107}},\ \bibinfo {pages} {195144} (\bibinfo {year} {2023}).~%
  \bibAnnoteFile{Stop}{NaimerT2023}%
\bibitem{ScammellHD2023a}%
  \BibitemOpen
  \bibfield{author}{%
  \bibinfo {author} {\bibfnamefont{H.~D.}\ \bibnamefont{Scammell}}\ and\ \bibinfo {author} {\bibfnamefont{M.~S.}\ \bibnamefont{Scheurer}},\ }%
  \Doi{10.1103/PhysRevLett.130.066001}{\emph{\bibinfo {title} {Tunable {{Superconductivity}} and {M\"obius} {{Fermi Surfaces}} in an {{Inversion-Symmetric Twisted}} van Der {{Waals Heterostructure}}}}},\ \bibinfo {journal} {Phys. Rev. Lett.}\ \textbf{\bibinfo {volume} {130}},\ \bibinfo {pages} {066001} (\bibinfo {year} {2023}).~%
  \bibAnnoteFile{Stop}{ScammellHD2023a}%
\bibitem{ScammellHD2024}%
  \BibitemOpen
  \bibfield{author}{%
  \bibinfo {author} {\bibfnamefont{H.~D.}\ \bibnamefont{Scammell}}\ and\ \bibinfo {author} {\bibfnamefont{M.~S.}\ \bibnamefont{Scheurer}},\ }%
  \Doi{10.1103/PhysRevB.109.035159}{\emph{\bibinfo {title} {Displacement Field Tunable Superconductivity in an Inversion-Symmetric Twisted van Der {{Waals}} Heterostructure}}},\ \bibinfo {journal} {Phys. Rev. B}\ \textbf{\bibinfo {volume} {109}},\ \bibinfo {pages} {035159} (\bibinfo {year} {2024}).~%
  \bibAnnoteFile{Stop}{ScammellHD2024}%
\bibitem{DavidA2019}%
  \BibitemOpen
  \bibfield{author}{%
  \bibinfo {author} {\bibfnamefont{A.}~\bibnamefont{David}}, \bibinfo {author} {\bibfnamefont{P.}~\bibnamefont{Rakyta}}, \bibinfo {author} {\bibfnamefont{A.}~\bibnamefont{Korm{\'a}nyos}},\ and\ \bibinfo {author} {\bibfnamefont{G.}~\bibnamefont{Burkard}},\ }%
  \Doi{10.1103/PhysRevB.100.085412}{\emph{\bibinfo {title} {Induced Spin-Orbit Coupling in Twisted Graphene--Transition Metal Dichalcogenide Heterobilayers: {{Twistronics}} Meets Spintronics}}},\ \bibinfo {journal} {Phys. Rev. B}\ \textbf{\bibinfo {volume} {100}},\ \bibinfo {pages} {085412} (\bibinfo {year} {2019}).~%
  \bibAnnoteFile{Stop}{DavidA2019}%
\bibitem{ChouYZ2022}%
  \BibitemOpen
  \bibfield{author}{%
  \bibinfo {author} {\bibfnamefont{Y.-Z.}\ \bibnamefont{Chou}}, \bibinfo {author} {\bibfnamefont{F.}~\bibnamefont{Wu}},\ and\ \bibinfo {author} {\bibfnamefont{S.}~\bibnamefont{Das~Sarma}},\ }%
  \Doi{10.1103/PhysRevB.106.L180502}{\emph{\bibinfo {title} {Enhanced Superconductivity through Virtual Tunneling in {{Bernal}} Bilayer Graphene Coupled to WSe$_2$}}},\ \bibinfo {journal} {Phys. Rev. B}\ \textbf{\bibinfo {volume} {106}},\ \bibinfo {pages} {L180502} (\bibinfo {year} {2022}).~%
  \bibAnnoteFile{Stop}{ChouYZ2022}%
\bibitem{wang_origin_2016}%
  \BibitemOpen
  \bibfield{author}{%
  \bibinfo {author} {\bibfnamefont{Z.}~\bibnamefont{Wang}}, \bibinfo {author} {\bibfnamefont{D.-K.}\ \bibnamefont{Ki}}, \bibinfo {author} {\bibfnamefont{J.~Y.}\ \bibnamefont{Khoo}}, \bibinfo {author} {\bibfnamefont{D.}~\bibnamefont{Mauro}}, \bibinfo {author} {\bibfnamefont{H.}~\bibnamefont{Berger}}, \bibinfo {author} {\bibfnamefont{L.~S.}\ \bibnamefont{Levitov}},\ and\ \bibinfo {author} {\bibfnamefont{A.~F.}\ \bibnamefont{Morpurgo}},\ }%
  \Doi{10.1103/PhysRevX.6.041020}{\emph{\bibinfo {title} {Origin and {Magnitude} of `{Designer}' {Spin}-{Orbit} {Interaction} in {Graphene} on {Semiconducting} {Transition} {Metal} {Dichalcogenides}}}},\ \bibinfo {journal} {Physical Review X}\ \textbf{\bibinfo {volume} {6}},\ \bibinfo {pages} {041020} (\bibinfo {year} {2016}).~%
  \bibAnnoteFile{Stop}{wang_origin_2016}%
\bibitem{wang_quantum_2019}%
  \BibitemOpen
  \bibfield{author}{%
  \bibinfo {author} {\bibfnamefont{D.}~\bibnamefont{Wang}}, \bibinfo {author} {\bibfnamefont{S.}~\bibnamefont{Che}}, \bibinfo {author} {\bibfnamefont{G.}~\bibnamefont{Cao}}, \bibinfo {author} {\bibfnamefont{R.}~\bibnamefont{Lyu}}, \bibinfo {author} {\bibfnamefont{K.}~\bibnamefont{Watanabe}}, \bibinfo {author} {\bibfnamefont{T.}~\bibnamefont{Taniguchi}}, \bibinfo {author} {\bibfnamefont{C.~N.}\ \bibnamefont{Lau}},\ and\ \bibinfo {author} {\bibfnamefont{M.}~\bibnamefont{Bockrath}},\ }%
  \Doi{10.1021/acs.nanolett.9b02445}{\emph{\bibinfo {title} {Quantum {Hall} {Effect} {Measurement} of {Spin}–{Orbit} {Coupling} {Strengths} in {Ultraclean} {Bilayer} {Graphene}/{WSe2} {Heterostructures}}}},\ \bibinfo {journal} {Nano Letters}\ \textbf{\bibinfo {volume} {19}},\ \bibinfo {pages} {7028} (\bibinfo {year} {2019}).~%
  \bibAnnoteFile{Stop}{wang_quantum_2019}%
\bibitem{LinJX2022}%
  \BibitemOpen
  \bibfield{author}{%
  \bibinfo {author} {\bibfnamefont{J.-X.}\ \bibnamefont{Lin}}, \bibinfo {author} {\bibfnamefont{Y.-H.}\ \bibnamefont{Zhang}}, \bibinfo {author} {\bibfnamefont{E.}~\bibnamefont{Morissette}}, \bibinfo {author} {\bibfnamefont{Z.}~\bibnamefont{Wang}}, \bibinfo {author} {\bibfnamefont{S.}~\bibnamefont{Liu}}, \bibinfo {author} {\bibfnamefont{D.}~\bibnamefont{Rhodes}}, \bibinfo {author} {\bibfnamefont{K.}~\bibnamefont{Watanabe}}, \bibinfo {author} {\bibfnamefont{T.}~\bibnamefont{Taniguchi}}, \bibinfo {author} {\bibfnamefont{J.}~\bibnamefont{Hone}},\ and\ \bibinfo {author} {\bibfnamefont{J.~I.~A.}\ \bibnamefont{Li}},\ }%
  \Doi{10.1126/science.abh2889}{\emph{\bibinfo {title} {Spin-Orbit{\textendash}Driven Ferromagnetism at Half Moir{\'e} Filling in Magic-Angle Twisted Bilayer Graphene}}},\ \bibinfo {journal} {Science}\ \textbf{\bibinfo {volume} {375}},\ \bibinfo {pages} {437} (\bibinfo {year} {2022}).~%
  \bibAnnoteFile{Stop}{LinJX2022}%
\bibitem{BhowmikS2023}%
  \BibitemOpen
  \bibfield{author}{%
  \bibinfo {author} {\bibfnamefont{S.}~\bibnamefont{Bhowmik}}, \bibinfo {author} {\bibfnamefont{B.}~\bibnamefont{Ghawri}}, \bibinfo {author} {\bibfnamefont{Y.}~\bibnamefont{Park}}, \bibinfo {author} {\bibfnamefont{D.}~\bibnamefont{Lee}}, \bibinfo {author} {\bibfnamefont{S.}~\bibnamefont{Datta}}, \bibinfo {author} {\bibfnamefont{R.}~\bibnamefont{Soni}}, \bibinfo {author} {\bibfnamefont{K.}~\bibnamefont{Watanabe}}, \bibinfo {author} {\bibfnamefont{T.}~\bibnamefont{Taniguchi}}, \bibinfo {author} {\bibfnamefont{A.}~\bibnamefont{Ghosh}}, \bibinfo {author} {\bibfnamefont{J.}~\bibnamefont{Jung}},\ and\ \bibinfo {author} {\bibfnamefont{U.}~\bibnamefont{Chandni}},\ }%
  \Doi{10.1038/s41467-023-39855-x}{\emph{\bibinfo {title} {Spin-Orbit Coupling-Enhanced Valley Ordering of Malleable Bands in Twisted Bilayer Graphene on {{WSe$_2$}}}}},\ \bibinfo {journal} {Nat Commun}\ \textbf{\bibinfo {volume} {14}},\ \bibinfo {pages} {4055} (\bibinfo {year} {2023}).~%
  \bibAnnoteFile{Stop}{BhowmikS2023}%
\bibitem{FangC2012}%
  \BibitemOpen
  \bibfield{author}{%
  \bibinfo {author} {\bibfnamefont{C.}~\bibnamefont{Fang}}, \bibinfo {author} {\bibfnamefont{M.~J.}\ \bibnamefont{Gilbert}},\ and\ \bibinfo {author} {\bibfnamefont{B.~A.}\ \bibnamefont{Bernevig}},\ }%
  \Doi{10.1103/PhysRevB.86.115112}{\emph{\bibinfo {title} {Bulk Topological Invariants in Noninteracting Point Group Symmetric Insulators}}},\ \bibinfo {journal} {Phys. Rev. B}\ \textbf{\bibinfo {volume} {86}},\ \bibinfo {pages} {115112} (\bibinfo {year} {2012}).~%
  \bibAnnoteFile{Stop}{FangC2012}%
\bibitem{yuan_magic_2019}%
  \BibitemOpen
  \bibfield{author}{%
  \bibinfo {author} {\bibfnamefont{N.~F.~Q.}\ \bibnamefont{Yuan}}, \bibinfo {author} {\bibfnamefont{H.}~\bibnamefont{Isobe}},\ and\ \bibinfo {author} {\bibfnamefont{L.}~\bibnamefont{Fu}},\ }%
  \Doi{10.1038/s41467-019-13670-9}{\emph{\bibinfo {title} {Magic of high-order van {Hove} singularity}}},\ \bibinfo {journal} {Nature Communications}\ \textbf{\bibinfo {volume} {10}},\ \bibinfo {pages} {5769} (\bibinfo {year} {2019}).~%
  \bibAnnoteFile{Stop}{yuan_magic_2019}%
\bibitem{huertas-hernando_spin-orbit_2006}%
  \BibitemOpen
  \bibfield{author}{%
  \bibinfo {author} {\bibfnamefont{D.}~\bibnamefont{Huertas-Hernando}}, \bibinfo {author} {\bibfnamefont{F.}~\bibnamefont{Guinea}},\ and\ \bibinfo {author} {\bibfnamefont{A.}~\bibnamefont{Brataas}},\ }%
  \Doi{10.1103/PhysRevB.74.155426}{\emph{\bibinfo {title} {Spin-orbit coupling in curved graphene, fullerenes, nanotubes, and nanotube caps}}},\ \bibinfo {journal} {Physical Review B}\ \textbf{\bibinfo {volume} {74}},\ \bibinfo {pages} {155426} (\bibinfo {year} {2006}).~%
  \bibAnnoteFile{Stop}{huertas-hernando_spin-orbit_2006}%
\bibitem{min_intrinsic_2006}%
  \BibitemOpen
  \bibfield{author}{%
  \bibinfo {author} {\bibfnamefont{H.}~\bibnamefont{Min}}, \bibinfo {author} {\bibfnamefont{J.~E.}\ \bibnamefont{Hill}}, \bibinfo {author} {\bibfnamefont{N.~A.}\ \bibnamefont{Sinitsyn}}, \bibinfo {author} {\bibfnamefont{B.~R.}\ \bibnamefont{Sahu}}, \bibinfo {author} {\bibfnamefont{L.}~\bibnamefont{Kleinman}},\ and\ \bibinfo {author} {\bibfnamefont{A.~H.}\ \bibnamefont{MacDonald}},\ }%
  \Doi{10.1103/PhysRevB.74.165310}{\emph{\bibinfo {title} {Intrinsic and {Rashba} spin-orbit interactions in graphene sheets}}},\ \bibinfo {journal} {Physical Review B}\ \textbf{\bibinfo {volume} {74}},\ \bibinfo {pages} {165310} (\bibinfo {year} {2006}).~%
  \bibAnnoteFile{Stop}{min_intrinsic_2006}%
\bibitem{konschuh_tight-binding_2010}%
  \BibitemOpen
  \bibfield{author}{%
  \bibinfo {author} {\bibfnamefont{S.}~\bibnamefont{Konschuh}}, \bibinfo {author} {\bibfnamefont{M.}~\bibnamefont{Gmitra}},\ and\ \bibinfo {author} {\bibfnamefont{J.}~\bibnamefont{Fabian}},\ }%
  \Doi{10.1103/PhysRevB.82.245412}{\emph{\bibinfo {title} {Tight-binding theory of the spin-orbit coupling in graphene}}},\ \bibinfo {journal} {Physical Review B}\ \textbf{\bibinfo {volume} {82}},\ \bibinfo {pages} {245412} (\bibinfo {year} {2010}).~%
  \bibAnnoteFile{Stop}{konschuh_tight-binding_2010}%
\bibitem{zhumagulov_emergent_2024}%
  \BibitemOpen
  \bibfield{author}{%
  \bibinfo {author} {\bibfnamefont{Y.}~\bibnamefont{Zhumagulov}}, \bibinfo {author} {\bibfnamefont{D.}~\bibnamefont{Kochan}},\ and\ \bibinfo {author} {\bibfnamefont{J.}~\bibnamefont{Fabian}},\ }%
  \Doi{10.1103/PhysRevLett.132.186401}{\emph{\bibinfo {title} {Emergent correlated phases in rhombohedral trilayer graphene induced by proximity spin-orbit and exchange coupling}}},\ \bibinfo {journal} {Physical Review Letters}\ \textbf{\bibinfo {volume} {132}},\ \bibinfo {pages} {186401} (\bibinfo {year} {2024}).~%
  \bibAnnoteFile{Stop}{zhumagulov_emergent_2024}%
\bibitem{zhumagulov_swapping_2023}%
  \BibitemOpen
  \bibfield{author}{%
  \bibinfo {author} {\bibfnamefont{Y.}~\bibnamefont{Zhumagulov}}, \bibinfo {author} {\bibfnamefont{D.}~\bibnamefont{Kochan}},\ and\ \bibinfo {author} {\bibfnamefont{J.}~\bibnamefont{Fabian}},\ }%
  {\bibinfo {title} {Swapping exchange and spin-orbit induced correlated phases in ex-so-tic van der {Waals} heterostructures},}\  (\bibinfo {year} {2023}),\ \bibinfo {note} {arXiv:2307.16025 [cond-mat]}~%
  \bibAnnoteFile{NoStop}{zhumagulov_swapping_2023}%
\bibitem{koh_correlated_2024}%
  \BibitemOpen
  \bibfield{author}{%
  \bibinfo {author} {\bibfnamefont{J.~M.}\ \bibnamefont{Koh}}, \bibinfo {author} {\bibfnamefont{J.}~\bibnamefont{Alicea}},\ and\ \bibinfo {author} {\bibfnamefont{E.}~\bibnamefont{Lantagne-Hurtubise}},\ }%
  \Doi{10.1103/PhysRevB.109.035113}{\emph{\bibinfo {title} {Correlated {Phases} in {Spin}-{Orbit}-{Coupled} {Rhombohedral} {Trilayer} {Graphene}}}},\ \bibinfo {journal} {Physical Review B}\ \textbf{\bibinfo {volume} {109}},\ \bibinfo {pages} {035113} (\bibinfo {year} {2024}).~%
  \bibAnnoteFile{Stop}{koh_correlated_2024}%
\bibitem{ghiasi_charge--spin_2019}%
  \BibitemOpen
  \bibfield{author}{%
  \bibinfo {author} {\bibfnamefont{T.~S.}\ \bibnamefont{Ghiasi}}, \bibinfo {author} {\bibfnamefont{A.~A.}\ \bibnamefont{Kaverzin}}, \bibinfo {author} {\bibfnamefont{P.~J.}\ \bibnamefont{Blah}},\ and\ \bibinfo {author} {\bibfnamefont{B.~J.}\ \bibnamefont{van Wees}},\ }%
  \Doi{10.1021/acs.nanolett.9b01611}{\emph{\bibinfo {title} {Charge-to-{Spin} {Conversion} by the {Rashba}–{Edelstein} {Effect} in {Two}-{Dimensional} van der {Waals} {Heterostructures} up to {Room} {Temperature}}}},\ \bibinfo {journal} {Nano Letters}\ \textbf{\bibinfo {volume} {19}},\ \bibinfo {pages} {5959} (\bibinfo {year} {2019}).~%
  \bibAnnoteFile{Stop}{ghiasi_charge--spin_2019}%
\bibitem{Csaba2022}%
  \BibitemOpen
  \bibfield{author}{%
  \bibinfo {author} {\bibfnamefont{C.~G.}\ \bibnamefont{P\'eterfalvi}}, \bibinfo {author} {\bibfnamefont{A.}~\bibnamefont{David}}, \bibinfo {author} {\bibfnamefont{P.}~\bibnamefont{Rakyta}}, \bibinfo {author} {\bibfnamefont{G.}~\bibnamefont{Burkard}},\ and\ \bibinfo {author} {\bibfnamefont{A.}~\bibnamefont{Korm\'anyos}},\ }%
  \Doi{10.1103/PhysRevResearch.4.L022049}{\emph{\bibinfo {title} {Quantum interference tuning of spin-orbit coupling in twisted van der Waals trilayers}}},\ \bibinfo {journal} {Phys. Rev. Res.}\ \textbf{\bibinfo {volume} {4}},\ \bibinfo {pages} {L022049} (\bibinfo {year} {2022}).~%
  \bibAnnoteFile{Stop}{Csaba2022}%
\bibitem{ChouYZ2022a}%
  \BibitemOpen
  \bibfield{author}{%
  \bibinfo {author} {\bibfnamefont{Y.-Z.}\ \bibnamefont{Chou}}, \bibinfo {author} {\bibfnamefont{F.}~\bibnamefont{Wu}}, \bibinfo {author} {\bibfnamefont{J.~D.}\ \bibnamefont{Sau}},\ and\ \bibinfo {author} {\bibfnamefont{S.}~\bibnamefont{Das~Sarma}},\ }%
  \Doi{10.1103/PhysRevB.106.024507}{\emph{\bibinfo {title} {Acoustic-Phonon-Mediated Superconductivity in Moir\'eless Graphene Multilayers}}},\ \bibinfo {journal} {Phys. Rev. B}\ \textbf{\bibinfo {volume} {106}},\ \bibinfo {pages} {024507} (\bibinfo {year} {2022}).~%
  \bibAnnoteFile{Stop}{ChouYZ2022a}%
\bibitem{cakmak_continuous_2019}%
  \BibitemOpen
  \bibfield{author}{%
  \bibinfo {author} {\bibfnamefont{G.}~\bibnamefont{Çakmak}}\ and\ \bibinfo {author} {\bibfnamefont{T.}~\bibnamefont{Öztürk}},\ }%
  \Doi{10.1016/j.diamond.2019.05.002}{\emph{\bibinfo {title} {Continuous synthesis of graphite with tunable interlayer distance}}},\ \bibinfo {journal} {Diamond and Related Materials}\ \textbf{\bibinfo {volume} {96}},\ \bibinfo {pages} {134} (\bibinfo {year} {2019}).~%
  \bibAnnoteFile{Stop}{cakmak_continuous_2019}%
\bibitem{guo_electrochemical_2013}%
  \BibitemOpen
  \bibfield{author}{%
  \bibinfo {author} {\bibfnamefont{G.~F.}\ \bibnamefont{Guo}}, \bibinfo {author} {\bibfnamefont{H.}~\bibnamefont{Huang}}, \bibinfo {author} {\bibfnamefont{F.~H.}\ \bibnamefont{Xue}}, \bibinfo {author} {\bibfnamefont{C.~J.}\ \bibnamefont{Liu}}, \bibinfo {author} {\bibfnamefont{H.~T.}\ \bibnamefont{Yu}}, \bibinfo {author} {\bibfnamefont{X.}~\bibnamefont{Quan}},\ and\ \bibinfo {author} {\bibfnamefont{X.~L.}\ \bibnamefont{Dong}},\ }%
  \Doi{10.1016/j.surfcoat.2012.07.016}{\emph{\bibinfo {title} {Electrochemical hydrogen storage of the graphene sheets prepared by {DC} arc-discharge method}}},\ \bibinfo {journal} {Surface and Coatings Technology}\ ,\ \bibinfo {series} {Proceedings of the 8th {Asian}-{European} {International} {Conference} on {Plasma} {Surface} {Engineering} ({AEPSE} 2011)}\textbf{\bibinfo {volume} {228}},\ \bibinfo {pages} {S120} (\bibinfo {year} {2013}).~%
  \bibAnnoteFile{Stop}{guo_electrochemical_2013}%
\bibitem{zhao_unconventional_2020}%
  \BibitemOpen
  \bibfield{author}{%
  \bibinfo {author} {\bibfnamefont{B.}~\bibnamefont{Zhao}}, \bibinfo {author} {\bibfnamefont{B.}~\bibnamefont{Karpiak}}, \bibinfo {author} {\bibfnamefont{D.}~\bibnamefont{Khokhriakov}}, \bibinfo {author} {\bibfnamefont{A.}~\bibnamefont{Johansson}}, \bibinfo {author} {\bibfnamefont{A.~M.}\ \bibnamefont{Hoque}}, \bibinfo {author} {\bibfnamefont{X.}~\bibnamefont{Xu}}, \bibinfo {author} {\bibfnamefont{Y.}~\bibnamefont{Jiang}}, \bibinfo {author} {\bibfnamefont{I.}~\bibnamefont{Mertig}},\ and\ \bibinfo {author} {\bibfnamefont{S.~P.}\ \bibnamefont{Dash}},\ }%
  \Doi{10.1002/adma.202000818}{\emph{\bibinfo {title} {Unconventional {Charge}–{Spin} {Conversion} in {Weyl}‐{Semimetal} {WTe} $_{\textrm{2}}$}}},\ \bibinfo {journal} {Advanced Materials}\ \textbf{\bibinfo {volume} {32}},\ \bibinfo {pages} {2000818} (\bibinfo {year} {2020}).~%
  \bibAnnoteFile{Stop}{zhao_unconventional_2020}%
\bibitem{veneri_twist_2022}%
  \BibitemOpen
  \bibfield{author}{%
  \bibinfo {author} {\bibfnamefont{A.}~\bibnamefont{Veneri}}, \bibinfo {author} {\bibfnamefont{D.~T.~S.}\ \bibnamefont{Perkins}}, \bibinfo {author} {\bibfnamefont{C.~G.}\ \bibnamefont{Péterfalvi}},\ and\ \bibinfo {author} {\bibfnamefont{A.}~\bibnamefont{Ferreira}},\ }%
  \Doi{10.1103/PhysRevB.106.L081406}{\emph{\bibinfo {title} {Twist angle controlled collinear {Edelstein} effect in van der {Waals} heterostructures}}},\ \bibinfo {journal} {Physical Review B}\ \textbf{\bibinfo {volume} {106}},\ \bibinfo {pages} {L081406} (\bibinfo {year} {2022}).~%
  \bibAnnoteFile{Stop}{veneri_twist_2022}%
\bibitem{ChouYZ2020}%
  \BibitemOpen
  \bibfield{author}{%
  \bibinfo {author} {\bibfnamefont{Y.-Z.}\ \bibnamefont{Chou}}, \bibinfo {author} {\bibfnamefont{F.}~\bibnamefont{Wu}},\ and\ \bibinfo {author} {\bibfnamefont{S.}~\bibnamefont{Das~Sarma}},\ }%
  \Doi{10.1103/PhysRevResearch.2.033271}{\emph{\bibinfo {title} {Hofstadter Butterfly and {{Floquet}} Topological Insulators in Minimally Twisted Bilayer Graphene}}},\ \bibinfo {journal} {Phys. Rev. Res.}\ \textbf{\bibinfo {volume} {2}},\ \bibinfo {pages} {033271} (\bibinfo {year} {2020}).~%
  \bibAnnoteFile{Stop}{ChouYZ2020}%
\bibitem{biswas_su2-invariant_2011}%
  \BibitemOpen
  \bibfield{author}{%
  \bibinfo {author} {\bibfnamefont{R.~R.}\ \bibnamefont{Biswas}}, \bibinfo {author} {\bibfnamefont{L.}~\bibnamefont{Fu}}, \bibinfo {author} {\bibfnamefont{C.~R.}\ \bibnamefont{Laumann}},\ and\ \bibinfo {author} {\bibfnamefont{S.}~\bibnamefont{Sachdev}},\ }%
  \Doi{10.1103/PhysRevB.83.245131}{\emph{\bibinfo {title} {{SU}(2)-invariant spin liquids on the triangular lattice with spinful {Majorana} excitations}}},\ \bibinfo {journal} {Physical Review B}\ \textbf{\bibinfo {volume} {83}},\ \bibinfo {pages} {245131} (\bibinfo {year} {2011}).~%
  \bibAnnoteFile{Stop}{biswas_su2-invariant_2011}%
\bibitem{shtyk_electrons_2017}%
  \BibitemOpen
  \bibfield{author}{%
  \bibinfo {author} {\bibfnamefont{A.}~\bibnamefont{Shtyk}}, \bibinfo {author} {\bibfnamefont{G.}~\bibnamefont{Goldstein}},\ and\ \bibinfo {author} {\bibfnamefont{C.}~\bibnamefont{Chamon}},\ }%
  \Doi{10.1103/PhysRevB.95.035137}{\emph{\bibinfo {title} {Electrons at the monkey saddle: {A} multicritical {Lifshitz} point}}},\ \bibinfo {journal} {Physical Review B}\ \textbf{\bibinfo {volume} {95}},\ \bibinfo {pages} {035137} (\bibinfo {year} {2017}).~%
  \bibAnnoteFile{Stop}{shtyk_electrons_2017}%
\bibitem{efremov_multicritical_2019}%
  \BibitemOpen
  \bibfield{author}{%
  \bibinfo {author} {\bibfnamefont{D.~V.}\ \bibnamefont{Efremov}}, \bibinfo {author} {\bibfnamefont{A.}~\bibnamefont{Shtyk}}, \bibinfo {author} {\bibfnamefont{A.~W.}\ \bibnamefont{Rost}}, \bibinfo {author} {\bibfnamefont{C.}~\bibnamefont{Chamon}}, \bibinfo {author} {\bibfnamefont{A.~P.}\ \bibnamefont{Mackenzie}},\ and\ \bibinfo {author} {\bibfnamefont{J.~J.}\ \bibnamefont{Betouras}},\ }%
  \Doi{10.1103/PhysRevLett.123.207202}{\emph{\bibinfo {title} {Multicritical {Fermi} {Surface} {Topological} {Transitions}}}},\ \bibinfo {journal} {Physical Review Letters}\ \textbf{\bibinfo {volume} {123}},\ \bibinfo {pages} {207202} (\bibinfo {year} {2019}).~%
  \bibAnnoteFile{Stop}{efremov_multicritical_2019}%
\end{thebibliography}%

\end{document}